\documentclass[webpdf,contemporary,large,namedate]{oup-authoring-template}

\onecolumn 

\usepackage{graphicx}
\usepackage{amsmath}
\usepackage{natbib}
\usepackage{amssymb}
\usepackage{lmodern,bm} 

\usepackage{enumitem}
\usepackage{relsize}
\usepackage{exscale}
\newlist{steps}{enumerate}{1}
\setlist[steps, 1]{label = Step \arabic*:}
\usepackage[caption=false]{subfig}
\usepackage{float}
\usepackage{bigints}
\usepackage{multicol, caption}
\usepackage{graphicx}
\usepackage[]{subfig}
\usepackage{float} 
\usepackage{lipsum}
\usepackage{relsize}

\usepackage[bottom=1in,right=1.25in]{geometry}

\newcommand{\norm}[1]{\left\lVert#1\right\rVert}

\theoremstyle{thmstyleone}%
\theoremstyle{thmstyletwo}%
\theoremstyle{thmstylethree}%

\begin{document}

\journaltitle{Journal of the Royal Statistical Society Series C: Applied Statistics}
\DOI{DOI HERE}
\copyrightyear{2023}
\pubyear{2023}
\access{Advance Access Publication Date: Day Month Year}
\appnotes{Paper}

\firstpage{1}


\title[Data Fusion with INLA-SPDE]{A Data Fusion Model for Meteorological Data using the INLA-SPDE method}

\author[1,4,$\ast$]{Stephen Jun Villejo \ORCID{0000-0002-0510-3143}}
\author[2]{Sara Martino}
\author[3]{Finn Lindgren}
\author[1]{Janine B Illian}
\authormark{Villejo et al.}

\address[1]{\orgdiv{School of Mathematics and Statistics}, \orgname{University of Glasgow}, 
\orgaddress{\state{Scotland}, \country{United Kingdom}}}
\address[2]{\orgdiv{Department of Mathematical Sciences}, \orgname{Norwegian University of Science and Technology}, \orgaddress{\state{Trondheim}, \country{Norway}}}
\address[3]{\orgdiv{School of Mathematics, College of Science and Engineering}, \orgname{University of Edinburgh}, \orgaddress{\state{Scotland}, \country{United Kingdom}}}
\address[4]{\orgdiv{School of Statistics}, \orgname{University of the Philippines Diliman}, \orgaddress{\state{Quezon City}, \country{Philippines}}}

\corresp[$\ast$]{Corresponding author. \href{email:email-id.com}{stephen.villejo@glasgow.ac.uk}}

\received{Date}{0}{Year}
\revised{Date}{0}{Year}
\accepted{Date}{0}{Year}



\abstract{
 We present a data fusion model designed to address the problem of sparse observational data by incorporating numerical forecast models as an additional data source to improve predictions of key variables. This model is applied to two main meteorological data sources in the Philippines. The data fusion approach assumes that different data sources are imperfect representations of a common underlying process. Observations from weather stations follow a classical error model, while numerical weather forecasts involve both a constant multiplicative bias and an additive bias, which is spatially structured and time-varying.
To perform inference, we use a Bayesian model averaging technique combined with integrated nested Laplace approximation (INLA). The model’s performance is evaluated through a simulation study, where it consistently results in better predictions and more accurate parameter estimates than models using only weather stations data or regression calibration, particularly in cases of sparse observational data. In the meteorological data application, the proposed data fusion model also outperforms these benchmark approaches, as demonstrated by leave-group-out cross-validation (LGOCV).}

\maketitle

\section{Introduction}
A common aim in spatial statistical modelling is to predict values of a spatial variable at unsampled locations, based on measurements taken at a finite  -- and often relatively small --  number of locations. In the context of environmental sciences and meteorology in particular, data on e.g.\ air-quality, environmental pollution, surface temperature, or rainfall are collected through a network of monitoring stations \citep{lawson2016handbook, lee2017rigorous, blangiardo2016two, greven2011approach, chien2014impact, arab2014modelling, jaya2022spatiotemporal, lee2015impact}.
These data are used for prediction, and to improve the understanding of the spatio-temporal dynamics of the underlying processes and the impact of these variables on potential outcomes of interest, such as health outcomes. However, due to high maintenance costs, monitoring networks are typically spatially sparse \citep{lawson2016handbook}.  Increasingly,  data from additional sources derived from satellite images or outcomes of numerical models with a high spatial resolution are available. These can be used jointly with the monitoring stations data to improve the accuracy of predictions in a process that combines information from different data sources and is often referred to as \textit{data fusion} or \textit{data assimilation} \citep{bauer2015quiet,gettelman2022future,lawson2016handbook}. The goal is to exploit the better spatial resolution of the additional data  to fill gaps in areas only sparsely covered by the monitoring stations in order to predict and map the variables into space with more accuracy at smaller scales than based on the monitoring stations data alone. However, a general issue that is common in attempts to combine data from more than one source is that the various data streams differ in their quality. Numerical or satellite data, specifically, are often biased due to calibration issues, and these biases have to be accounted for in the modelling process.  
Motivated by a data application for meteorological data from the Philippines, we develop a joint modelling methodology that not only flexibly accounts for differences in data quality, but also allows us to gauge the quality of the different data sources.

This work combines two primary meteorological data sources for the Philippines: observational data from a sparse network of weather synoptic stations and simulated outputs from a numerical weather forecast model called the \textit{Global Spectrum Model} (GSM) \citep{PHClimate_PAGASA}. While the latter provides broad spatial coverage, it is typically biased due to sensitivity to model initialization and parameterization. In contrast, the weather stations data, which are likely to be less biased, provide limited spatial coverage, leaving key areas under-sampled, as shown in Figure 1. To address these limitations, we propose using both data sources together through data fusion. Unlike most existing methods discussed in Section \ref{subsec:curretapproaches}, our approach explicitly accounts for calibration biases, incorporates a flexible bias structure, defines a single interpretable latent process across data sources, and includes a measurement error model for the stations data.

A simulation study compares the performance of our proposed data fusion model to two benchmark methods: the stations-only model and the regression calibration model. In the data application, we compare the predictions  from the  data fusion approach and the two  benchmark approaches using leave-group-out cross-validation \citep{liu2022leave,adin2023automatic}. Although our application only considers two data sources, the model and  framework can easily extend to more than two data sources and are relevant beyond this specific context.

\subsection{Meteorological data from the Philippines }\label{sec:data_application}
\begin{figure}
        \centering
        \includegraphics[scale=.3]{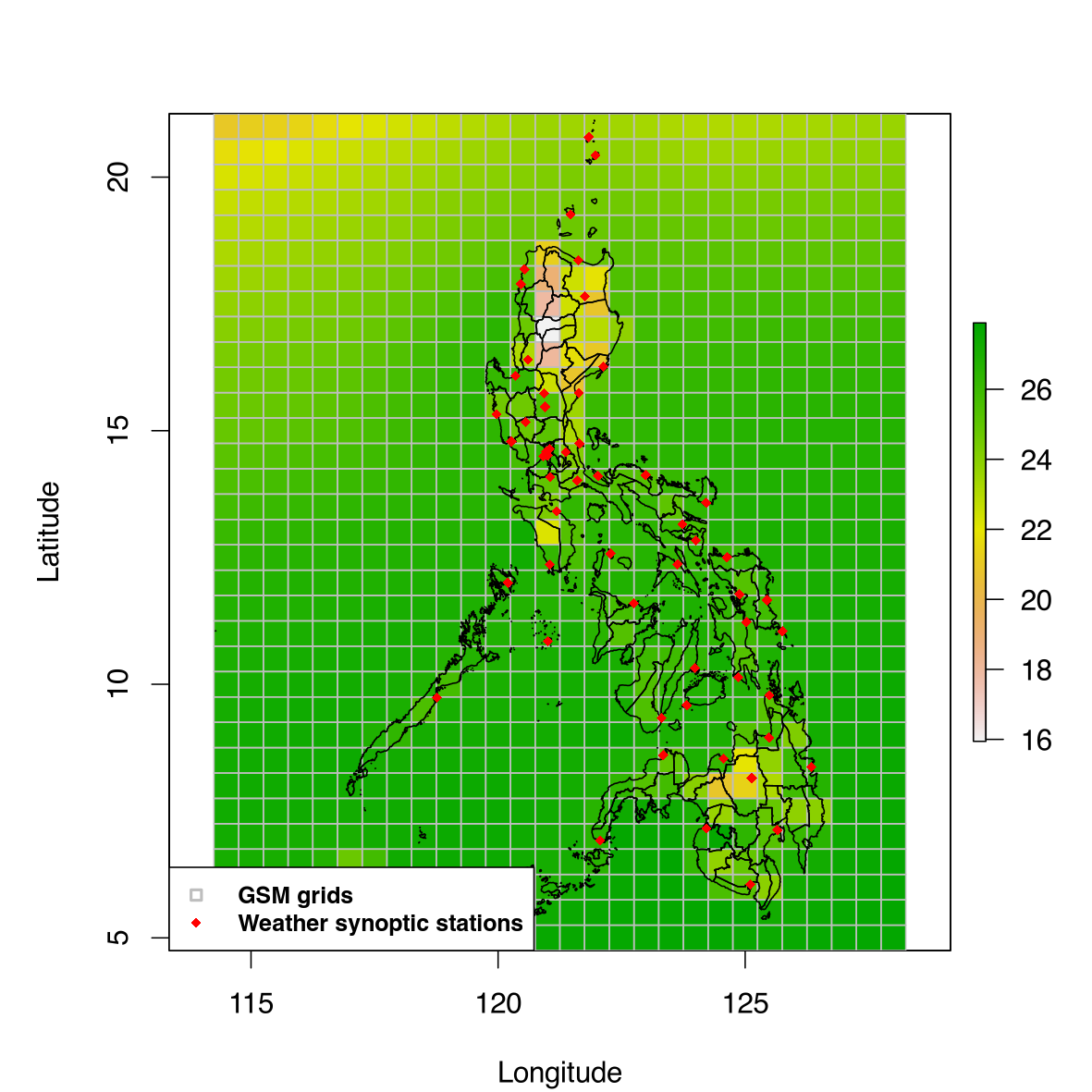}
        \caption{Meteorological data sources for the Philippines: a sparse network of weather synoptic stations and an outcome of a numerical weather forecast model called \textit{Global Spectrum Model}. The measurements are monthly aggregated values of temperature for August 2019. }
        \label{fig:climate_data}
    \end{figure}

The Philippines is an archipelagic country, covering an area of ca.\ 300 thousand km$^2$ (see Figure \ref{fig:climate_data}), situated in tropical Southeast Asia. The eastern part and some southern parts of the country are mostly classified as tropical rainforest, and is characterized by the lack of a distinct wet or dry season with relatively high rainfall all year round. On the other hand, most of the country's western section is classified as tropical monsoon or tropical Savannah, characterized by pronounced dry and wet seasons \citep{coronas1920climate, kintanar_climate, PHClimate_PAGASA}. The rainy season of the country, which also coincides with the hot episode of a year, lasts from June to November, while the rest of the year is generally considered dry. The dry season is further categorized into either a cool dry or a hot dry period, where the former lasts from December to February and the latter from March to May \citep{PHClimate_PAGASA}. 




The \textit{Philippine Atmospheric, Geophysical and Astronomical Services Administration} (PAGASA) maintains a network of 57 weather synoptic stations that regularly record several meteorological variables including temperature, relative humidity, and rainfall. 
The spatial distribution of the weather stations, shown in Figure \ref{fig:climate_data}, is a sparse network relative to the country's total surface area, with some regions,  especially in the north, being heavily  undersampled. Consequently, reconstructing meteorological variable surfaces based only on the weather stations would therefore result in high uncertainty in many parts of the country. To remedy this problem of data sparsity, PAGASA utilizes outcomes from the \textit{Global Spectrum Model} (GSM), a numerical weather forecast model maintained by the Japanese Meteorological Agency. The GSM provides forecast outputs of up to 132 hours four times a day (with initial times 0000, 0600, 1200, and 1800 UTC) within 4 hours of the initial time, and up to 264 hours twice a day (with initial time 0000 and 1200 UTC) within 7 hours of the initial time.  
As an illustration, Figure \ref{fig:climate_data} shows the map of mean temperature from the GSM for August 2019 at a spatial resolution of 0.5 degrees (approximately 55km $\times$ 55 km) corresponding to  924 grid cells. Although the GSM outcomes are gridded,  PAGASA interpret them as as point-referenced at the centroids \citep{PHClimate_PAGASA}. 

PAGASA provided the aggregated monthly data from both the weather stations and GSM for  January 2019 to December 2020, and for the following meteorological variables: mean temperature (in $^\circ\text{C}$), mean relative humidity (in \%), and total rainfall (in $mm$). The GSM outcomes were first simulated daily, using the 0000 UTC initial time, to produce forecasts at 3-hour intervals and up to eight forecast horizons. The simulated outcomes were then aggregated at the monthly level, yielding averages for temperature and relative humidity and totals for rainfall. The use of a monthly temporal scale is motivated by its relevance to future work, where the model predictions will serve as input to an epidemiological model for Dengue fever, as the case counts are typically available at the monthly level \citep{naish2014climate, abdullah2022association}. Therefore, the goal of this study is to reconstruct monthly surfaces for meteorological variables, specifically temperature, relative humidity, and rainfall. Here, the focus is on improving the accuracy of spatial predictions and mapping of these variables in space, rather than on forecasting future outcomes.

\begin{figure}[h!]
    \centering
    \includegraphics[scale=.23]{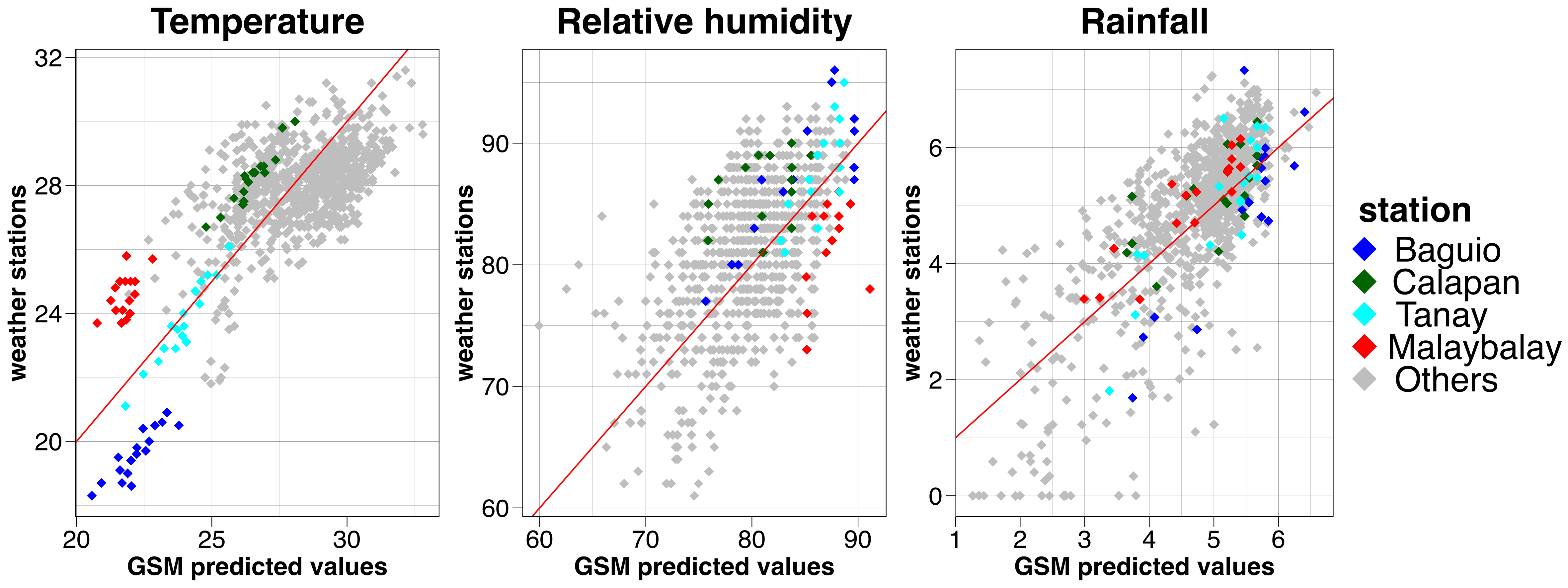}
    \caption{Scatterplot of the observed values at the weather stations versus interpolated outcomes of the GSM for three meteorological variables: temperature, relative humidity, and log-transformed rainfall. The plot shows the discrepancies in the values between the two data sources.}
    \label{fig:GSMvsStations}
\end{figure}

The GSM outcomes have a wider spatial coverage than the weather stations, but numerical weather forecasts are known to be biased \citep{lawson2016handbook, bauer2015quiet}. To assess this, we initially interpolate the GSM values on the three meteorological variables of interest -- temperature, relative humidity, and log-transformed rainfall amounts -- and predict the values at the  weather stations' locations. This is done by fitting the following geostatistical model:
\begin{align*}
\text{w}(\mathbf{s},t) = \beta_0+\beta_1z(\mathbf{s},t) + \xi(\mathbf{s},t) + e(\mathbf{s},t).
\end{align*}
Here, $\text{w}(\mathbf{s},t)$ is the simulated value of the meteorological variable from the GSM at the grid cell with centroid $\mathbf{s}$, $z(\mathbf{s},t)$ is a known covariate, $\xi(\mathbf{s},t)$ is a spatio-temporal random effect, and $e(\mathbf{s},t)$ is a random noise, i.e., $e(\mathbf{s},t)\overset{\text{iid}}{\sim}\mathcal{N}(0,\sigma^2_e)$. For the covariates, we used elevation for temperature and relative humidity, and relative humidity for the log-transformed rainfall. We assume that $\xi(\mathbf{s},t)$ evolves in time as an autoregressive process of order 1, i.e.,
\begin{equation*}
    \xi(\mathbf{s},t)=\phi\xi(\mathbf{s},t-1)+\omega(\mathbf{s},t),
\end{equation*}
where $|\phi|<1$ and  $\omega(\mathbf{s},t)$ is a time-independent Gaussian process with a Mat\'ern covariance function specified in Equations \eqref{eq:materncovariance1} and \eqref{eq:materncovariance2} and which follows the stationary distribution of the process at time $t=1$. We fit this model using INLA and the SPDE method \citep{rue2009approximate, lindgren2011explicit} and then predict the values of $\mathbb{E}[\text{w}(\mathbf{s},t)]$ at the weather stations' locations using the posterior predictive mean. We then compare the predicted values with the observed ones (see Figure \ref{fig:GSMvsStations}). The results show a general agreement between the two sets of values, but a clear bias is visible. Figure \ref{fig:GSMvsStations} highlights specific weather stations, where it becomes clear, especially for temperature, that there is a spatially-varying additive bias but no multiplicative bias, since the GSM outcomes at a specific weather station seem parallel to the $x=y$ line. For the other two meteorological variables, it is also clear that there is a spatially-varying additive bias, but accounting for a multiplicative bias parameter with a complex structure might be necessary. 

Figure \ref{fig:GSMvsStations} also shows that the quality of the GSM outcomes varies among the three variables. The discrepancy in the outcomes between the GSM and weather stations for the rainfall data is bigger than for the other two meteorological variables. The proposed data fusion model, which is discussed in Section \ref{sec:frameworkmodel}, is able to gauge the relative quality of the GSM outcomes for the three meteorological variables.


\subsection{Current data fusion approaches}\label{subsec:curretapproaches}

One statistical model for data fusion called \textit{Bayesian melding} was proposed by \cite{fuentes2005model} and discussed in \cite{lawson2016handbook}. The primary motivation in their work is modelling the concentration of air pollutants and the resulting public health implication. The Bayesian melding model is based on the assumption that all data sources are error-prone realizations of a common latent spatial process. Suppose $\text{w}_1(\mathbf{s})$ is the observed value of the quantity of interest  from a station with spatial location $\mathbf{s}$, and $\text{w}_2(B)$ is the realized outcome at a grid cell $B$ for the numerical forecast model. The Bayesian melding model assumes the following structure:
\begin{align}
    &x(\mathbf{s}) = \mu(\mathbf{s}) + \xi(\mathbf{s})\label{eq:melding_latent}\\
    &\text{w}_1(\mathbf{s}) = x(\mathbf{s})+e(\mathbf{s}) \label{eq:melding_w1}\\
    &\text{w}_2(\mathbf{s}) = \alpha_0(\mathbf{s}) + \alpha_1(\mathbf{s})x(\mathbf{s})+\delta(\mathbf{s})\label{eq:melding_w2s}\\
    &\text{w}_2(B) = \dfrac{1}{|B|}\int_B \text{w}_2(\mathbf{s})d\mathbf{s} .\label{eq:melding_w2B}
\end{align}
In Equation \eqref{eq:melding_latent}, $x(\mathbf{s})$ is the latent process of interest, $\mu(\mathbf{s})$ is the mean of the process, which is typically a function of fixed covariates, and $\xi(\mathbf{s})$ is a residual component, which can be spatially correlated. Equation \eqref{eq:melding_w1} links the observed outcomes at the stations $\text{w}_1(\mathbf{s})$ and the latent process $x(\mathbf{s})$. It follows the classical error model, i.e., $e(\mathbf{s})\overset{\text{iid}}{\sim} \mathcal{N}(0,\sigma^2_{e})$, where $e(\mathbf{s})$ is the error component with variance $\sigma^2_e$. Data from the numerical forecast model is treated as a gridded data, where $\text{w}_2(B)$ is the spatial average over grid cell $B$ of a conceptual point-referenced model $\text{w}_2(\mathbf{s})$ as shown in Equation \eqref{eq:melding_w2B}. 
Equation \eqref{eq:melding_w2s} specifies the biases in the numerical forecast model via the model components $\alpha_0(\mathbf{s})$ and $\alpha_1(\mathbf{s})$, which are interpreted as additive and multiplicative biases, respectively. It also involves another random noise component $\delta(\mathbf{s})\overset{\text{iid}}{\sim} \mathcal{N}(0,\sigma^2_{\delta})$, where $\sigma^2_{\delta}$ is the error variance. The main advantage of this specification is that it allows the two spatially misaligned data to jointly inform about the latent process $x(\mathbf{s})$. To avoid issues with identifiability, $\alpha_0(\mathbf{s})$ and $\alpha_1(\mathbf{s})$ are typically parameterized as fixed effects instead of spatial random fields. In particular, $\alpha_0(\mathbf{s})$ is a polynomial function of $\mathbf{s}$, while $\alpha_1(\mathbf{s})$ is an unknown constant. 

A similar idea was proposed in \cite{moraga2017geostatistical}. Given a zero-mean process $\xi(\mathbf{s})$ with a stationary covariance function, the model for the data outcomes $\text{w}_1(\mathbf{s})$ and $\text{w}_2(B)$ is given by 
\begin{align}\label{eq:moraga}
   \begin{split}
        \text{w}_1(\mathbf{s})|\xi(\mathbf{s}) &\sim \mathcal{N}\Big(\mu(\mathbf{s})+\xi(\mathbf{s}),\sigma^2_e\Big)\\
    \text{w}_2(B) &= \dfrac{1}{|B|}\int_B\Big(\mu(\mathbf{s})+\xi(\mathbf{s})\Big)d\bm{s}. 
   \end{split}
\end{align}
However, the model in Equation \eqref{eq:moraga} does not account for the measurement error in $\text{w}_2(B)$. The same model specification is used in \cite{zhong2023bayesian}. Although it uses the same idea as the Bayesian melding model in Equations $\eqref{eq:melding_latent}$ to $\eqref{eq:melding_w2B}$, where both $\text{w}_1(\mathbf{s})$ and $\text{w}_2(B)$ have a common latent process, it does not incorporate bias parameters such as $\alpha_0(\mathbf{s})$ and $\alpha_1(\mathbf{s})$. In a joint modelling framework, the data coming from $\text{w}_2(B)$ can dominate the parameter estimation since there are considerably more outcomes from this data source compared to $\text{w}_1(\mathbf{s})$ \citep{lawson2016handbook}. The calibration parameters in Equation \eqref{eq:melding_w2s} impose a restriction on this by accounting for a higher measurement error from this data source.  
These calibration parameters were accounted for in \cite{villejo2023data} using the same estimation strategy in \cite{moraga2017geostatistical} and \cite{cameletti2019bayesian}. However,  \cite{villejo2023data} assumes  that both $\alpha_0(\mathbf{s})$ and $\alpha_1(\mathbf{s})$ are constant in space and time. This assumption may limit the flexibility needed for effectively calibrating the numerical forecast model.

\cite{forlani2020joint} proposed another data fusion model, but instead of assuming a single latent process for the observed outcomes, they assumed several latent processes, which are shared across all the data sources. These different latent processes do not have a clear interpretation; in fact, these processes were simply referred to as spatial random effects. For instance, suppose $\text{w}_1(\mathbf{s})$, $\text{w}_2(\mathbf{s})$, and $\text{w}_3(\mathbf{s})$ are three data sources with mean $\mu_1(\mathbf{s})$, $\mu_2(\mathbf{s})$, and $\mu_3(\mathbf{s})$, respectively. Then the proposed model is as follows:
\begin{align}\label{eq:forlani}
   \begin{split}
        \mu_1(\mathbf{s}) &= \beta_1 + \xi_1(\mathbf{s})\\
    \mu_2(\mathbf{s}) &= \beta_2 + \lambda_2\xi_1(\mathbf{s}) + \xi_2(\mathbf{s})\\
     \mu_3(\mathbf{s}) &= \beta_3 + \lambda_3\xi_1(\mathbf{s}) + \lambda_4\xi_2(\mathbf{s}) + \xi_3(\mathbf{s})
   \end{split}
\end{align}
The parameters $\beta_1$, $\beta_2$, and $\beta_3$ are fixed effects while $\xi_i(\mathbf{s}), i=1,2,3$ are spatial random effects which are shared among the three data sources, and with $\lambda_j, j=2,3,4$ as unknown scaling parameters. This approach is also closely related to the so-called coregionalization model \citep{schmidt2003bayesian}. The Bayesian melding model in Equations \eqref{eq:melding_latent} to \eqref{eq:melding_w2B} assumes a single latent process which has a clear interpretation as the true process, and that the different data sources are error-prone realizations of the true process with varying levels of accuracy. However, the model in Equations \eqref{eq:forlani} deviates from this general principle.

One bottleneck with the Bayesian melding model is that it requires considerable computational effort because of the change-of-support integral in Equation \eqref{eq:melding_w2B}. A model which tries to overcome this difficulty was proposed in \cite{sahu2010fusing}. The model is specified as follows:
\begin{align}
\label{eq:sahu_model}
    \begin{split}
        x(\mathbf{s}) &= \tilde{x}(B)+\nu(\mathbf{s})\\
    \text{w}_1(\mathbf{s}) &= x(\mathbf{s})+e(\mathbf{s})\\
    \text{w}_2(B) &= \alpha_0 + \alpha_1\tilde{x}(B) + \psi(B),
    \end{split}
\end{align}
where $\tilde{x}(B)$ is considered as the true areal process and is defined on the same grid resolution as $\text{w}_2(B)$, $\nu(\mathbf{s})$ is a Gaussian error, and $\psi(B)$ is a discrete spatial effect which is commonly estimated using a conditionally autoregressive model \citep{besag1974spatial}. \cite{mcmillan2010combining} added further simplification by eliminating $x(\mathbf{s})$ in Equation \eqref{eq:sahu_model}. A limitation of both models is that they consider the underlying true process as discrete and that they can only provide gridded predictions of the true process. This approach is also referred to as \textit{upscaling} since the point-referenced data are coarsened to areal level \citep{lawson2016handbook}.

Another statistical approach for doing data fusion, called the regression calibration approach, fits a regression model using the outcomes of the numerical forecast model as a predictor and the observational data as the response variable. The model is given by
\begin{equation}\label{eq:regressioncalib}
    \text{w}_1(\mathbf{s}) = \alpha_0(\mathbf{s}) + \alpha_1(\mathbf{s})\text{w}_2(B_{\mathbf{s}}) + e(\mathbf{s}), \;\; e(\mathbf{s}) \overset{\text{iid}}{\sim}  \mathcal{N}(0,\sigma^2_e).
\end{equation}
Here, $\alpha_0(\mathbf{s})$ and $\alpha_1(\mathbf{s})$ are spatially-varying additive and multiplicative biases, respectively, while $\text{w}_2(B_{\mathbf{s}})$ is the value of the numerical forecast model at the grid $B$ which contains the point location $\mathbf{s}$. \cite{chen2021novel} used this approach to estimate chlorophyll-A concentration over eutrophic lakes,  \cite{lee2017rigorous} to model air quality, and \cite{berrocal2010spatio} to model ozone concentration. Obviously, there is a computational advantage with this approach since it only uses the values $\text{w}_2(B_{\mathbf{s}})$ linked to a stations data $\text{w}_1(\mathbf{s})$ to fit the model, and then uses the full data $\text{w}_2(B)$ for predictions. A limitation of the regression calibration approach is that it assumes that the observed measurements $\text{w}_1(\mathbf{s})$ are the gold standard, even though it is very likely to have instrumental errors. Also, since $\text{w}_2(B)$ is used as a predictor, then it cannot contain missing values \citep{lawson2016handbook}. However, remote-sensed data can be missing due to cloud cover and highly reflective surfaces. In addition, the resolution of the predicted latent surface for the quantity of interest depends on the resolution of $\text{w}_2(B)$. This approach is also referred to as \textit{downscaling} since it allows point-level predictions even if $w_2(B)$ represents an areal average. \cite{forlani2020joint} performed a comparison between the Bayesian melding model and the regression calibration approach in a specific data application. The results show that the Bayesian melding model gave better model predictions compared to the regression calibration approach. 

Unlike the approaches discussed so far, this work accounts for calibration biases in the data fusion model, specifies a flexible bias formulation, defines a single and interpretable latent process for the different data sources, and specifies a measurement error for the stations data. In particular, we specify a random field for the additive bias $\alpha_0(\mathbf{s})$ which we call an \textit{error field}. We use a flexible specification by assuming that the error field is a time-varying random field. Moreover, we assume that $\alpha_1$ is constant in space and time. Finally, our proposed model treats the outcomes of the numerical forecast model as point-referenced at the centroids of the grid cells $B$, which is the same strategy used in \cite{forlani2020joint}, \cite{villejo2023data}, and \cite{lee2017rigorous}. 

Monte Carlo methods have been the standard approach in doing inference with the Bayesian melding model \citep{falk2010estimating, vsevvcikova2011uncertain, liu2011empirical, poole2000inference}. In recent years, the use of the integrated nested Laplace approximation (INLA) paired with the stochastic partial differential equations (SPDE) approach has been widely used \citep{rue2009approximate, lindgren2011explicit}. The INLA method is a deterministic approach for doing Bayesian inference for latent Gaussian models. Moreover, the SPDE method is an efficient computational method to estimate Gaussian fields of the Mat\'ern class \citep{lindgren2011explicit}. A key challenge in the proposed model, which is discussed in Section \ref{subsec:model}, is the presence of non-linear components in the predictor expression, mainly due to the $\alpha_1$ multiplicative bias parameter. To address this complexity, we employ Bayesian model averaging with INLA for model inference \citep{gomez2020bayesian}.

Section \ref{sec:frameworkmodel} discusses the framework and the proposed data fusion model, while section \ref{sec:estimation} discusses the model estimation approaches.  Sections \ref{sec:simulation} and \ref{sec:application} discuss the results from a simulation study and the application to real data, respectively. 

\section{Data Fusion Framework and Model}\label{sec:frameworkmodel}
\subsection{Framework}\label{subsec:framework}
We assume that the latent process, at a spatial location $\mathbf{s}$ and time $t$, $\mathbf{s}\in\mathcal{S}$, $t=1,\dots,T$,   is denoted by $x(\mathbf{s},t)$.  This is observed via two different data sets. The first one is
\begin{equation*}
\textbf{\text{w}}_{1t}^\intercal = \begin{pmatrix} \text{w}_1(\mathbf{s}_1,t) & \text{w}_1(\mathbf{s}_2,t)& \ldots & \text{w}_1(\mathbf{s}_{n_M},t) \end{pmatrix},\;\;\;   \mathbf{s}_i\in\mathcal{S},
\end{equation*}
 which are observations  from a set of $n_M$ stations in locations $\mathbf{s}_i$, $i = 1,\dots,n_M$,
 at times $t = 1,\dots,T$. The second one is
\begin{equation*}
    \textbf{\text{w}}_{2t}^\intercal = \begin{pmatrix} \text{w}_2(\mathbf{g}_1,t) & \text{w}_2(\mathbf{g}_2,t)& \ldots & \text{w}_2(\mathbf{g}_{n_G},t) \end{pmatrix},\;\;\;    \mathbf{g}_j\in\mathcal{S}, 
\end{equation*}
which are gridded outcomes of a numerical model such as the GSM model. Here, $\text{w}_2(\mathbf{g}_j,t)$ is the value at the grid cell with centroid $\mathbf{g}_j$, $j = 1, \dots, n_G$, and time $t$.
We assume that the two data sources are aligned in time. If this is not the case, it is always possible to aggregate the data with higher time resolution. 
We further assume that $\textbf{\text{w}}_{2t}$  has a much wider spatial coverage than $\textbf{\text{w}}_{1t}$ ($n_M \ll n_G$) but more biased, and that both $\textbf{\text{w}}_{1t}$ and $\textbf{\text{w}}_{2t}$ are error-prone realizations of the same process of interest $\bm{x}_t$. This implies that:
\begin{equation}\label{eq:Data_genform}
\begin{split}
\textbf{\text{w}}_{1t} &= f_1(\bm{x}_t,\bm{\theta}_1) + \bm{e}_{1t} \\
     \textbf{\text{w}}_{2t} &= f_2(\bm{x}_t,\bm{\theta}_2) + \bm{e}_{2t},
\end{split}
\end{equation}
where $f_1(\cdot)$ and $f_2(\cdot)$ are some deterministic functions of the process $\bm{x}_t$ with bias parameters  $\bm{\theta}_1$ and  $\bm{\theta}_2$, respectively. The terms $\bm{e}_{1t}$ and $\bm{e}_{2t}$ are assumed independent error components. Typically, some simplifying assumptions are made on $f_1(\cdot)$ and $f_2(\cdot)$ to facilitate model inference. For instance, in the classical INLA approach \citep{rue2009approximate}, it is a requirement for the predictor to be a linear (deterministic) function of the latent Gaussian parameters; although recently, the class of models that can be fitted using INLA has been extended to those which are non-linear in the latent parameters using an iterative INLA approach and which can be easily implemented using the \texttt{inlabru} package in \texttt{R} \citep{lindgren2024inlabru}. Extending the above framework to more than two data sources is straightforward as a new data set would be treated as yet another error-prone realization of the latent process of interest.

\subsection{Proposed model}\label{subsec:model}
Using Equations \eqref{eq:Data_genform} to represent the two data sources, we propose the following data fusion model:
\begin{align}
\label{eq:model1}
    x(\mathbf{s},t) &=  \bm{\beta}^\intercal \bm{z}(\mathbf{s},t)+\xi(\mathbf{s},t)  \\
    \label{eq:model2}
    \text{w}_1(\mathbf{s}_i,t) &= x(\mathbf{s}_i,t) + e_1(\mathbf{s}_i,t),\;\;\; i=1,\ldots,n_M, \\
    \label{eq:model3}
    \text{w}_2(\mathbf{g}_j,t) &=  \alpha_0(\mathbf{g}_j,t) + \alpha_1 x(\mathbf{g}_j,t) + e_2(\mathbf{g}_j,t),\;\;\; j=1,\ldots,n_G. 
\end{align}

 The latent process of interest, $x(\mathbf{s},t)$, is modelled as a linear function of some known covariates $\bm{z}(\mathbf{s},t)$ (including an intercept) and a random field $\xi(\mathbf{s},t)$. The observed data $\text{w}_1(\mathbf{s}_i,t)$ from the weather stations are assumed to be unbiased realizations of the latent process $x(\mathbf{s},t)$ with an additive error term $e_1(\mathbf{s}_i,t) \overset{\text{iid}}{\sim} \mathcal{N}(0,\sigma^2_{e_1}), i=1,\ldots,n_M$, as in Equation \eqref{eq:model2}. The iid assumption is justified by the fact that the weather stations are sparsely located in the spatial domain and operate independently of each other. The data $\text{w}_2(\mathbf{g}_j,t)$ from the numerical weather forecast model are assumed to be biased realizations of the latent values $x(\mathbf{g}_j,t)$, $ j=1,\ldots,n_G$. We specify both an additive and a multiplicative bias for $\text{w}_2(\mathbf{g}_j,t)$. We assume that the multiplicative bias $\alpha_0(\mathbf{g}_j,t)$ varies both in space and time, and refer to is as the \textit{error field}. On the other hand, we assume the multiplicative bias $\alpha_1$ to be constant. For temperature, this is justified based on Figure \ref{fig:GSMvsStations}. For the other two meteorological variables, a spatially-varying multiplicative bias could better fit the data. Notice, however, that having a spatially-varying multiplicative bias in the model would pose a greater computational challenge as the model would contain a product of two random fields. For this reason, in this work, we have chosen to consider a constant multiplicative bias.
 Finally, the model for $\text{w}_2(\mathbf{g}_j,t)$ contains another unstructured error $e_2(\mathbf{g}_j,t) \overset{\text{iid}}{\sim} \mathcal{N}(0,\sigma^2_{e_2})$. 

Both the spatio-temporal field $\xi(\cdot,t)$ in Equation \eqref{eq:model1} and the error field $\alpha_0(\cdot,t)$ in Equation \eqref{eq:model3} are modelled using a Mat\'ern Gaussian space-time field. In particular, we assume that
\begin{equation}\label{eq:modelAR1}
\begin{split}
&\xi(\mathbf{s},t) = \phi_{1}\ \xi(\mathbf{s},t-1) + \omega_1(\mathbf{s},t)\\
    &\alpha_0(\mathbf{s},t)  = \phi_{2}\ \alpha_0(\mathbf{s},t-1) +  \omega_2(\mathbf{s},t),
\end{split}
\end{equation}
where $|\phi_{1}|<1$ and $|\phi_{2}|<1$ model the temporal dependence in a first order autoregressive (AR1) fashion,  while  $\omega_1(\mathbf{s},t)$ and $\omega_2(\mathbf{s},t)$ are time-independent Gaussian   innovation processes with Mat\'ern  covariance structure, i.e.,
\begin{align}
	&\text{Cov}\Big(\omega_h(\mathbf{s}_i,t),\omega_h(\mathbf{s}_j,u)\Big) = \begin{cases} 
            0 & t \neq u \\
  			\Sigma^{(h)}_{i,j}& t = u
  		\end{cases} \label{eq:materncovariance1}\\
    &\Sigma^{(h)}_{i,j}= \frac{\sigma^2_{h}}{2^{\nu_h-1}\Gamma(\nu_h)}\big(\kappa_h\norm{\mathbf{s}_i-\mathbf{s}_j}\big)^{\nu_h}K_{\nu_h}\big(\kappa_h\norm{\mathbf{s}_i-\mathbf{s}_j}\big), \label{eq:materncovariance2}
\end{align}
for $h = 1,2$.
Here, $\norm{\cdot}$ is the Euclidean distance in $\mathbb{R}^2$ between two locations  $\mathbf{s}_i$ and $\mathbf{s}_j$, and $K_{\nu_h}(\cdot)$ is the modified Bessel function of the second kind and order $\nu_h>0$.  The Mat\'ern field is parameterized by the marginal variance $\sigma^2_{h}$, a scaling parameter $\kappa_h$, and a smoothness parameter $\nu_h$ which is related to the mean-square differentiability of the process. The smoothness parameter is typically fixed at some value because it is poorly identified in many applications \citep{lindgren2011explicit}. The scaling parameter is related to the range parameter $\rho_h$ via the empirically derived relationship $\rho_{h} \approx \dfrac{\sqrt{8\nu_{h}}}{\kappa_{h}}$. 
We use the stochastic partial differential equations (SPDE) approach to represent the Mat\'ern fields in the model as Gaussian Markov random fields, which yields a sparse precision matix consequently making the computation efficient \citep{lindgren2011explicit}.

Given the model structure in Equations \eqref{eq:modelAR1} to \eqref{eq:materncovariance2}, we have $\bm{\xi}_t|\bm{\xi}_{t-1}\sim \mathcal{N}(\phi_1\bm{\xi}_{t-1},\bm{\Sigma}^{(1)})$ and $\bm{\alpha}_{0_t}|\bm{\alpha}_{0_{t-1}}\sim \mathcal{N}(\phi_2\bm{\alpha}_{0_{t-1}},\bm{\Sigma}^{(2)})$ for $t=2,\ldots,T$, where $\bm{\xi}_t = \begin{pmatrix}
    \xi(\mathbf{s}_1,t) & \ldots & \xi(\mathbf{s}_{n_M},t)
\end{pmatrix}^\intercal$, $\bm{\alpha}_{0_t} = \begin{pmatrix}
    \alpha_0(\mathbf{g}_1,t) & \ldots & \alpha_0(\mathbf{g}_{n_G},t)
\end{pmatrix}^\intercal$, and that  $\bm{\Sigma}^{(1)}$ and $\bm{\Sigma}^{(2)}$ are dense covariance matrices whose elements are given in Equation \eqref{eq:materncovariance2}. Both $\xi(\mathbf{s},t)$ and $\alpha_0(\mathbf{s},t)$ follow the stationary distribution at $t=1$, i.e., $\xi(\mathbf{s},1)\sim \mathcal{N}\Big(0,\sigma^2_{1}/(1-\phi_1^2)\Big)$ and $\alpha_0(\mathbf{g},1)\sim \mathcal{N}\Big(0,\sigma^2_{2}/(1-\phi_2^2)\Big)$. 

We compare the performance of the proposed model to two benchmark approaches. The first one, denoted  \textit{stations-only model}, uses only the data from the stations. This model essentially fits Equations \eqref{eq:model1} and \eqref{eq:model2} only. The second benchmark model is the regression calibration model shown in Equation \eqref{eq:regressioncalib} in Section \ref{subsec:curretapproaches}. In the spatio-temporal scenario, we assume that the additive and multiplicative biases in the regression calibration model are both spatially and temporally-varying. 

\section{Model Estimation}\label{sec:estimation} 
For estimation, we use the integrated nested Laplace approximation (INLA), a method for  deterministic Bayesian inference for latent Gaussian models \citep{rue2009approximate}. The stations-only model and the regression calibration model are standard spatial models and are straightforward to estimate using INLA \citep{cameletti2013spatio}. On the other hand, the proposed model in Equations \eqref{eq:model1} to \eqref{eq:model3} can be tricky. It is useful to rewrite the model, in vector form, as follows:
\begin{equation}\label{eq:estimationmodel}
\begin{split}
\textbf{\text{w}}_{1t} &= \bm{Z}_t\bm{\beta} + \bm{\xi}_t + \bm{e}_{1t},\;\;\; \bm{e}_{1t}\sim \mathcal{N}(\bm{0},\sigma^2_{e_1}\mathbb{I}) \\
  &\;\;\; \bm{\xi}_t = \phi_1\bm{\xi}_{t-1} + \bm{\omega}_{1t}\\
    \textbf{\text{w}}_{2t} &=  {\bm{\alpha}}_{0_t} + \alpha_1(\bm{Z}_t\bm{\beta} + \bm{\xi}_t) + \bm{e}_{2t}, \;\;\; \bm{e}_{2t}\sim \mathcal{N}(\bm{0},\sigma^2_{e_2}\mathbb{I})\\ &\;\;\; \bm{\alpha}_{0_t} = \phi_2\bm{\alpha}_{0_{t-1}} + \bm{\omega}_{2t}.
\end{split}
\end{equation}
The model specification involves two likelihood components: $\textbf{\text{w}}_{1t}$ and $\textbf{\text{w}}_{2t}$. The latent part of the model includes the fixed effects $\bm{\beta}$, the space-time effects $\bm{\xi}_t$, and the error field ${\bm{\alpha}}_{0_t}$, $t=1,\ldots, T$. The fixed effects $\bm{\beta}$ are given a non-informative Gaussian prior, while the random fields $\bm{\xi}_t$ and ${\bm{\alpha}}_{0_t}$ follow a Gaussian autoregressive structure as described in Equations \eqref{eq:modelAR1} to \eqref{eq:materncovariance2}. The hyperparameters include the multiplicative bias $\alpha_1$, the parameters linked to  $\bm{\xi}_{1t}$ ($\sigma_{1},\rho_1,$ and $\phi_1$), the parameters linked to $\bm{\alpha}_{0_t}$ ($\sigma_{2},\rho_2$, and $\phi_2$), and the measurement error variance parameters $\sigma^2_{e_1}$ and $ \sigma^2_{e_2}$. The model hyperparameters, except for $\alpha_1$, are given penalized complexity (PC) priors \citep{fuglstad2019constructing, simpson2017penalising}. PC priors are weakly informative priors which penalize the complexity of Gaussian random fields by shrinking the range towards infinity and the marginal variance towards zero. These are defined and expressed through probability statements of the type $\mathbb{P}(\sigma>\sigma_\text{o}) = \zeta_1$ and $\mathbb{P}(\rho<\rho_\text{o})= \zeta_2$, where $\zeta_1,\zeta_2 \in (0,1)$ are probability values chosen by the user, while $\sigma_\text{o}$ and $\rho_\text{o}$ are user-defined values of the standard deviation and range parameter, respectively.


The proposed data fusion model falls in the class of models that can be fitted using INLA since, given the hyperparameters, the latent field is Gaussian. However, estimating $\alpha_1$, which acts as a scaling parameter for the Gaussian field  $\bm{Z}_t\bm{\beta}+\bm{\xi}_t$, can be difficult as the optimizer could run into numerical issues. Hence, we explore the use of a Bayesian model averaging (BMA) approach with INLA \citep{gomez2020bayesian}. This approach fits the data fusion model conditional on $\alpha_1$, and then averages all the conditional INLA models to obtain the final posterior estimates.
In addition, it is easy to determine a reasonable set of values for $\alpha_1$: a value of 1 implies that the numerical model has no multiplicative bias, and the further the value of $\alpha_1$  from 1, the more serious the multiplicative bias. In the data application, we do not expect $\alpha_1$ to be very far from 1, particularly for temperature and relative humidity; thus, we defined a grid of $\alpha_1$ values from 0.5 to 1.5 and with length step of 0.1.

The Bayesian model averaging with INLA is described as follows. Suppose all observed data is denoted by $\mathbf{Y}$, a latent parameter is denoted by $x_j$, and the hyperparameters are denoted by $\bm{\theta}$. INLA computes the posterior marginals based on the following integrals:
\begin{equation*}\label{eq:INLA_posterior}
\begin{split}
\pi(\theta_i|\mathbf{Y}) &= \int\pi(\boldsymbol{\theta}|\mathbf{Y})d\boldsymbol{\theta}_{-i} \\
         \pi(x_j|\mathbf{Y}) &= \int\pi(x_j|\boldsymbol{\theta},\mathbf{Y})\pi(\boldsymbol{\theta}|\mathbf{Y})d\boldsymbol{\theta},
\end{split}
\end{equation*}
where $\bm{\theta}_{-i}$ denotes the vector of hyperparameters excluding $\theta_i$. We express $\bm{\theta} = \begin{pmatrix}
    \alpha_1 & \bm{\theta}_{-\alpha_1} 
\end{pmatrix}^{\intercal}$, where $\bm{\theta}_{-\alpha_1}$ includes all hyperparameters excluding $\alpha_1$. In this regard, the posterior marginals of $\bm{x}$ and $\bm{\theta}_{-\alpha_1}$ can be expressed as
\begin{align}\label{eq:INLA_BMA}
         \pi(\cdot|\mathbf{Y}) &= \int \pi(\cdot,\alpha_1|\mathbf{Y})d\alpha_1 =\int\pi(\cdot|\alpha_1,\mathbf{Y})\pi(\alpha_1|\mathbf{Y})d\alpha_1.
\end{align}
The probability density $\pi(\cdot|\alpha_1,\mathbf{Y})$ is a conditional marginal posterior which can be easily estimated using INLA for a fixed $\alpha_1$. The density $\pi(\alpha_1|\mathbf{Y})$ in Equation \eqref{eq:INLA_BMA} can be expressed as $\pi(\alpha_1|\mathbf{Y}) \propto \pi(\mathbf{Y}|\alpha_1)\pi(\alpha_1)$, where $\pi(\mathbf{Y}|\alpha_1)$ is the conditional marginal likelihood and $\pi(\alpha_1)$ is the prior for $\alpha_1$. The computation is done by specifying a grid of values for $\alpha_1$, say $\alpha_1^{(k)}, k=1,\ldots,K$, and then estimating the model conditional on each $\alpha_1^{(k)}$. Given this ensemble of INLA models, the weights for model averaging are computed as:
\begin{equation}\label{eq:weightsBMA}
    w_k = \dfrac{\pi\Big(\mathbf{Y}|\alpha_1^{(k)}\Big) \pi\Big(\alpha_1^{(k)}\Big)}{\sum_{k=1}^K \pi\Big(\mathbf{Y}|\alpha_1^{(k)}\Big) \pi\Big(\alpha_1^{(k)}\Big)}.
\end{equation}
The marginal posteriors given in Equation \eqref{eq:INLA_BMA} are then computed as
\begin{equation}\label{eq:BMAformula}
    \pi\Big(\cdot|\mathbf{Y}\Big) \approx \sum_{k=1}^K \pi\Big(\cdot|\alpha_1^{(k)},\mathbf{Y}\Big)w_k. 
\end{equation}
All other posterior quantities of interest are computed using model averaging. For example, the predicted $x(\mathbf{s},t)$ field is given by the following
\begin{equation}\label{eq:pred_field_x}
    \hat{x}(\mathbf{s},t) = \sum_{k=1}^K \Bigg\{\mathbb{E}\Big[\bm{\beta}|\alpha_1^{(k)},\mathbf{Y}\Big]^{\intercal}\bm{z}(\mathbf{s},t) + \mathbb{E}\Big[\bm{\xi}|\alpha_1^{(k)},\mathbf{Y}\Big]\Bigg\}w_k,
\end{equation}
where $\mathbb{E}\Big[\cdot|\alpha_1^{(k)},\mathbf{Y}\Big]$ is evaluated with respect to the conditional marginal posteriors $\pi\Big(\bm{\beta}|\alpha_1^{(k)},\mathbf{Y}\Big)$ and $\pi\Big(\bm{\xi}|\alpha_1^{(k)},\mathbf{Y}\Big)$. Equation \eqref{eq:pred_field_x} is equivalent to
\begin{equation}\label{eq:pred_field_x_v2}
    \begin{split}
        \hat{x}(\mathbf{s},t) &= \sum_{k=1}^K \Bigg\{\int\bm{\beta}\pi\Big(\bm{\beta}|\alpha_1^{(k)},\mathbf{Y}\Big)w_kd\bm{\beta}\Bigg\}^{\intercal}\bm{z}(\mathbf{s},t) + \sum_{k=1}^K \Bigg\{\int\bm{\xi}\pi\Big(\bm{\xi}|\alpha_1^{(k)},\mathbf{Y}\Big)w_kd\bm{\xi}\Bigg\}\\
    &= \Bigg\{\int\bm{\beta}\Big[\sum_{k=1}^K \pi\Big(\bm{\beta}|\alpha_1^{(k)},\mathbf{Y}\Big)w_k\Big]d\bm{\beta}\Bigg\}^{\intercal}\bm{z}(\mathbf{s},t) + \int\bm{\xi}\Big[\sum_{k=1}^K\pi\Big(\bm{\xi}|\alpha_1^{(k)},\mathbf{Y}\Big)w_k\Big]d\bm{\xi}\\
    &= \Bigg\{\int\bm{\beta}\pi(\bm{\beta}|\mathbf{Y})d\bm{\beta}\Bigg\}^{\intercal}\bm{z}(\mathbf{s},t) +  \int\bm{\xi}\pi(\bm{\xi}|\mathbf{Y})w_kd\bm{\xi} \\
    &= \mathbb{E}[\bm{\beta}|\mathbf{Y}]^\intercal\bm{z}(\mathbf{s},t) + \mathbb{E}[\bm{\xi}|\mathbf{Y}],
    \end{split}
\end{equation}
where $\mathbb{E}[\cdot|\mathbf{Y}]$ is evaluated with respect to the marginal posteriors which are approximated via Equation \eqref{eq:BMAformula}. Equation \eqref{eq:pred_field_x} implies that the predicted field $\hat{x}(\mathbf{s},t)$ can be computed by evaluating the predictor expression for $x(\mathbf{s},t)$ given in Equation \eqref{eq:model1} using the mean of the conditional marginal posteriors and then getting the weighted average using the weights $w_k$, while Equation \eqref{eq:pred_field_x_v2} implies that it is equivalent to directly evaluating Equation \eqref{eq:model1} using the mean of the marginal posteriors in Equation \eqref{eq:BMAformula}. A disadvantage of the model averaging approach is that it requires us to fit the models conditional on each $\alpha_1$ value which can be inefficient especially for large spatio-temporal datasets.

\section{Simulation Study}\label{sec:simulation}
In this section, we perform a simulation study to assess the performance of the proposed data fusion model in comparison to two benchmark approaches: stations-only model and regression calibration model. We use the Belo horizonte region in Brazil as our study domain whose shapefile is available in the \texttt{R} package \texttt{spdep} \citep{bivand2015comparing}. This region has a total surface area of ca. 330 $\text{km}^2$. We perform the simulation study in a purely spatial context and simulate the process of interest as:
\begin{equation*}
    x(\mathbf{s}) = \beta_0 + \beta_1 z(\mathbf{s})+\xi(\mathbf{s}) 
\end{equation*}
where $\xi(\mathbf{s})$ is a Mat\'ern random field with the following parameters: effective range $\rho_{\xi} = 2$ degrees, marginal standard deviation $\sigma_{\xi} = 3.16$, and smoothness parameter $\nu_{\xi}=1$. The value of the range is such that the spatial correlation becomes negligible at a distance of ca 222 km which corresponds to half of the maximum distance in the study region.  Moreover, $z(\mathbf{s})$ is a known covariate and is simulated from a Mat\'ern process with effective range of 3 degrees, a marginal variance of 1, and a smoothness parameter equal to 1. The fixed effects are $\beta_0=10$ and $\beta_1=3$. 


The two observed datasets are simulated using Equations \eqref{eq:model2} and \eqref{eq:model3}, respectively, but without the time component. We set $e_1(\mathbf{s}_i)\overset{\text{iid}}{\sim} \mathcal{N}\big(0, \sigma^2_{e_1} = .25\big)$ and $e_2(\mathbf{g}_j)\overset{\text{iid}}{\sim} \mathcal{N}\big(0, \sigma^2_{e_2} = .01\big)$. The values of $\sigma^2_{e_1}$ and $\sigma^2_{e_2}$ are chosen based on the empirical results from the temperature model in Section \ref{subsec:res_meantemp}. This implies that the noise in $\text{w}_2(\mathbf{g}_j)$ is mainly attributed to the error field $\alpha_0(\mathbf{g})$ which is simulated from a Mat\'ern process with range $\rho_{\alpha_0} = 1$, marginal standard deviation $\sigma_{\alpha_0} = 1$, and smoothness parameter $\nu_{\alpha_0}=1$. The range parameter and marginal variance of $\xi(\mathbf{s})$ is chosen to be higher compared to $\alpha_0(\mathbf{g})$ based on the empirical results from the real data application discussed in Section \ref{sec:application}. Finally, the constant multiplicative bias parameter is $\alpha_1=1.1$.

\begin{figure}[h]
     \centering
     \subfloat[][Simulation grid]{\includegraphics[clip,scale=0.16]{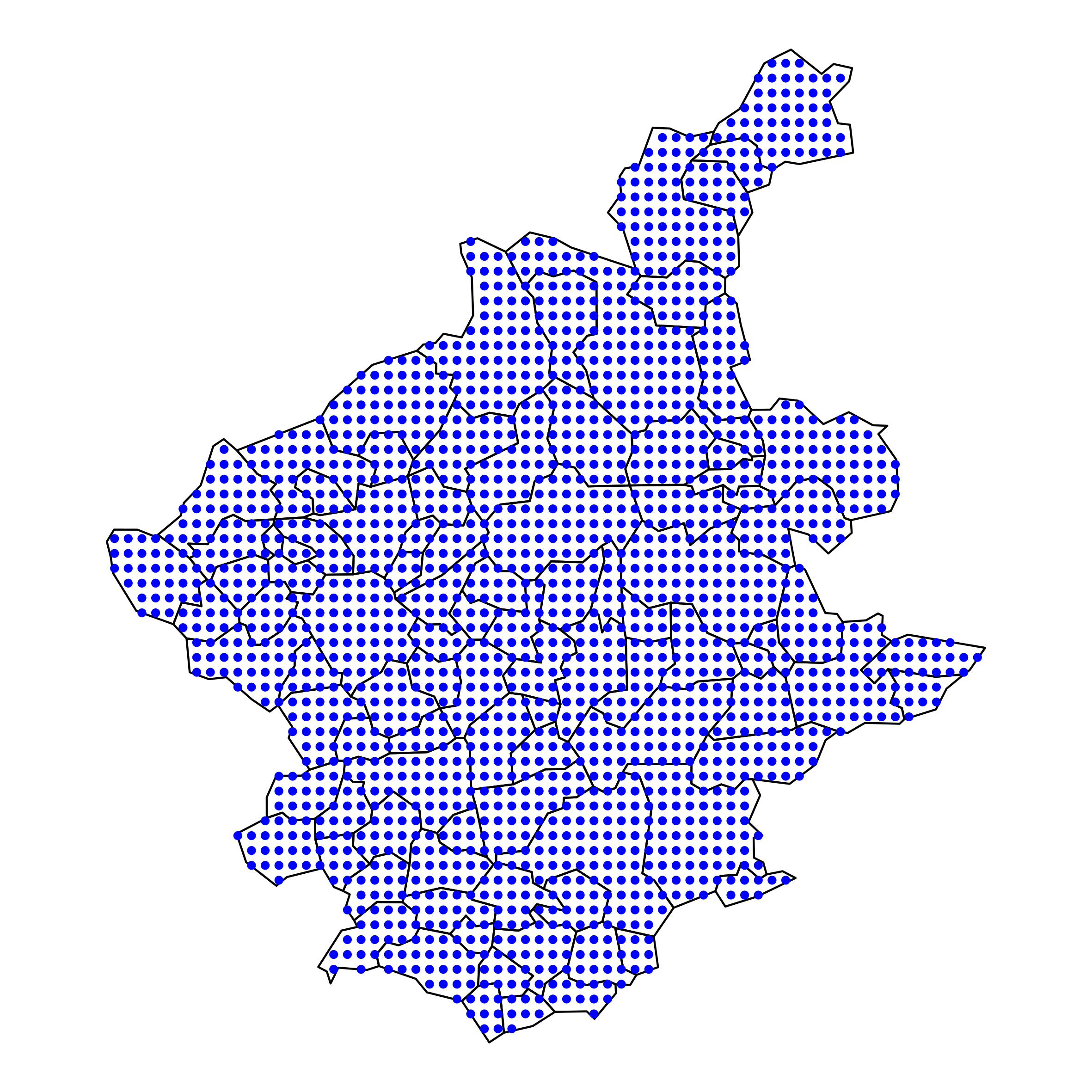}\label{fig:sim_grid}}
     \subfloat[][$x(\mathbf{s})$]{\includegraphics[clip,scale=0.16]{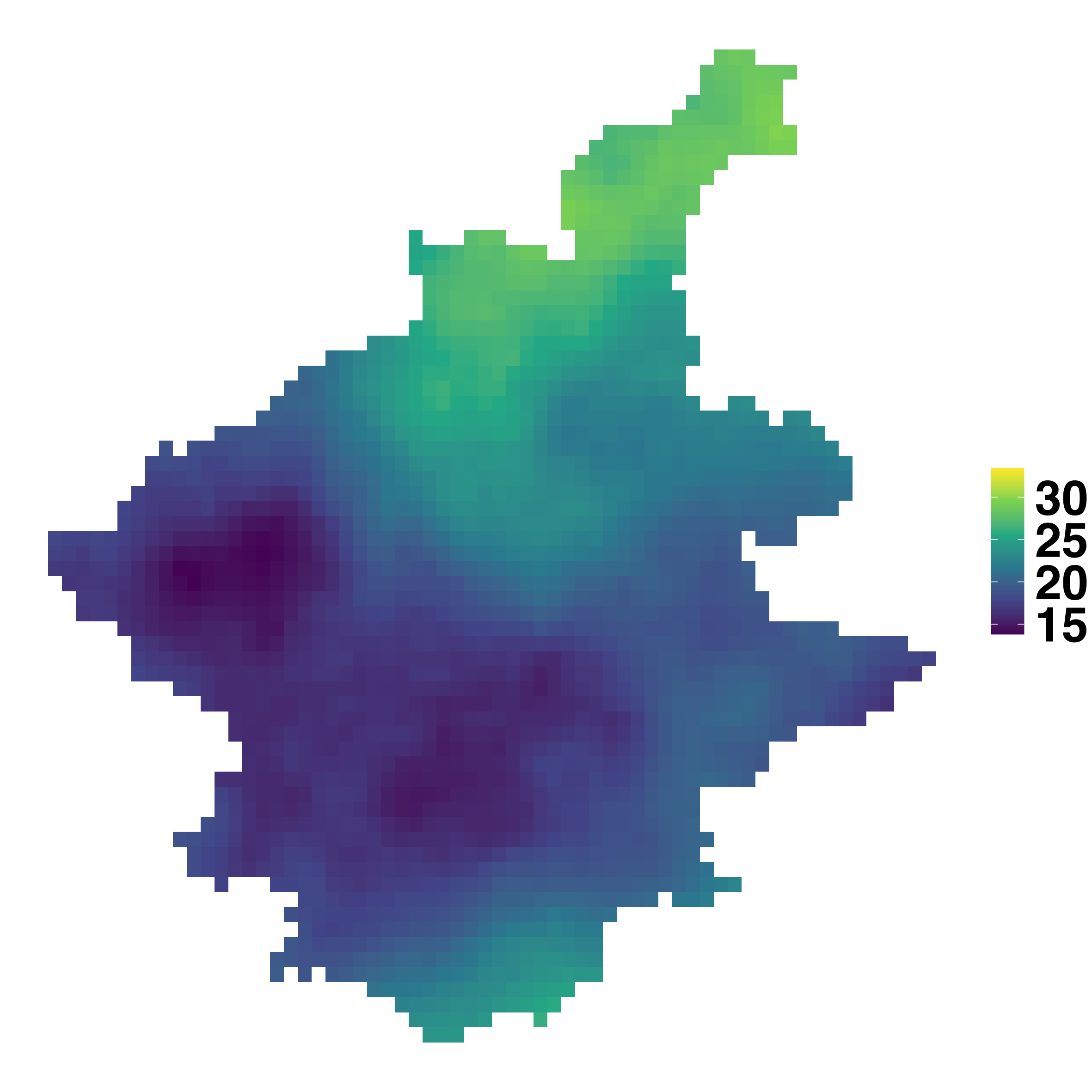}\label{fig:sim_dgp_x}}
     \subfloat[][$\text{w}_2(\mathbf{g}_j)$]{\includegraphics[,clip,scale=0.16]{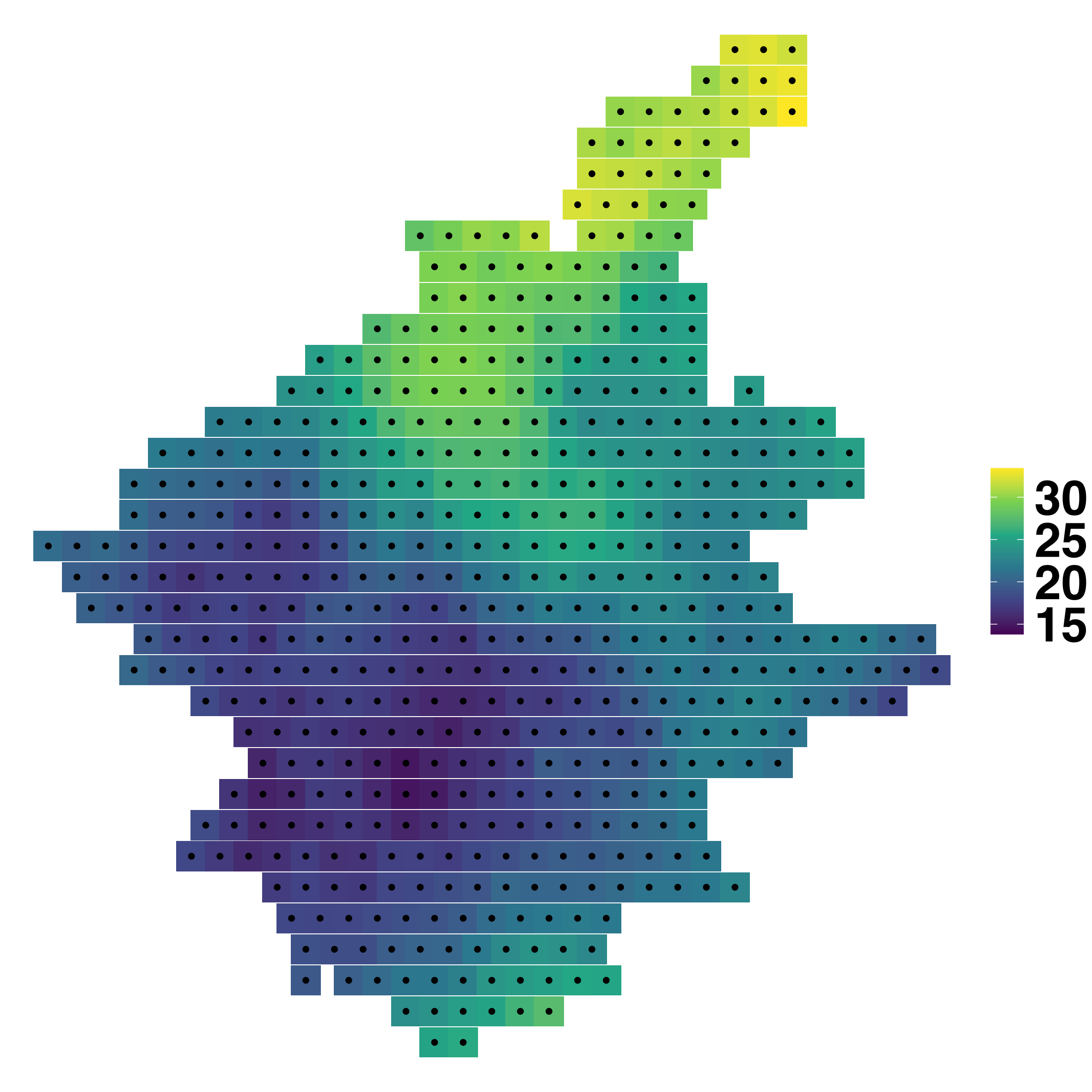}\label{fig:sim_dgp_w2}} 
     \subfloat[][ $\alpha_0(\mathbf{g})$]{\includegraphics[clip,scale=0.16]{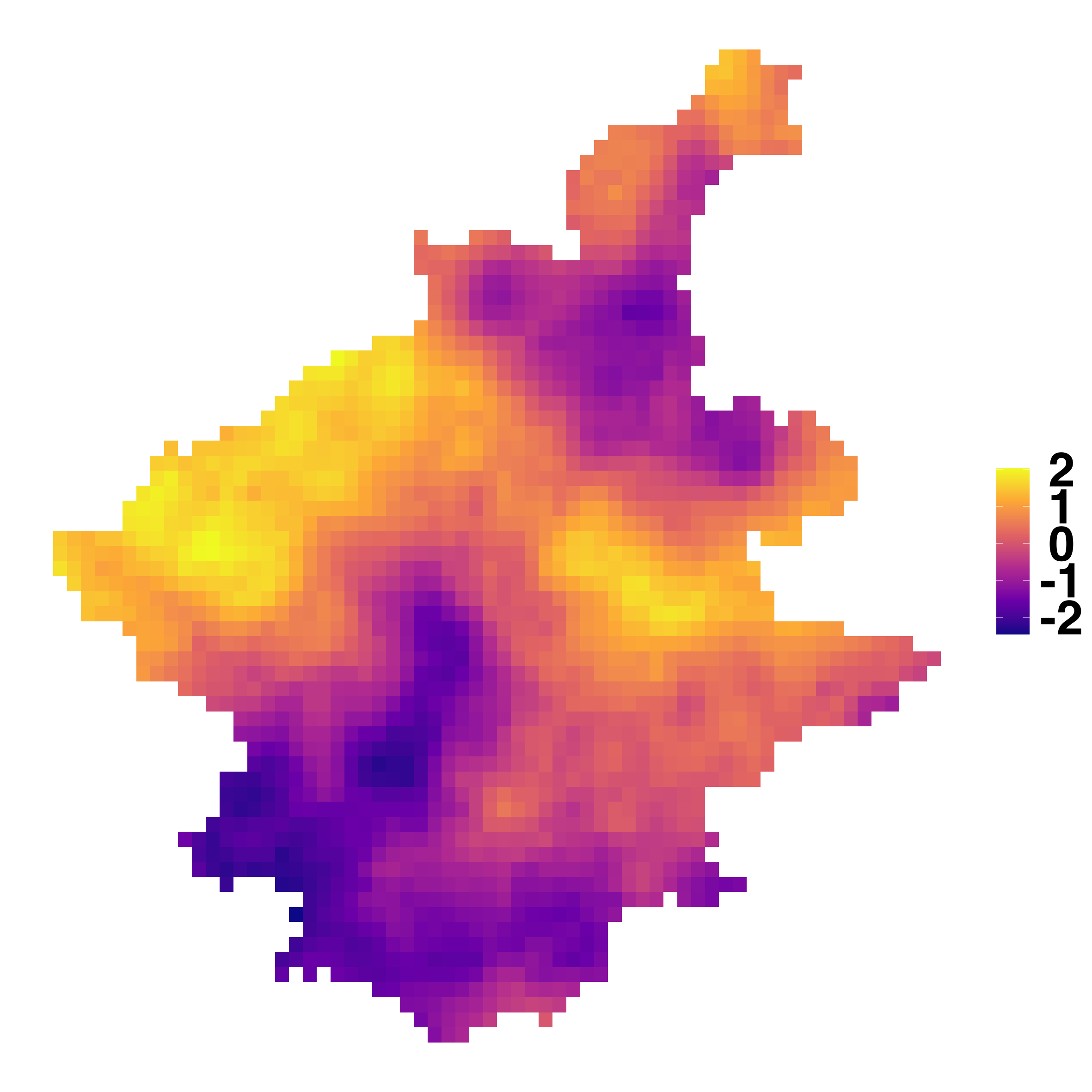}\label{fig:sim_dgp_error}}
    \caption{(a) dense simulation grid, (b) a simulated true field $x(\mathbf{s})$, (c) a simulated outcome from the numerical model $\text{w}_2(\mathbf{g}_j)$, (d) a simulated error field $\alpha_0(\mathbf{g})$}
     \label{fig:sim_sample_DGP}
\end{figure}
Figure \ref{fig:sim_grid} shows the dense simulation grid. Figure \ref{fig:sim_dgp_x} shows a simulated $x(\mathbf{s})$ field, while Figure \ref{fig:sim_dgp_w2} shows a simulated outcome from the numerical forecast model which is at a coarser resolution than the simulation grid. In particular, the centroids of the grid cells in Figure \ref{fig:sim_dgp_w2} is a coarse subset of the points in Figure \ref{fig:sim_grid}. The corresponding simulated error field $\alpha_0(\mathbf{g})$ is shown in Figure \ref{fig:sim_dgp_error}. 

\begin{figure}[t]
     \centering
     \subfloat[][$\text{w}_1(\mathbf{s}_i)$ vs $x(\mathbf{s}_i)$]{\includegraphics[clip,scale=0.2]{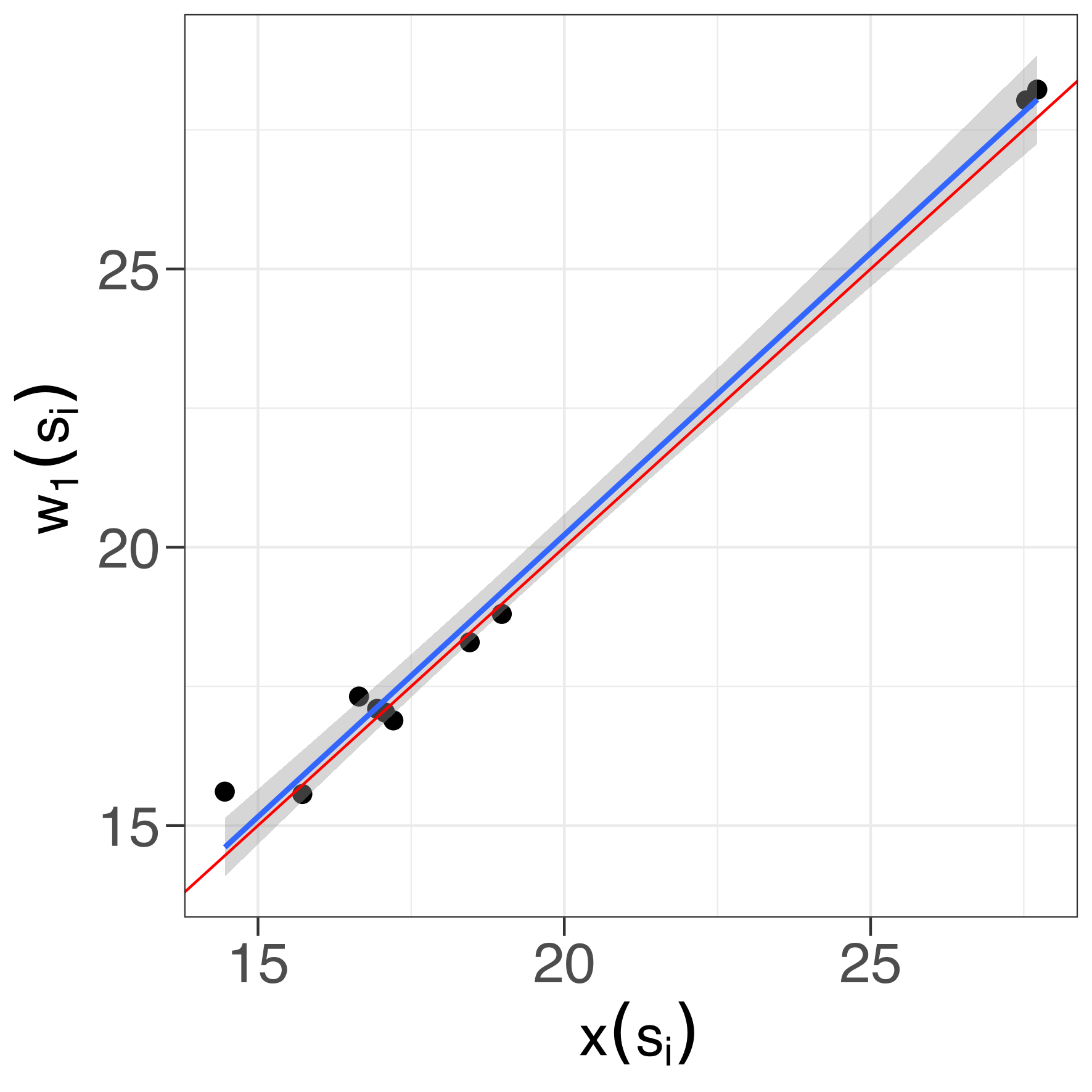}\label{fig:sim_dgp_scatter1}}
     \hspace{5mm}
     \subfloat[][$\text{w}_2(\mathbf{g}_j)$ vs $x(\mathbf{g}_j)$]{\includegraphics[clip,scale=0.2]{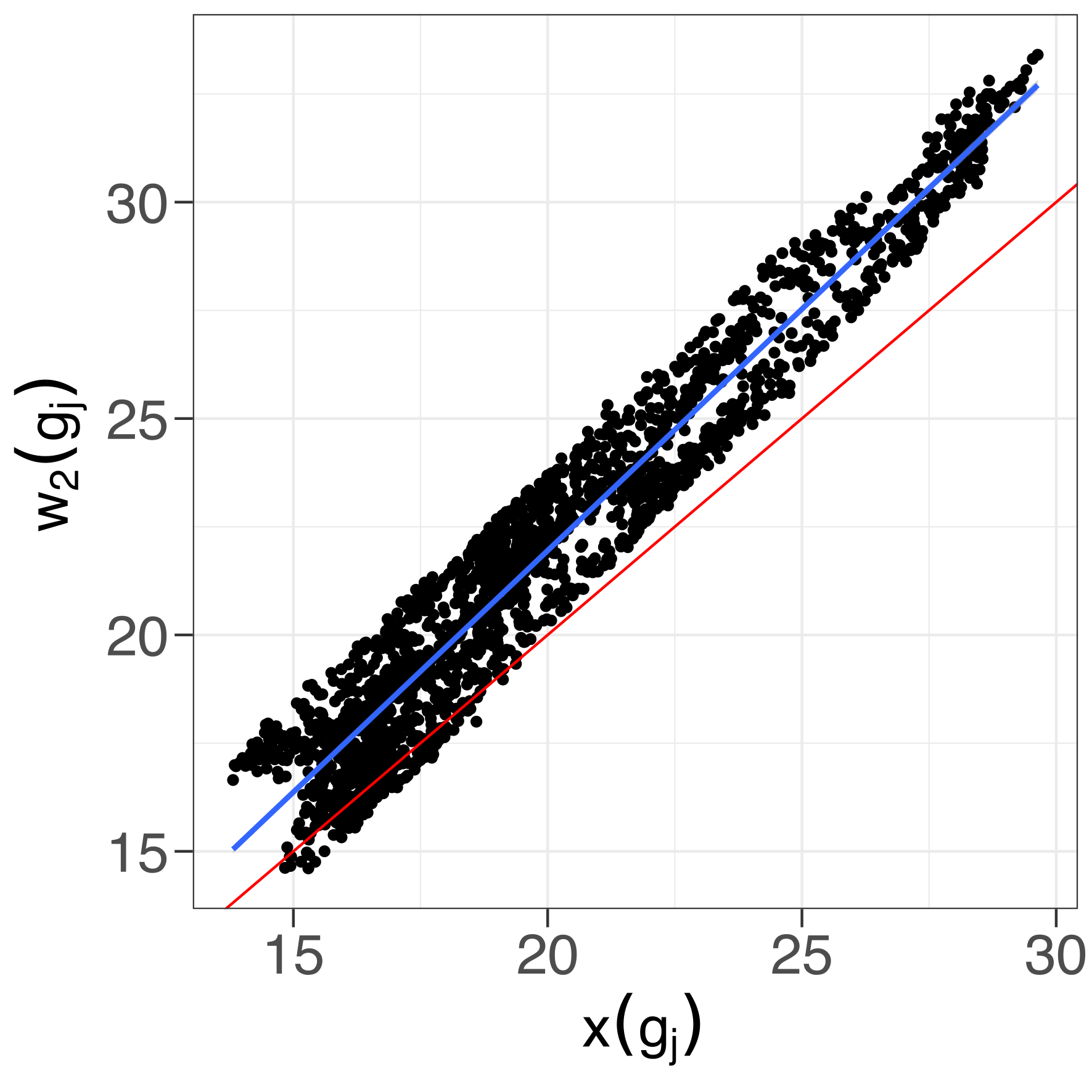}\label{fig:sim_dgp_scatter2}}
     \hspace{5mm}
     \subfloat[][$\text{w}_1(\mathbf{s}_i)$ vs $\text{w}_2(\mathbf{s}_i)$]{\includegraphics[clip,scale=0.2]{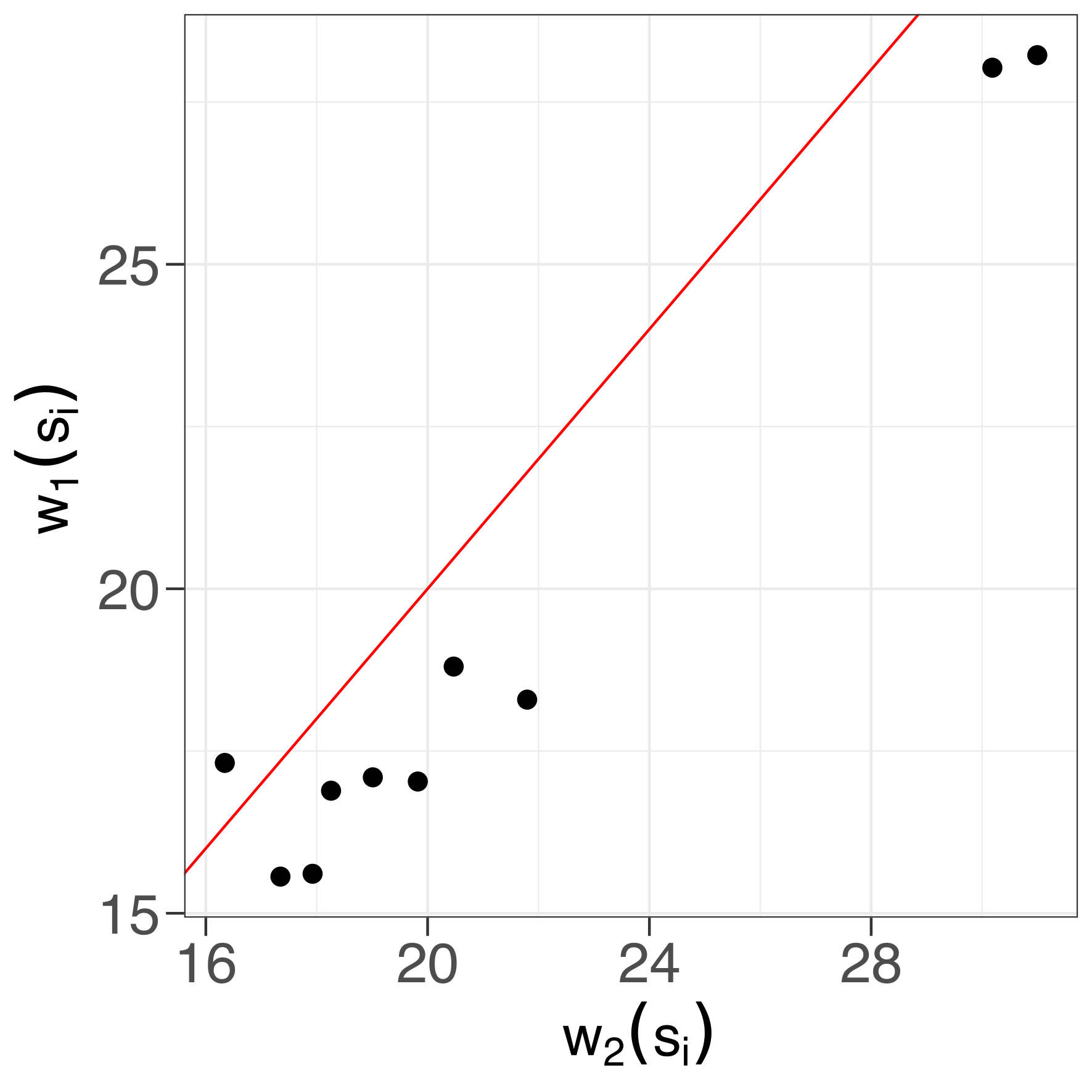}\label{fig:sim_dgp_scatter3}}
    \caption{(a) simulated observed values at 10 stations versus true values, (b) simulated outcomes from numerical forecast model versus true values, (c) simulated observed values at 10 stations versus outcomes from numerical forecast model.}
     \label{fig:sim_sample_DGP_scatter}
\end{figure}

\begin{figure}[t]
     \centering
     \subfloat[][$n_M$ = 10]{\includegraphics[trim={3cm 3cm 3cm 2cm},clip,scale=0.21]{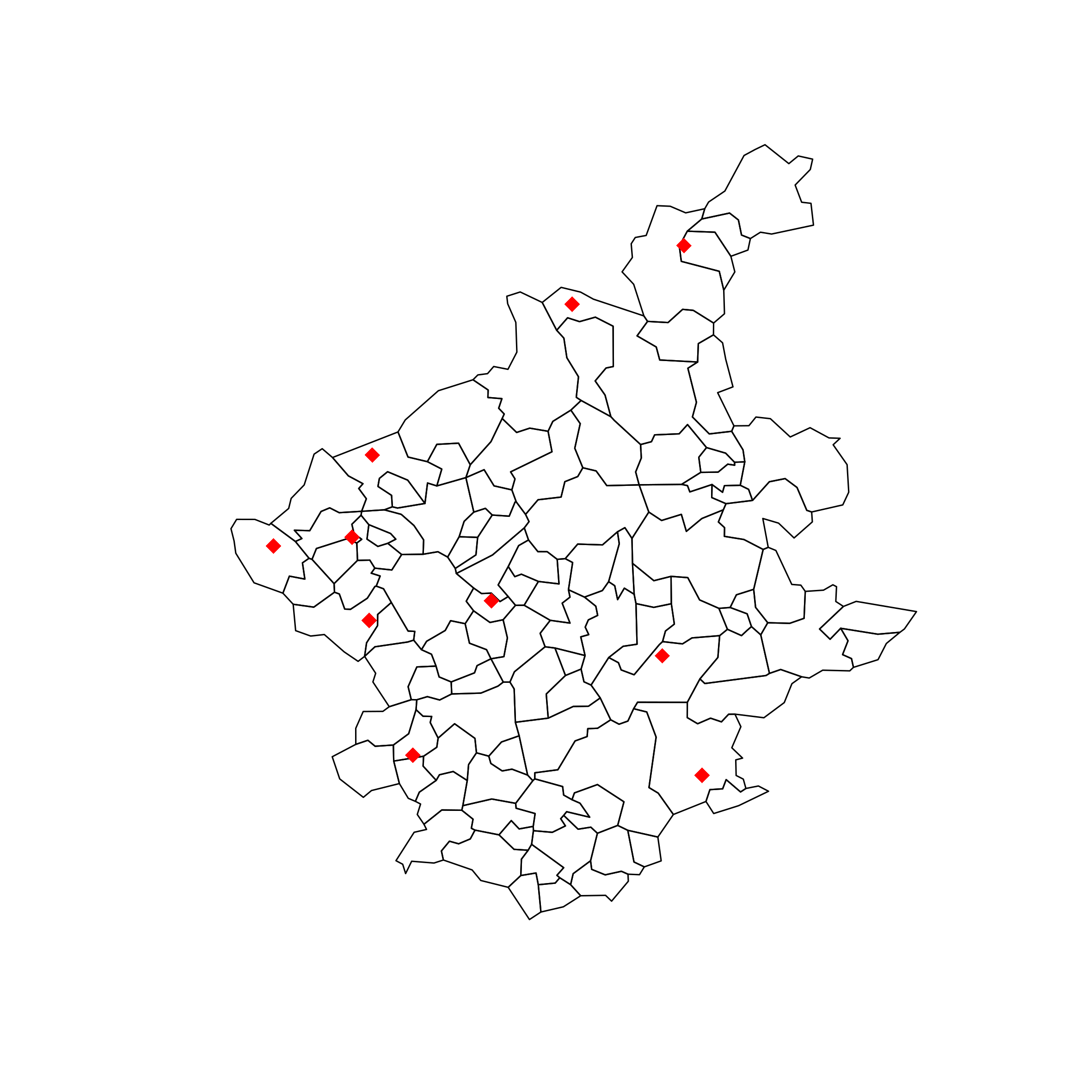}\label{fig:simN10}}
     \hspace{5mm}
     \subfloat[][$n_M$ = 25]{\includegraphics[trim={3cm 3cm 3cm 2cm},clip,scale=0.21]{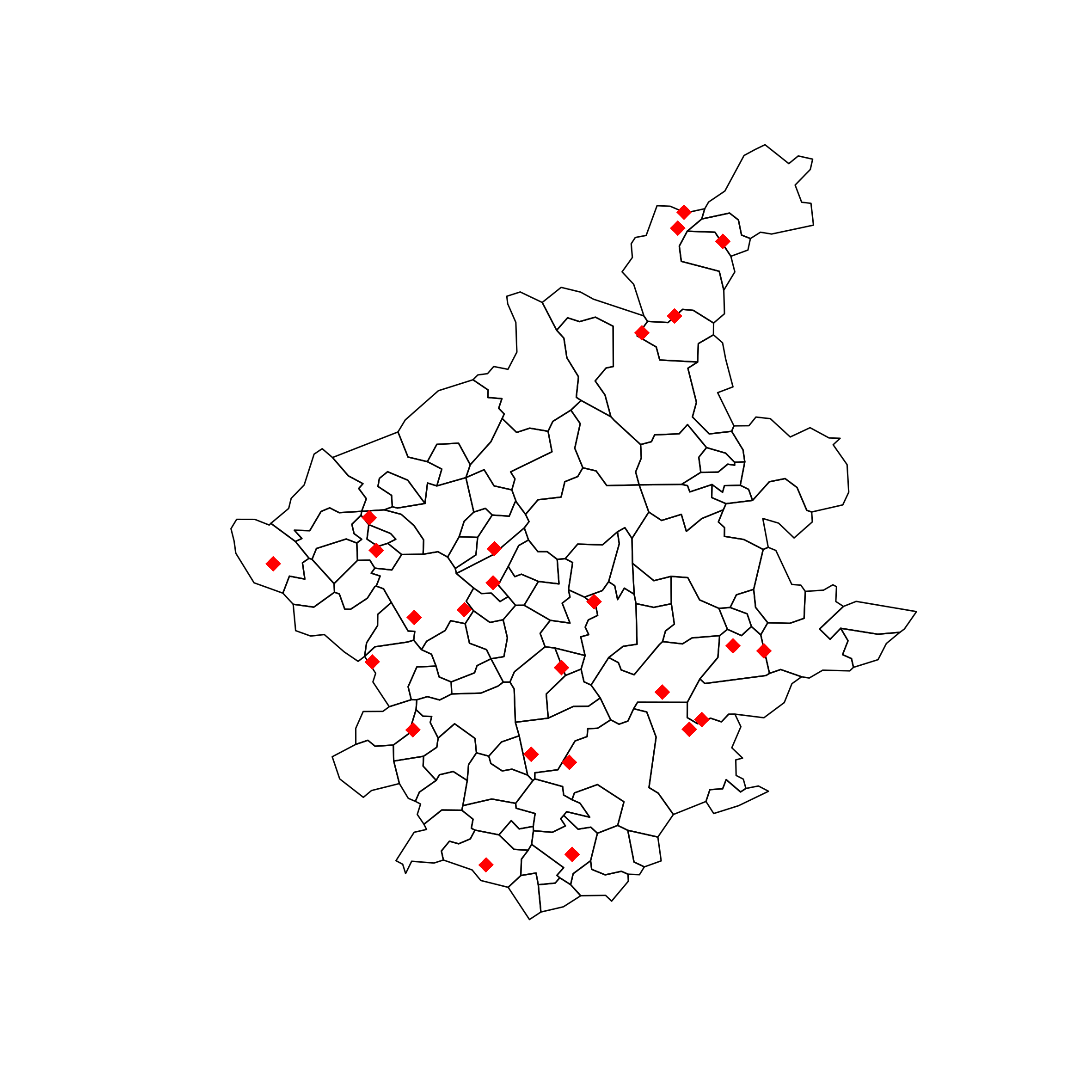}\label{fig:simN25}}
     \hspace{5mm}
     \subfloat[][$n_M$ = 40]{\includegraphics[trim={3cm 3cm 3cm 2cm},clip,scale=0.21]{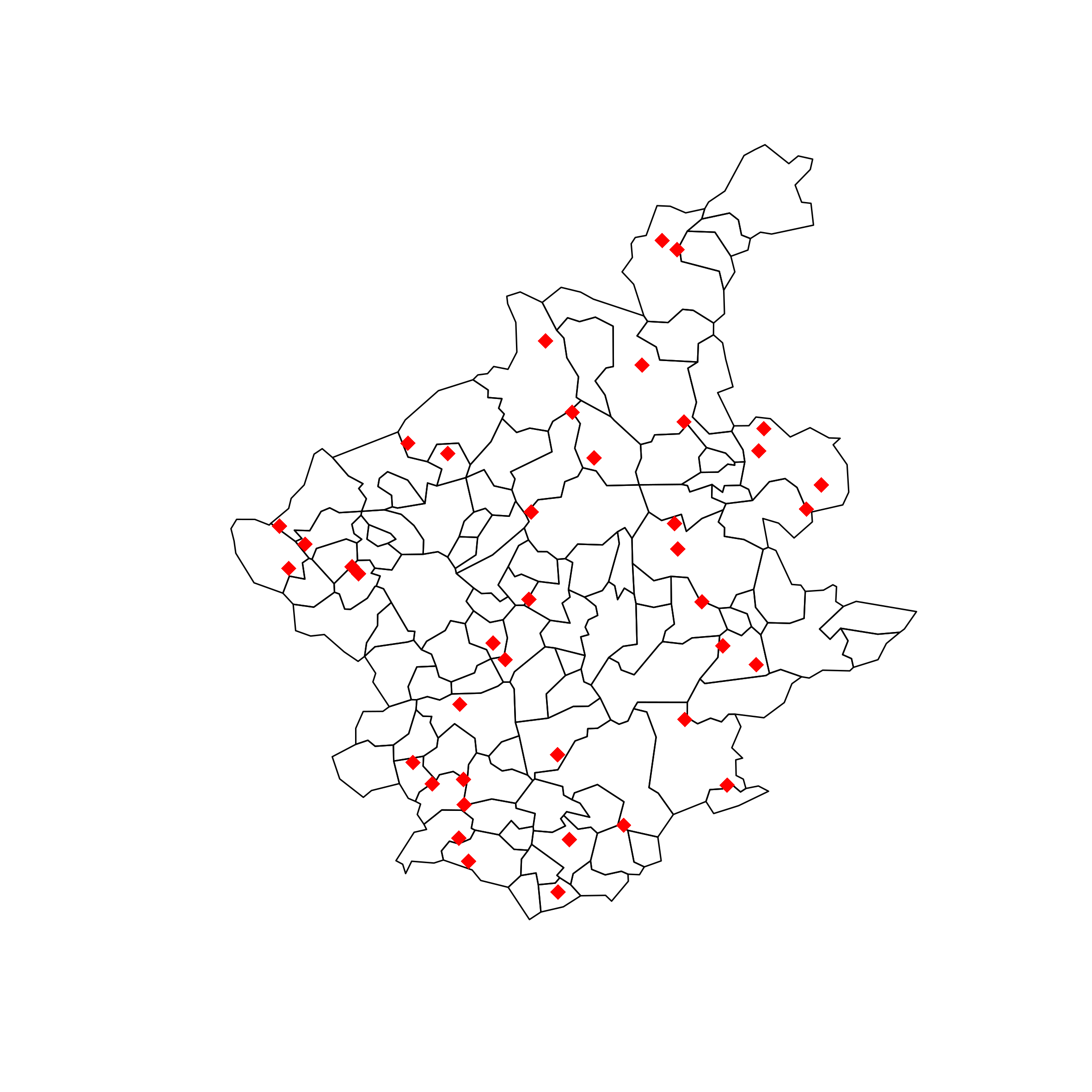}\label{fig:simN30}}
     \caption{Spatial location of stations: (a) a sparse network, (b) a denser network but with an undersampled region, (c) a dense uniformly distributed network.}
     \label{fig:sim_sparsity}
\end{figure}

Figure \ref{fig:sim_dgp_scatter3} highlights a significant discrepancy between the stations data and the numerical forecast model outcomes at the stations' locations for a simulated data with $n_M=10$ stations whose spatial locations are shown in Figure \ref{fig:simN10}. Figure \ref{fig:sim_dgp_scatter3} is similar to Figure \ref{fig:GSMvsStations} which shows discrepancies between the two data sources in the real data application. Moreover, the difference in bias severity between $\text{w}_1(\mathbf{s}_i)$ and $\text{w}_2(\mathbf{g}_j)$ is illustrated in Figures \ref{fig:sim_dgp_scatter1} and \ref{fig:sim_dgp_scatter2}, respectively. The data from the 10 stations closely align with the true values. On the other hand, $\text{w}_2(\mathbf{g}_j)$ exhibits more bias and an overestimation of the true values due to the multiplicative bias parameter $\alpha_1 > 1$. 
 
Our main interest is to understand if jointly modelling the two data sources offers advantages and how these change with the sparsity of the stations data. We therefore consider three different scenarios for the number of stations. The first scenario consists of only $n_M = 10$ stations  with areas severely undersampled (Figure \ref{fig:simN10}). In the second case, we have $n_M = 25$ stations, and a large area that is undersampled (Figure \ref{fig:simN25}). The third case has $n_M = 40$ stations uniformly distributed over the study area (Figure \ref{fig:simN30}). The spatial locations are held constant for all the data replicates so that the configuration of the stations does not influence the results. 


\subsection{Model definition and estimation}
We compare three modelling approaches: the stations-only model given by: $\text{w}_1(\mathbf{s}_i) = \beta_0 + \beta_1 z(\mathbf{s}_i) + \xi(\mathbf{s}_i) + e_1(\mathbf{s}_i), e_1(\mathbf{s}_i)\overset{\text{iid}}{\sim}\mathcal{N}\big(0,\sigma_{e_1}^2\big), i = 1,\dots, n_M$,   the regression calibration method (discussed in Section \ref{subsec:curretapproaches} and expressed in Equation \eqref{eq:regressioncalib}), and 
the proposed data fusion model: 
\begin{align*}
    &\text{w}_1(\mathbf{s}_i) = \beta_0 + \beta_1 z(\mathbf{s}_i) + \xi(\mathbf{s}_i)  + e_1(\mathbf{s}_i), &e_1(\mathbf{s}_i)\overset{\text{iid}}{\sim}\mathcal{N}\big(0,\sigma_{e_1}^2\big), \qquad i = 1,\dots, n_M\\
    &\text{w}_2(\mathbf{g}_j) = \alpha_0(\mathbf{g}_j)  + \alpha_1 \Big(\beta_0 + \beta_1 z(\mathbf{g}_j) + \xi(\mathbf{g}_j)\Big)  + e_2(\mathbf{g}_j), &e_2(\mathbf{g}_j)\overset{\text{iid}}{\sim}\mathcal{N}\big(0,\sigma_{e_2}^2\big), \qquad   j = 1,\dots, n_G.
\end{align*}
We are mainly interested in the unknown field $x(\mathbf{s})=\beta_0+\beta_1z(\mathbf{s})+\xi(\mathbf{s})$. We assign $\beta_0$, $\beta_1$, and $\beta_2$ vague, zero mean,  Gaussian priors. We use penalized complexity (PC) priors for the variance parameters  $\sigma^2_{e_1}$ and $\sigma^2_{e_2}$ and the parameters of the random fields $\xi(\cdot)$ and $\alpha_0(\cdot)$: $\rho_{\xi}$, $\sigma_{\xi}$, $\rho_{\alpha_0}$, and $\sigma_{\alpha_0}$. We define two scenarios for how we specify the PC priors. The first scenario, which we call \textit{matching priors}, uses the actual values used to generate the data. The second scenario, which we call \textit{non-matching priors}, uses $\sigma_{{e_1}_\text{o}}=1.5,  \sigma_{{e_2}_\text{o}}=0.5, {\sigma}_{\xi_\text{o}} = 1$, ${\rho}_{\xi_\text{o}} = 0.5$, ${\sigma}_{\alpha_{0\text{o}}} = .5$, ${\rho}_{\alpha_{0\text{o}}} = .5$, which are arbitrarily chosen values. The probability value in the PC priors is set to $\zeta_1=\zeta_2=0.5$ for all the parameters. As an example, when defining the prior for $\sigma^2_{e_1}$, we have $\mathbb{P}(\sigma_{e_1}>0.25)=0.5$ for the matching prior scenario, and $\mathbb{P}(\sigma_{e_1}>1.5)=0.5$ for the non-matching prior scenario. Note that the matching priors are not necessarily more informative than the non-matching priors and that both cases are weakly informative.

We use the {\tt inlabru} library \citep{lindgren2024inlabru} to fit the models.  For the proposed model, we use the Bayesian model averaging approach with INLA, which is discussed in Section \ref{sec:estimation}. We define a regular grid of $\alpha_1$ values centered on 1 and use a uniform prior for $\alpha_1$ in computing the weights. The {\tt R} code to implement the simulation study are available in \texttt{\hyperlink{https://github.com/StephenVillejo/DataFusionClimatePH}{https://github.com/StephenVillejo/DataFusionClimatePH}}. 

\subsection{Model assessment}\label{sec:model_assessment}
We compare the performance of the three modelling approaches, considering the accuracy in the predicted field $\hat{x}(\mathbf{s})$ and the estimates of model parameters. To predict the field, we consider the  posterior mean $\mathbb{E}[x(\mathbf{s})|\mathbf{Y}]$ with the corresponding uncertainty given by the posterior standard deviation $\sqrt{\mathbb{V}[x(\mathbf{s})|\mathbf{Y}]}$, both evaluated over the grid shown in Figure \ref{fig:sim_grid}. We use the following metrics for model assessment:
\begin{enumerate}
    \item Average squared error of the estimated field over $\mathcal{S}$: $\dfrac{1}{|\mathcal{S}|}\mathop{\mathlarger{\int}_{\mathcal{S}}} \Big(x(\mathbf{s}) - \mathbb{E}[x(\mathbf{s})|\mathbf{Y}]\Big)^2 d\mathbf{s}$
    \item Average posterior uncertainty of the estimated field over $\mathcal{S}$: $\dfrac{1}{|\mathcal{S}|}\mathop{\mathlarger{\int}_{\mathcal{S}}} \sqrt{\mathbb{V}[x(\mathbf{s})|\mathbf{Y}]}d\mathbf{s}$ \item Average Dawid-Sebastiani (DS) score which is a measure of the closeness between an observed quantity of interest and the prediction distribution, say $\mathcal{F}$.  The DS score is based on a coherent design criterion and is appropriate for predictive decision problems \citep{dawid1999coherent}. Suppose $\mathbb{E}_{\mathcal{F}}[y]$ and $\mathbb{V}_{\mathcal{F}}[y]$ are the mean and variance, respectively, of the predictive distribution $\mathcal{F}(y)$.  The DS score for a prediction on $y$ is given by
    \begin{equation}\label{eq:DSformula}
        \dfrac{\Big(y-\mathbb{E}_{\mathcal{F}}[y]\Big)^2}{\mathbb{V}_{\mathcal{F}}[y]} + \log\big(\mathbb{V}_{\mathcal{F}}[y]\big).
    \end{equation}
    \item Relative error of each parameter estimate: e.g., $\bigg|\dfrac{\hat{\beta}-\beta}{\beta}\bigg|$, where $\hat{\beta}$ is the posterior mean of $\beta$, i.e., $\hat{\beta}=\mathbb{E}[\beta|\mathbf{Y}]$.
    \item Posterior uncertainty in the parameter estimates: e.g., $\sqrt{\mathbb{V}[\beta|\mathbf{Y}]}$ for $\beta$.
\end{enumerate}
The first metric is a measure of the average discrepancy between the estimated field $\hat{x}(\mathbf{s})$ and the true field $x(\mathbf{s})$ while the second metric assesses the average uncertainty in the estimated field. We approximate both integrals using the estimated values on the prediction grid. The third metric is another proper scoring rule which depends on the predictive mean and variance of the observed data \citep{gneiting2007strictly}. 
As for the first two metrics, a lower value for the average DS score is preferred. Finally, the last two metrics look at the bias and uncertainty in the parameter estimates. All simulation results are computed based on 500 independent data replicates.

\subsection{Simulation study results}
\begin{figure}[t]
    \centering
    \begin{minipage}{0.49\textwidth}
        \centering
        \includegraphics[width=.95\textwidth]{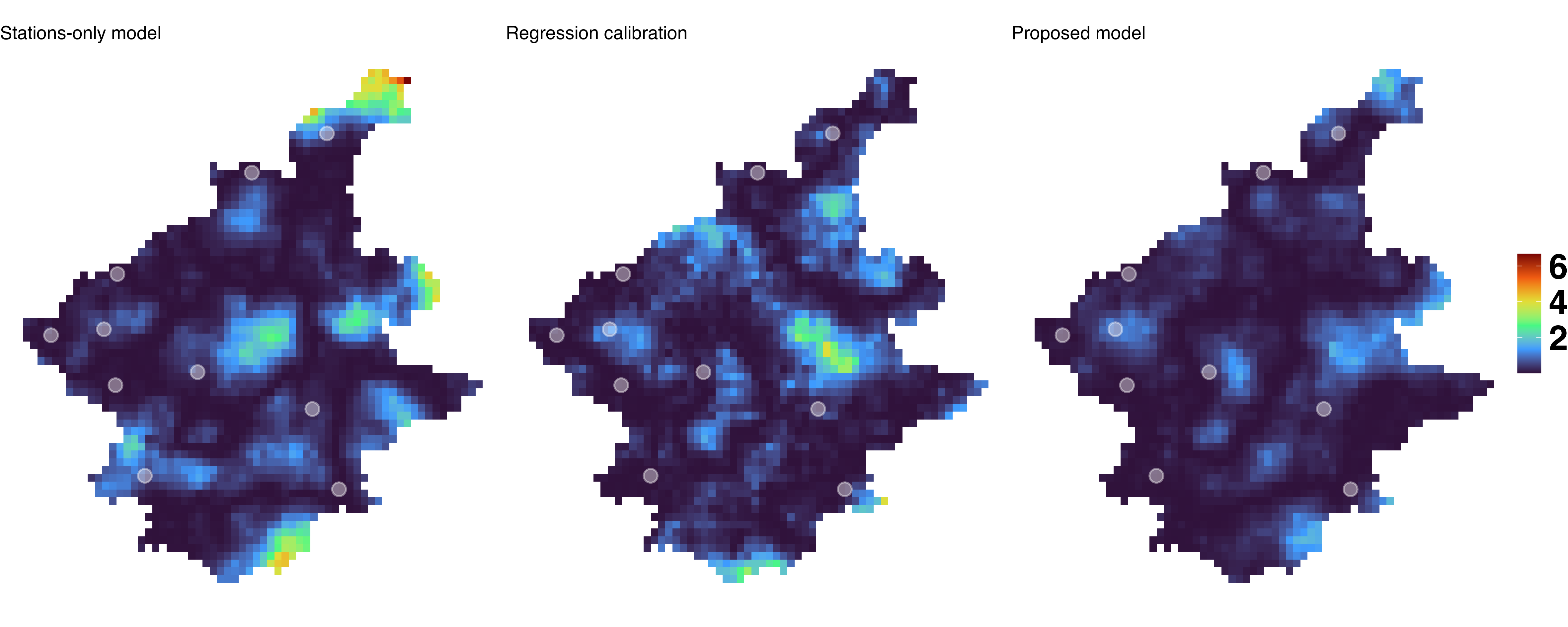}
        
        \captionof{figure}{Comparison of squared errors for the simulated data in Figures \ref{fig:sim_sample_DGP} and \ref{fig:sim_sample_DGP_scatter}. The errors from the proposed model are generally the smallest.}
        \label{fig:sim_sqerror_combined}
    \end{minipage}
    \hfill
    \begin{minipage}{0.49\textwidth}
        \centering
        \includegraphics[width=.95\textwidth]{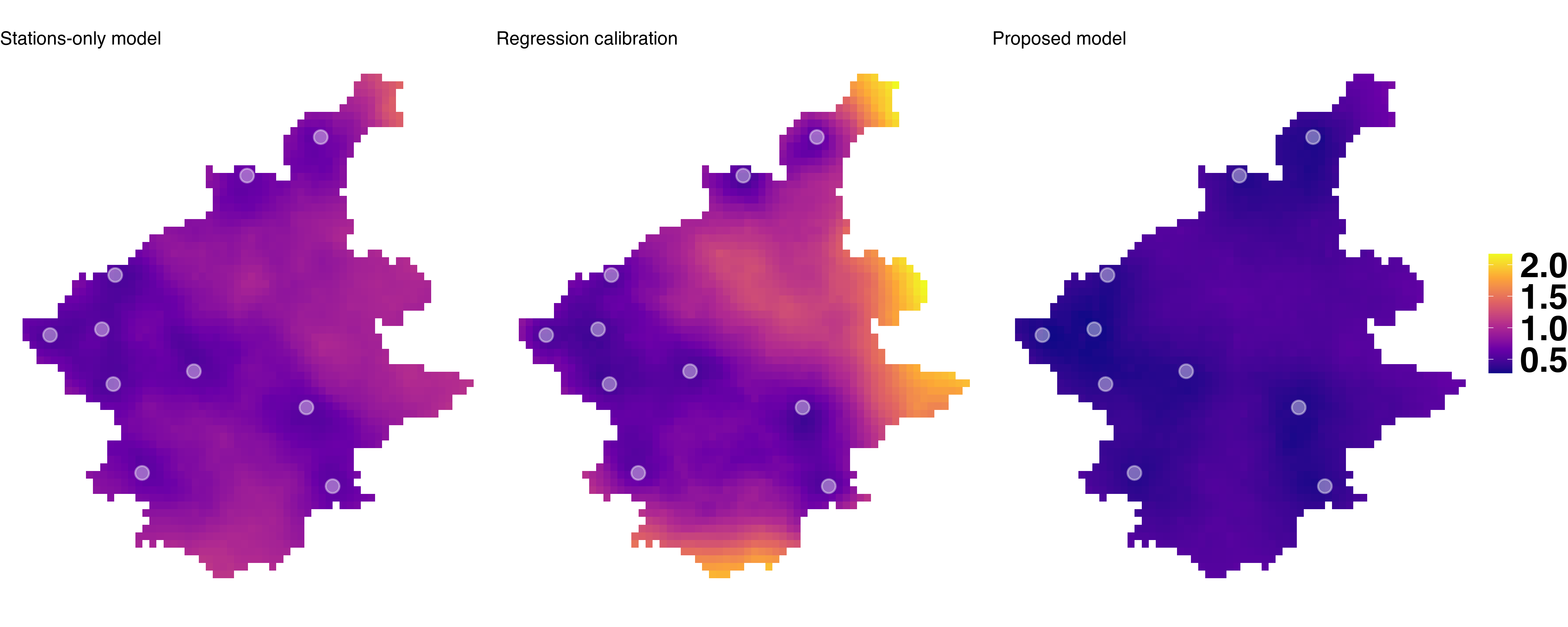}
        
        \captionof{figure}{Comparison of the posterior uncertainty for the simulated data in Figures \ref{fig:sim_sample_DGP} and \ref{fig:sim_sample_DGP_scatter}. The posterior uncertainty from the proposed model are the smallest.}
        \label{fig:sim_sd_combined}
    \end{minipage}
\end{figure}

\begin{figure}[]
    \centering
    \hspace*{\fill}
    \subfloat[][Log average squared error of $\hat{x}(\mathbf{s})$ ]
    {\includegraphics[scale=0.26]{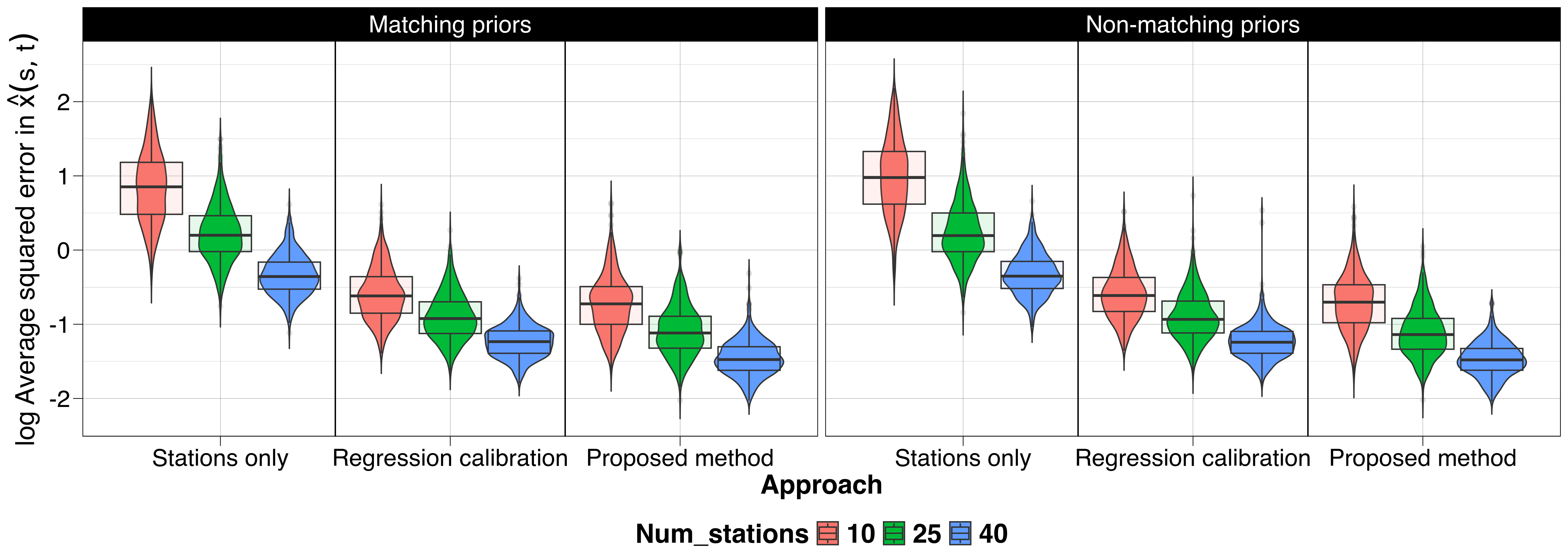}
    \label{fig:Ortho_I_x}}
    \hspace*{\fill}

    \hspace*{\fill}
    \subfloat[][Average posterior uncertainty of $\hat{x}(\mathbf{s})$]
    {\includegraphics[scale=0.26]{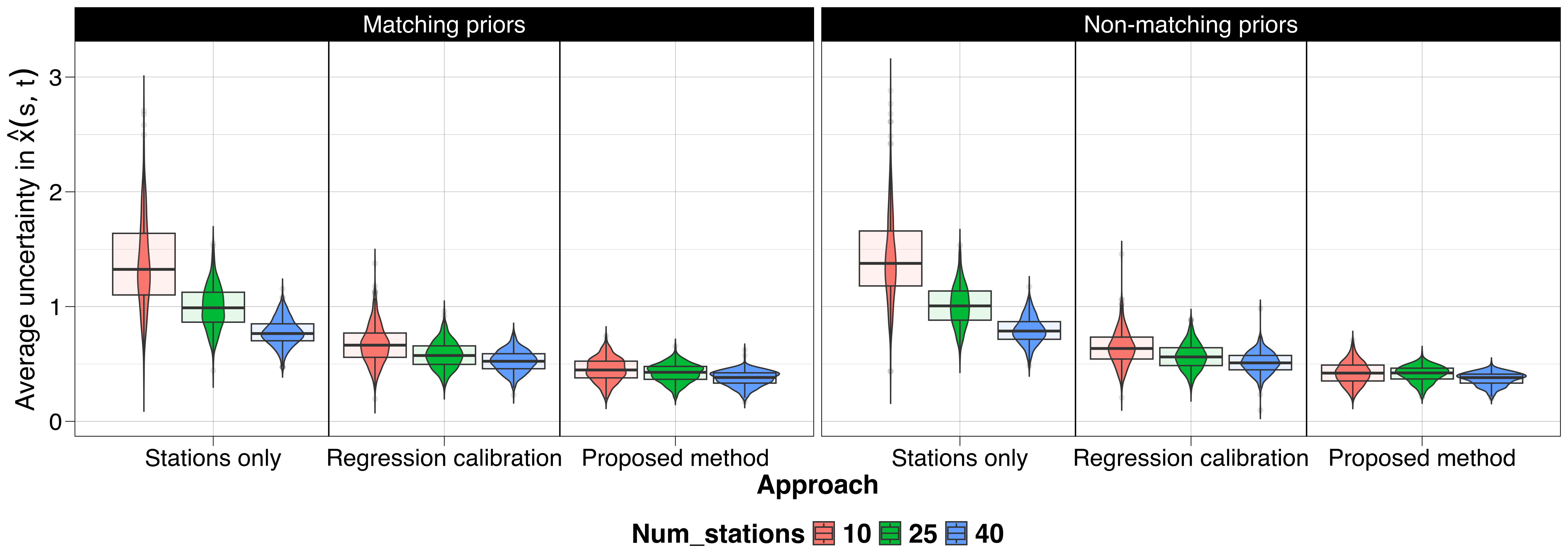}
    \label{fig:Ortho_SD_x}}
    \hspace*{\fill}
    
    \caption{Plots of the (a) log average squared errors and (b) average posterior uncertainty from 500 simulated datasets with respect the number of stations, the priors used, and the modelling approach: stations-only model, regression calibration model, and proposed data fusion model. The posterior uncertainty from the proposed model is smallest. The stations-only model has the highest log average squared error.}
    \label{fig:Ortho_I_x_all}
\end{figure}

Figures \ref{fig:sim_sqerror_combined} and \ref{fig:sim_sd_combined} show a comparison of the squared errors and the posterior uncertainty of the estimated fields, respectively, among the three different modelling approaches on the data example in Figures \ref{fig:sim_sample_DGP} and \ref{fig:sim_sample_DGP_scatter}. The stations' location are shown as white points. The squared errors are largest for the stations-only model and smallest for the proposed data fusion model (see Figure \ref{fig:sim_sqerror_combined}). The posterior uncertainty in the estimated field is also the smallest for the proposed model (see Figure \ref{fig:sim_sd_combined}). As expected, the posterior uncertainty is smallest at the stations' locations, which is very apparent for the stations-only model and the regression calibration model. The average squared errors for the stations-only model, regression calibration model, and the proposed model for this specific case are 0.53, 0.42, and 0.27, respectively, while the average posterior uncertainties are 0.78, 0.93, and 0.55, respectively. The average DS scores are 0.37, 0.08, and -0.23, respectively.

Figure \ref{fig:Ortho_I_x} shows a plot of the log average squared errors of the estimated field $\hat{x}(\mathbf{s})$ based on 500 data replicates for the different simulation scenarios. The proposed data fusion model generally gives smaller log average squared errors, especially when the data on the stations are very sparse. Moreover, the results show that there is no substantial difference with respect to the priors. Figure \ref{fig:Ortho_SD_x} shows the results for the average posterior uncertainty. It shows that the proposed model gives lower uncertainty estimates and that a higher number of stations is associated with lower posterior uncertainty. The same figure also shows that the specification of priors does not influence the results. 
The results for the DS scores are consistent with the insights from the previous two scores (see Figure \ref{fig:Ortho_I_x_v2} of Appendix \ref{subsec:app_simres}). 


Figure \ref{fig:Sim_summary_weights} shows a summary of the model averaging weights of the INLA models for each $\alpha_1$ value. The conditional INLA model with the highest weight corresponds to the true value $\alpha_1=1.1$, and with weights rapidly decreasing as $\alpha_1$ goes further away from 1.1. The BMA weights do not vary much between the use of matching and non-matching priors, and the sparsity of the stations data. Figure \ref{fig:sim_res_param_3} compares the average relative error and average posterior uncertainty for the measurement error standard deviation $\sigma_{e_1}$. The proposed method generally outperforms the other two approaches, especially when the data from the stations are sparse. Figures \ref{fig:sim_res_param}a and \ref{fig:sim_res_param}b of Appendix \ref{subsec:app_simres} show the results for the marginal standard deviation and the range parameter of the spatial field $\xi(\mathbf{s})$, respectively, while Figures \ref{fig:sim_res_param_2} of Appendix \ref{subsec:app_simres} show the results for the fixed effects $\beta_0$ and $\beta_1$. Note that for the aforementioned parameters, we can only make a comparison between the stations-only model and the proposed model since these parameters are not defined and specifed in the regression calibration model as shown in Equation \eqref{eq:regressioncalib}. The results show that the proposed method outperforms the stations-only model in terms of the relative error and posterior uncertainty. 



\begin{figure}[H]
     \centering
    \includegraphics[trim={0cm 0cm 0 0cm},clip,scale=.42]{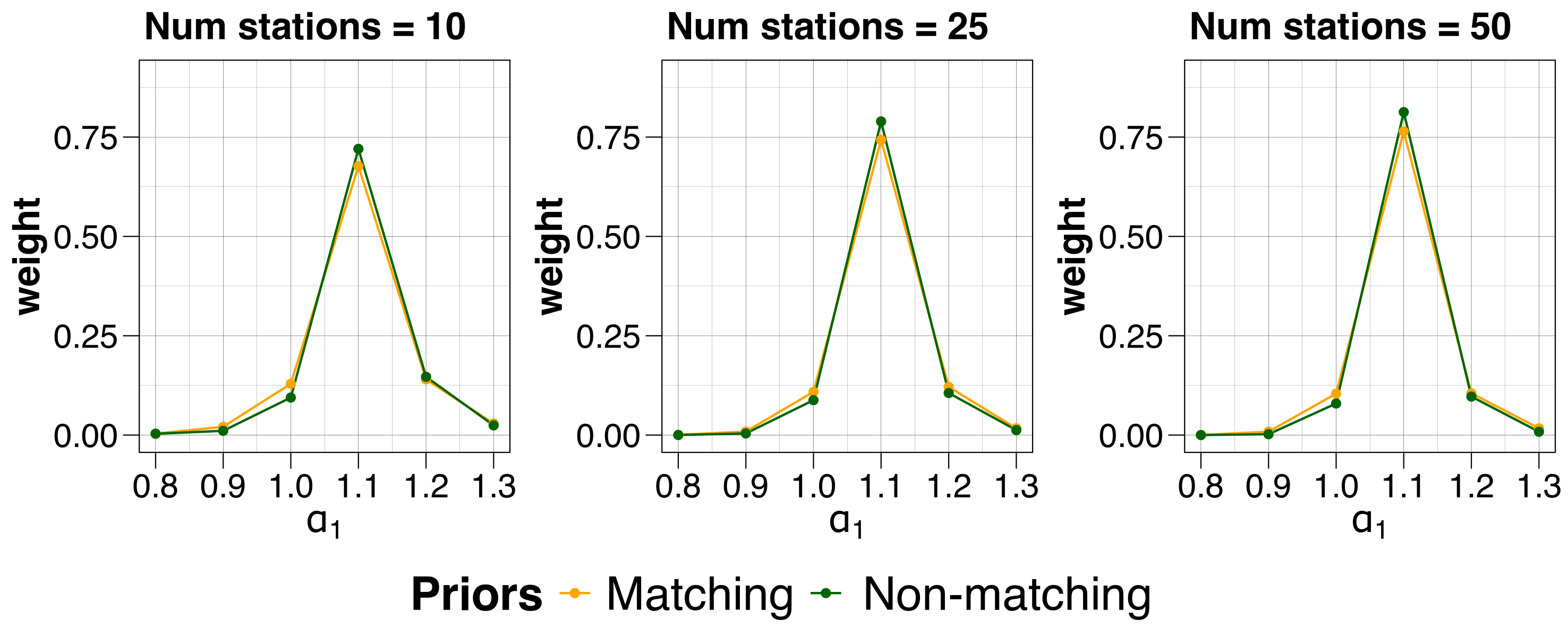}
    \caption{Average model averaging weights from 500 simulated datasets for different $\alpha_1$ values in fitting the proposed data fusion model with respect to the sparsity of the stations data and the priors used. The correct value of $\alpha_1$ has the highest weight.}
     \label{fig:Sim_summary_weights}
\end{figure}

\begin{figure}[H]
    \centering
    \includegraphics[scale=.39]{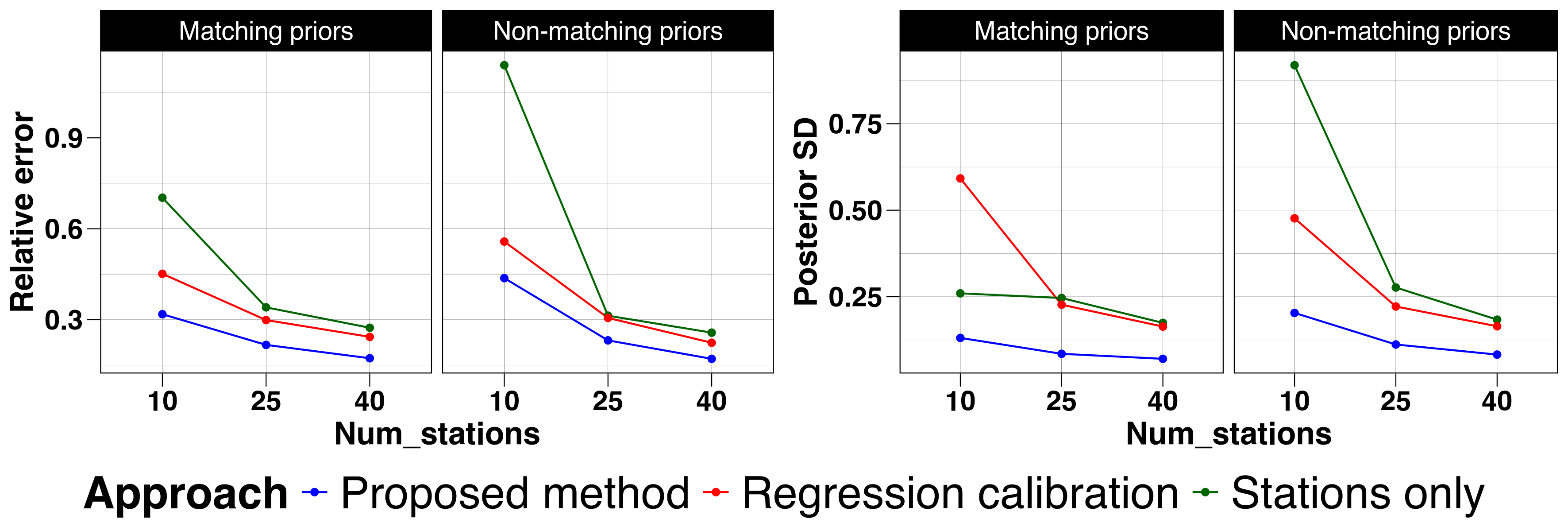}
    \caption{Plot of average relative errors and average posterior uncertainty from 500 simulated datasets for $\sigma_{e_1}$.}
    \label{fig:sim_res_param_3}
\end{figure}

\section{Results for Meteorological Data from the Philippines}\label{sec:application}
In this section, we present the results from applying the three modelling approaches on the meteorological data in the Philippines. Sections \ref{subsec:res_meantemp}, \ref{subsec:res_relhum}, and \ref{subsec:res_rainfall} discuss the results for temperature, relative humidity, and rainfall, respectively. Section \ref{subsec:LGOCV} presents the results of the leave-group-out cross-validation.

\subsection{Temperature}\label{subsec:res_meantemp}
The mean (standard deviation) temperature in the stations and GSM data is $27.67^{\circ}\text{C}$ (1.87) and $25.68^{\circ}\text{C}$ (1.96), respectively. The percentage of missing data from the stations is only 3.8\%; while the GSM data has no missing data. We have the following form for the fixed effects in the latent process:
\begin{equation}
    \label{eq:temp_latentprocess} \text{Temperature}(\mathbf{s},t) = \beta_0 + \beta_1  \log\Big(\text{Elevation}(\mathbf{s},t)\Big) + \beta_2  \text{Cool}(\mathbf{s},t) + \beta_3 \text{ClimateType}(\mathbf{s},t).
\end{equation}
The `Cool' variable in Equation \eqref{eq:temp_latentprocess} is a binary variable which takes a value of `1' for months December to February and a value of `0' for the other months. The climate type variable is also a binary variable which takes `1' for the eastern section of the country and `0' for the western section (see  Section \ref{sec:data_application} for details). 


 In defining the three models (the stations-only model,  the regression calibration model, and the proposed data fusion model), we 
 used penalized complexity (PC) priors for the Mat\'ern
field parameters \citep{fuglstad2019constructing, simpson2017penalising}. The parameter values for the Mat\'ern PC priors are as follows: $\rho_{1\text{o}}=\rho_{2\text{o}}=300 \;\text{km}, \sigma_{1\text{o}}=1.90$, and $\sigma_{2\text{o}}=0.01$. The value for the range parameters is one-third the maximum distance of the spatial domain. The value for $\sigma_{1\text{o}}$ is the standard deviation of the temperature values, while the value of $\sigma_{2\text{o}}$ is chosen to be some value smaller than $\sigma_{1\text{o}}$ based on preliminary model results. The variance parameters of $e_1(\mathbf{s}_i,t)$ and $e_2(\mathbf{g}_j,t)$ are also given PC priors, with $\sigma_{{e_1}_\text{o}}=0.2$ and $\sigma_{{e_2}_\text{o}}=0.01$. The probability value of all PC priors is set to be equal to 0.50. The rest of the model parameters are given default non-informative priors.

We defined a grid of values from 0.5 to 1.5 with a length step of 0.1 for the multiplicative bias parameter $\alpha_1$, and which we assigned a uniform prior. An ensemble of INLA models were fitted for a fixed $\alpha_1$, and the BMA weights are computed using Equation \eqref{eq:weightsBMA}. The marginal log-likelihoods $\log \pi\Big(\mathbf{Y}|\alpha_1^{(k)}\Big)$ and the corresponding BMA weights $w_k$ are shown in Table \ref{tab:TMEAN_mliksweights} of Appendix \ref{subsec:app_temperature}. The results show that the weight of the model with $\alpha_1=1$ is approximately equal to 1, while the weights for the other models are close to 0. This implies that there is no multiplicative bias which agrees with the insights from Figure \ref{fig:GSMvsStations}a.


\begin{table}[h]
\caption{\label{tab:TMEANestimates_fixed}Posterior estimates of fixed effects for the temperature model -- stations-only model versus proposed data fusion model} 
\centering
\begin{tabular}{|l|cccc|cccc|}
\hline\hline
 &  \multicolumn{4}{c|}{\textbf{Stations only}} & \multicolumn{4}{c|}{\textbf{Proposed model}}\\
 Parameter & Mean & SD & P2.5\% & P97.5\% & Mean & SD & P2.5\% & P97.5\% \\ 
  \hline\hline
$\beta_0$ & 28.664 & 2.6270 & 23.510 & 33.818 & 28.919 & 4.603 & 19.897 & 37.940 \\ 
  $\beta_1$, \color{brown}{log(Elevation)} & -0.631 & 0.094 & -0.815 & -0.446 & -0.709 & 0.051 & -0.808 & -0.609 \\ 
  $\beta_2$, \color{brown}{Cool} & -0.683 & 0.198 & -1.072 & -0.295 & -0.6178 & 0.177 & -0.965 & -0.271 \\ 
 $\beta_3$, \color{brown}{Climate Type} & 2.183 & 0.699 & 0.813 & 3.553 & 0.606 & 0.337 & -0.054 & 1.266 \\ 
   \hline\hline
\end{tabular}
\end{table}


Table \ref{tab:TMEANestimates_fixed} shows the posterior estimates of the fixed effects for the stations-only model and the proposed data fusion model. Note that these parameters are not explicitly specified and estimated using the regression calibration model, shown in Equation \eqref{eq:regressioncalib}. The two models agree on the conclusions: it is colder at higher elevation, cooler during December to February, and areas in the western section of the country are cooler. The climate type variable is not significant in the proposed data fusion model, while that of the stations-only model is significant. Finally, the uncertainty in the fixed effects, expect for $\beta_0$, are smaller for the proposed model.


The posterior estimates of the hyperparameters for the stations-only model and proposed are model are also quite similar (see Table \ref{tab:TMEANestimates_hyper} of Appendix \ref{subsec:app_temperature}). Note that only the estimates from the stations-only model and proposed model are comparable, since the regression calibration model does not specify the latent process in Equation \eqref{eq:model1}. In the proposed data fusion model, the estimated range of the spatial field, $\hat{\rho}_1$, is higher than the one for the error field, $\hat{\rho}_2$, indicating that $\xi(\mathbf{s},t)$ is smoother than $\alpha_0(\mathbf{g},t)$. The spatial correlation in the temperature spatial field and the error field becomes negligible at a distance of around 765 km and 113 km, respectively. Also, the estimated marginal standard deviation $\hat{\sigma}_{1}$ of the spatial field is much larger than that of the error field $\hat{\sigma_{2}}$. The estimated autocorrelation parameters $\hat{\phi}_1$ and $\hat{\phi}_2$ in both the spatial field and error field are close to 1 which suggests a high degree of dependence in time. Moreover, the posterior estimates of the regression calibration model are shown Table \ref{tab:TMEANestimates_regcalib} of Appendix \ref{subsec:app_temperature}. The results show that the $2.5^{\text{th}}$ and $97.5^{\text{th}}$ percentile of the multiplicative bias estimates of the regression calibration model are 0.9 and 1.1, respectively. The values do not change much in space and time, which justifies the assumption of a constant $\alpha_1$ in the proposed model.

\begin{figure}[]
	\centering
	\begin{minipage}{.54\columnwidth}
		\centering
		\subfloat[][Predicted temperature fields]
    {\includegraphics[scale=0.14]{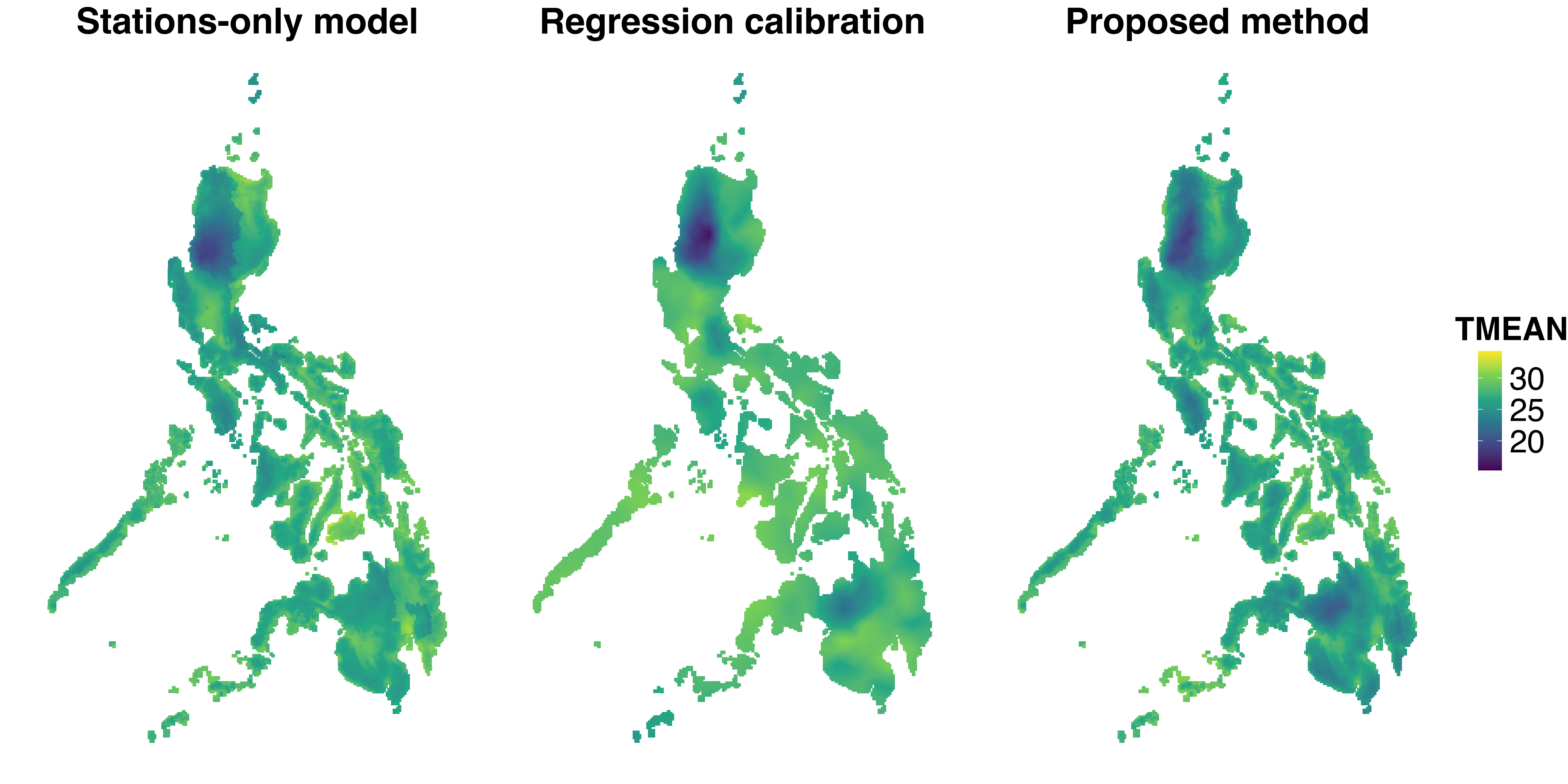}
    \label{fig:TMEAN_field}} \\
    \subfloat[][Posterior uncertainty (log scale)]
    {\includegraphics[scale=0.14]{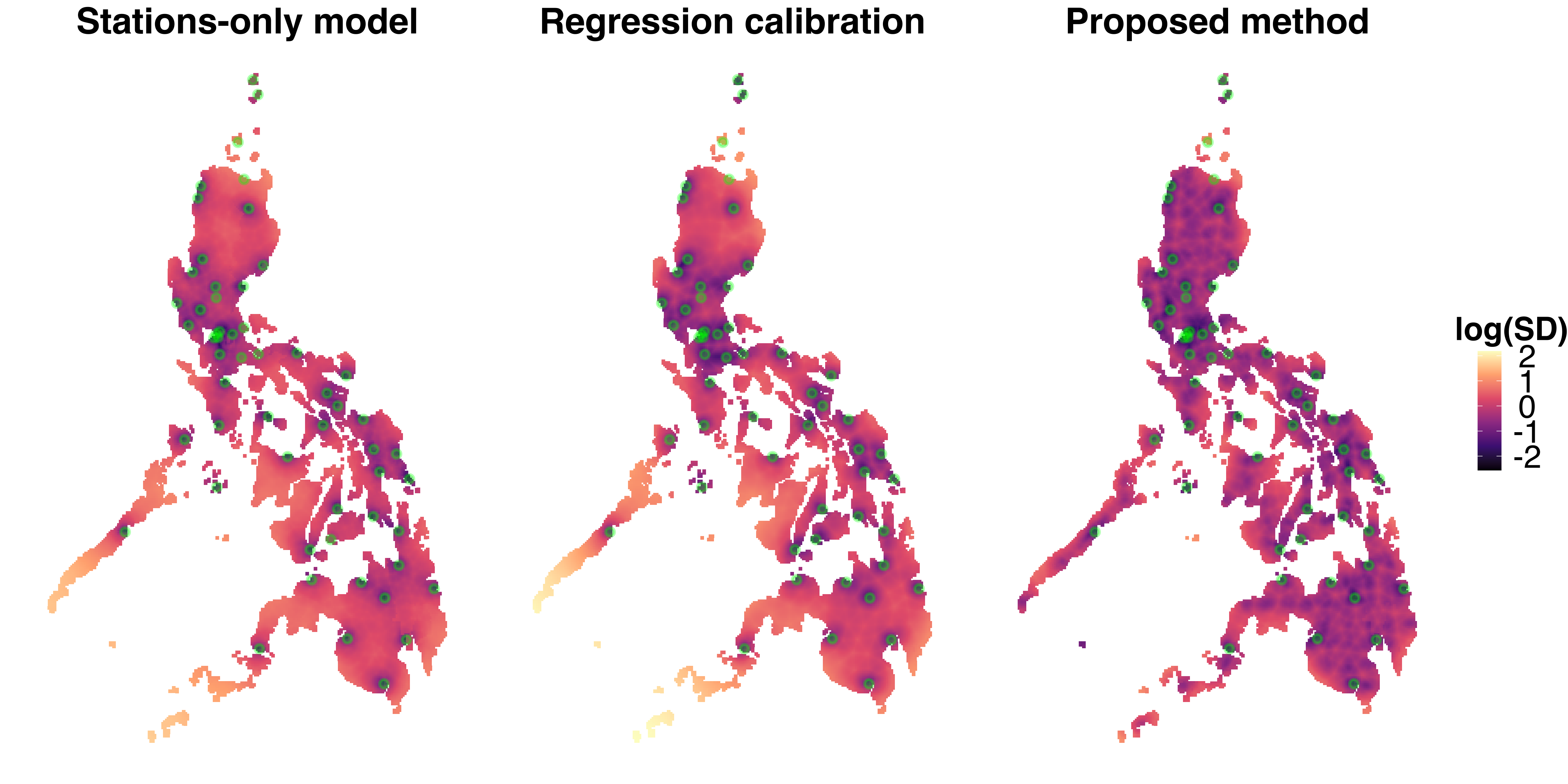}\label{fig:TMEAN_fieldsd}}
    \caption{Comparison of the estimated temperature fields and corresponding posterior uncertainties (log scale) for August 2019. The posterior uncertainties from the proposed data fusion model are smaller.}
    \label{fig:TMEANfields}
	\end{minipage}%
    \hspace{5mm}
	\begin{minipage}{.37\columnwidth}
		\centering
		\includegraphics[scale=.21]{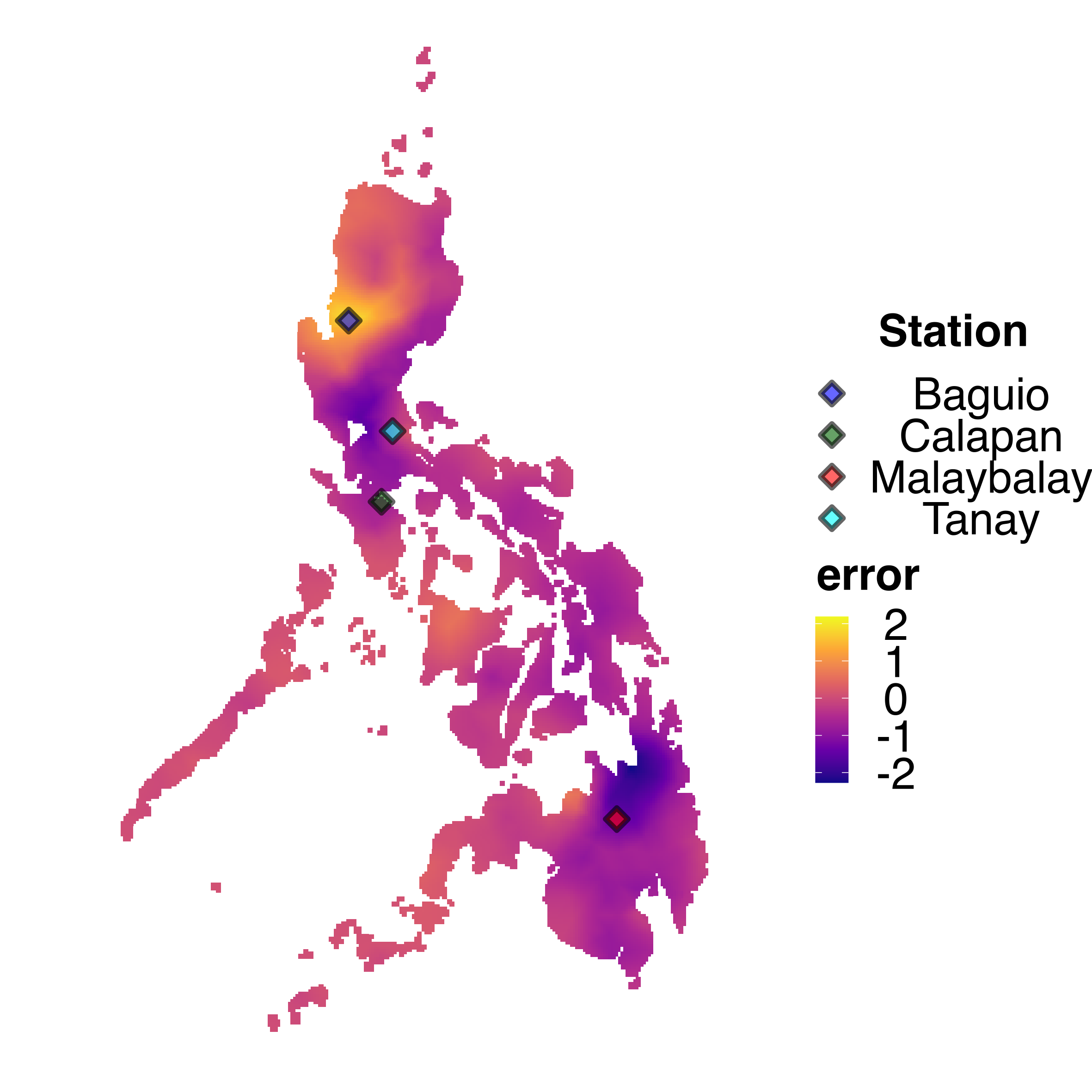}
        
        \captionof{figure}{Estimated error field for the temperature model for August 2019. The estimated error fields at the specific stations correspond to the additive bias shown in Figure \ref{fig:GSMvsStations}.}
        \label{fig:TMEAN_error}
	\end{minipage}
\end{figure}

Figure \ref{fig:TMEAN_field}  shows that the estimated temperature fields (for August 2019) among the three approaches are very similar. The dark spot in the northern part of the country is primarily mountainous which makes it cooler than the other parts of the country. Moreover, Figure \ref{fig:TMEAN_fieldsd} shows the corresponding uncertainty estimates or posterior standard deviation (SD) in log scale. The uncertainty is higher for the two benchmark approaches, especially in the islands in the lower left portion. The average  posterior SD of the estimated fields from the stations-only model, regression calibration model, and the  data fusion model is 1.239, 1.123, and 0.771, respectively. Moreover, the posterior SD is smaller at the stations' locations (green points) which is more apparent for the two benchmark approaches. Figure \ref{fig:TMEAN_spde} of Appendix \ref{subsec:app_temperature} shows the estimated spatial fields, $\hat{\xi}(\mathbf{s},t)$, for the three approaches and for the same month. The spatial structure looks quite similar as the estimated temperature fields in Figures \ref{fig:TMEAN_field}. 

Figure \ref{fig:TMEAN_error} shows the estimated error field $\hat{\alpha}_0(\mathbf{g}_j,t)$ for the GSM data in August 2019. This plot can be compared to Figure \ref{fig:GSMvsStations} which shows the discrepancies in the values between the weather stations and the GSM outcomes. In particular,  Figure \ref{fig:TMEAN_error} shows that the estimated additive bias around Baguio station (in blue) is the highest, indicating that the GSM overestimated the temperature in this region. This is consistent with Figure  \ref{fig:GSMvsStations} which shows that the GSM values exceed the observed data at Baguio station. Similarly, the estimated additive bias around Malaybalay (in red) is negative, which aligns with  the negative bias seen in the GSM outcomes for this area. For Tanay station (in cyan), the estimated additive bias is close zero, which is also consistent with Figure \ref{fig:GSMvsStations} which shows little to no bias in the GSM outcomes at this location.

\begin{figure}
     \centering
     \subfloat[][$\text{w}_1(\mathbf{s},t)$ vs $\hat{x}(\mathbf{s},t)$]{\includegraphics[trim={0cm 0cm 0 0cm},clip,scale=0.28]{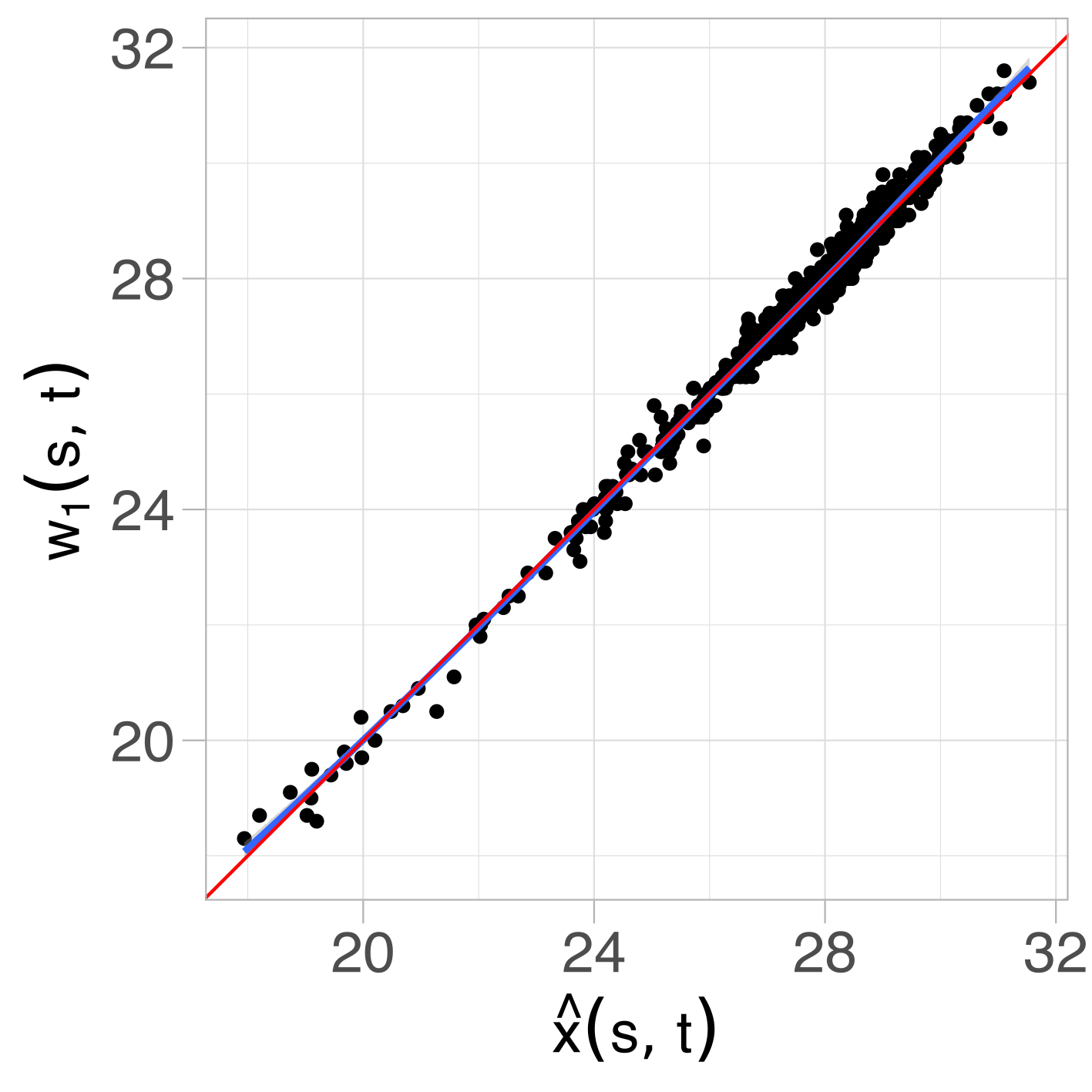}\label{fig:TMEAN_scatter1}}
     \subfloat[][$\text{w}_2(\mathbf{g},t)$ vs $\hat{x}(\mathbf{g},t)$]{\includegraphics[trim={0cm 0cm 0 0cm},clip,scale=0.28]{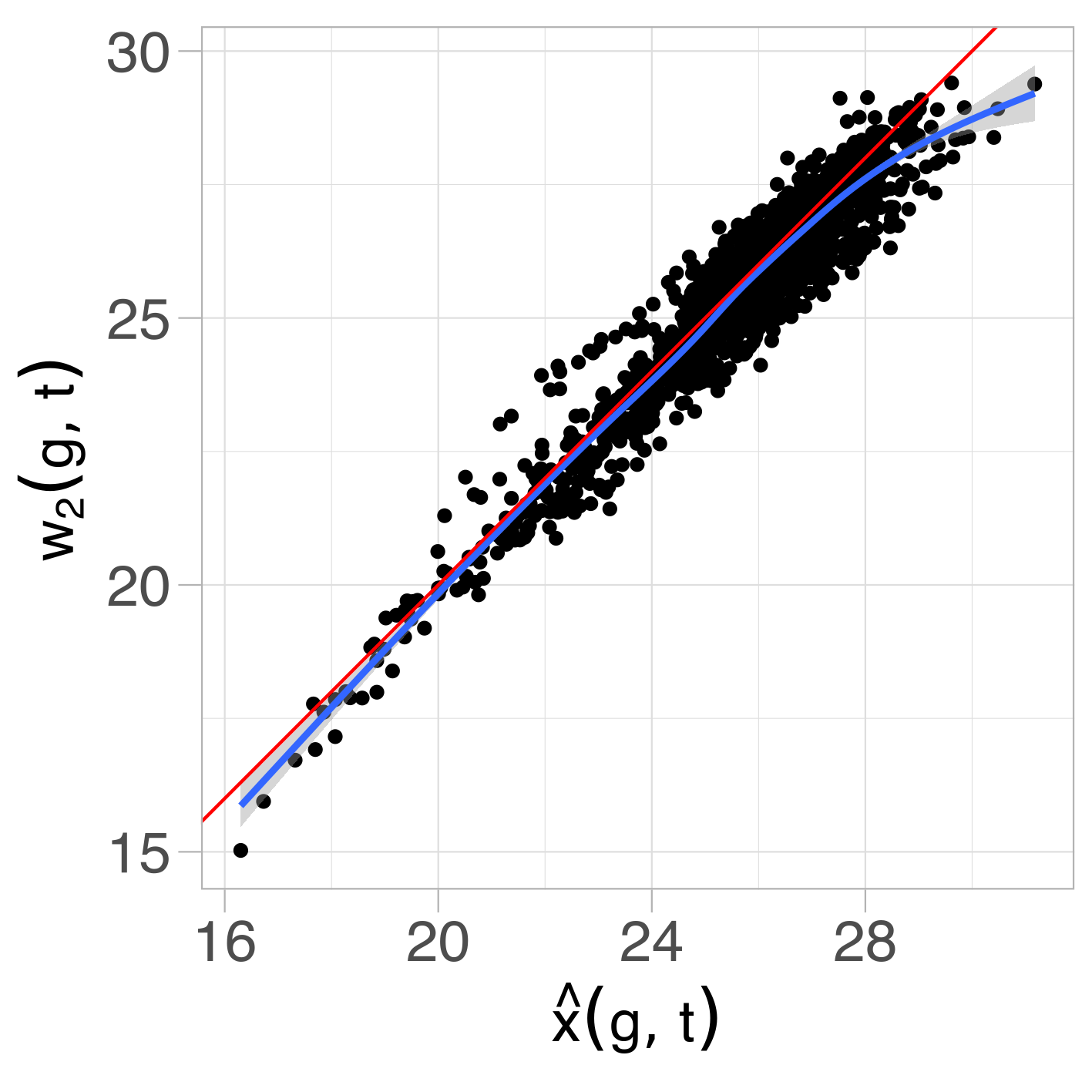}\label{fig:TMEAN_scatter2}}
     \subfloat[][$\text{w}_2(\mathbf{g},t)$ vs $\hat{\text{w}}_2(\mathbf{g},t)$]{\includegraphics[trim={0cm 0cm 0 0cm},clip,scale=0.28]{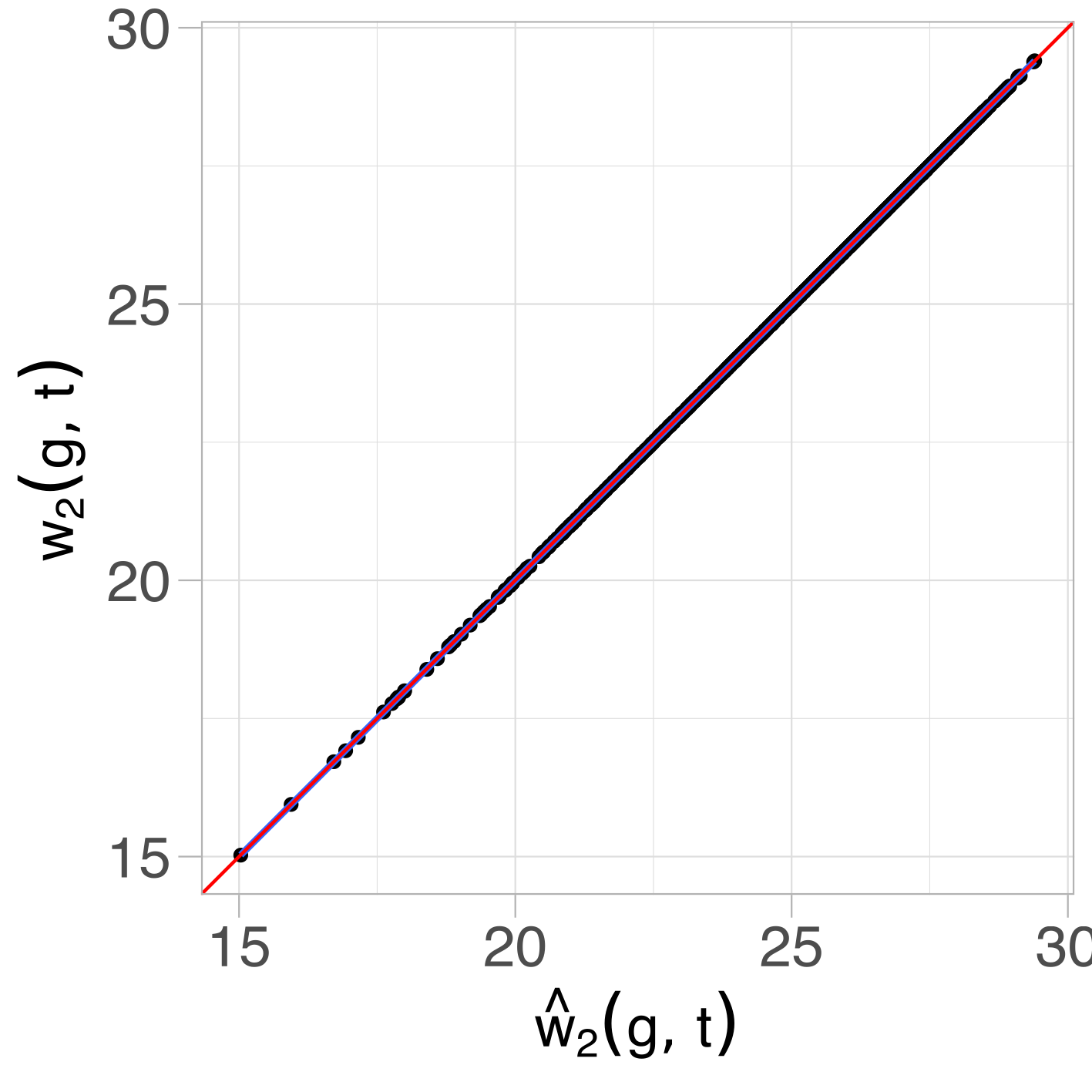}\label{fig:TMEAN_scatter3}}
     \caption{Plot of observed temperature values versus predicted values using the proposed data fusion model for (a) weather stations, (b) GSM data, and (c) calibrated GSM data. The blue line is the smooth local regression curve, while the red line is the $x=y$ line.}
     \label{fig:TMEANscatters}
\end{figure}

Figure \ref{fig:TMEAN_scatter1} shows a close correspondence  between the observed values at the stations $\text{w}_1(\mathbf{s}_i,t)$ and the corresponding predicted latent values $\hat{x}(\mathbf{s}_i,t)$ using the proposed data fusion model. Figure \ref{fig:TMEAN_scatter2}, which shows a scatterplot between the observed GSM values  $\text{w}_2(\mathbf{g}_j,t)$ and the corresponding predicted latent values $\hat{x}(\mathbf{g}_j,t)$, indicates a strong bias in the GSM values, with several points that are either overestimated or underestimated. Finally, Figure \ref{fig:TMEAN_scatter3} shows a very close correspondence between the GSM data values $\text{w}_2(\mathbf{g}_j,t)$ and the predicted values $\hat{\text{w}}_2(\mathbf{g}_j,t)=\hat{\alpha}_0(\mathbf{g}_j,t)+\hat{\alpha}_1\hat{x}(\mathbf{g}_j,t)$. This implies that the error field can be effectively used to calibrate the GSM outcomes for temperature via $\hat{x}(\mathbf{g}_j,t) = \Big(\text{w}_2(\mathbf{g}_j,t)-\hat{\alpha}_0(\mathbf{g}_j,t)\Big)/\hat{\alpha}_1.$

\subsection{Relative Humidity}\label{subsec:res_relhum}
The mean (standard deviation) of the observed relative humidity (RH) is 81.51 (6.06) for the stations and  83.33 (4.86) for the GSM. The percentage of missing values from the stations data is only 3.74\%. Although the RH variable is bounded from 0 to 100, we assumed a Gaussian likelihood for the response variable, which is reasonable since the range of the RH values in the data is from 61 to 94.36, i.e., none of the values are equal to the bounds. Also, all the predicted values from the models are well within the bounds. A model which properly constrains the values can be considered in a future work. 

The predictor expression for the fixed effects in the latent process is as follows:
\begin{equation}
\label{eq:rh_latentprocess}
\begin{split}
    \log\Big(\text{RH}(\mathbf{s},t)\Big) = \beta_0 + \beta_1  \log\text{Temperature}&(\mathbf{s},t) + \beta_2  \Big(\log\text{Temperature}(\mathbf{s},t)\Big)^2 \\
   & + \beta_3 \log\Big(\text{Elevation}(\mathbf{s},t)\Big) + \beta_4 \text{ClimateType}(\mathbf{s},t).
\end{split}
\end{equation}
 Elevation and climate type are also used as predictors. In addition, we used log temperature and its quadratic term, as recommended by \cite{PHClimate_PAGASA}. Such non-linear relationship between RH and temperature is also established in the atmospheric science literature \citep{goody1995principles}. A log transformation on temperature is possible since the Philippines is a tropical country with mean temperature ranging from 16$^{\circ}$C to 32$^{\circ}$C. We used the predictions generated from the temperature model in Section \ref{subsec:res_meantemp} as input in Equation \eqref{eq:rh_latentprocess}. 
 
PC priors are used for the Mat\'ern field parameters. For the range, we used the same values as in Section \ref{subsec:res_meantemp}. For the marginal standard deviation, we set $\sigma_{
1\text{o}}=0.08$ and $ \sigma_{2\text{o}}=0.01$.  The variance parameters of $e_1(\mathbf{s}_i,t)$ and $e_2(\mathbf{g}_j,t)$ are also given PC priors, with $\sigma_{{e_1}_\text{o}}=0.01$ and $\sigma_{{e_2}_\text{o}}=0.004$. The probability value in the PC priors are also set equal to 0.50. The rest of the model parameters are given the default non-informative priors.

We used the same grid of $\alpha_1$ values as Section \ref{subsec:res_meantemp} to fit the conditional INLA models. The results shows that the model with $\alpha_1=1$ gave the highest marginal log-likelihood value and with a weight close to 1, while the rest of the $\alpha_1$ values have weights close to 0. The marginal log-likelihoods $\log \pi\Big(\mathbf{Y}|\alpha_1^{(k)}\Big)$ and the corresponding BMA weights $w_k$ are shown in Table \ref{tab:RH_mliksweights} of Appendix \ref{subsec:app_relhum}. 

Table \ref{tab:RHestimates_fixed} shows the posterior estimates of the model fixed effects for the stations-only model and the proposed data fusion model. The estimates are quite similar, although the climate type variable is not significant in the stations-only model. The results show that there is a significant non-linear relationship between temperature and relative humidity, and that the elevation variable is negatively associated with relative humidity. Moreover, the climate type variable is positively related with relative humidity which means that areas in the eastern section of the country have higher relative humidity, on average, than the western part. 

\begin{table}[h]
\caption{\label{tab:RHestimates_fixed}Posterior estimates of fixed effects for the relative humidity model -- stations-only model versus proposed data fusion model}
\centering
\begin{tabular}{|l|cccc|cccc|}
\hline\hline
 &  \multicolumn{4}{c|}{\textbf{Stations only}} & \multicolumn{4}{c|}{\textbf{Proposed model}}\\
 Parameter & Mean & SD & P2.5\% & P97.5\% & Mean & SD & P2.5\% & P97.5\% \\ 
  \hline\hline
$\beta_0$ & 4.451 & 0.030 & 4.392 & 4.509 & 4.451 & 0.050 & 4.353 & 4.548 \\ 
  $\beta_1$, \color{brown}log(Temp) & 0.567 & 0.048 & 0.473 & 0.661 & 0.790 & 0.039 & 0.714 & 0.866 \\ 
  $\beta_2$, \color{brown}log(Temp)$^2$ & -0.173 & 0.014 & -0.201 & -0.145 & -0.237 & 0.012 & -0.260 & -0.215 \\ 
  $\beta_3$, \color{brown}log(Elevation)  & -0.008 & 0.003 & -0.014 & -0.002 & -0.013 & 0.002 & -0.017 & -0.010 \\ 
  $\beta_4$, \color{brown}Climate Type & 0.028 & 0.0180 & -0.007 & 0.063 & 0.026 & 0.009 & 0.009 & 0.043   \\ 
   \hline\hline
\end{tabular}
\end{table}


As in Section \ref{subsec:res_meantemp},  the range $\rho_{1}$ of the spatial field in the latent process $\xi(\mathbf{s},t)$ is estimated to be larger than the range $\rho_2$ of the error field $\alpha_0(\mathbf{s},t)$ (see Table \ref{tab:RHestimates_hyperpar} of Appendix  \ref{subsec:app_relhum}). 
Moreover, the estimated marginal variance of the spatial field is also larger than that of the error field. The estimates of the AR parameter are both close to 1, although the estimated value of the parameter for the spatial field is higher than that of the error field. Moreover, Table \ref{tab:RHestimates_regcalib} of Appendix \ref{subsec:app_relhum} shows the posterior estimates of the regression calibration model for relative humidity. The $2.5^{\text{th}}$ and $97.5^{\text{th}}$ percentile of the multiplicative bias estimates of the regression calibration model are 0.96 and 1.02, respectively. The values are approximately equal to 1, which justifies the assumption of a constant $\alpha_1$ in the proposed model, and agrees with the results from the proposed model.

\begin{figure}[h]
    \centering
 \subfloat[][Stations-only model]      {\includegraphics[scale=0.16]{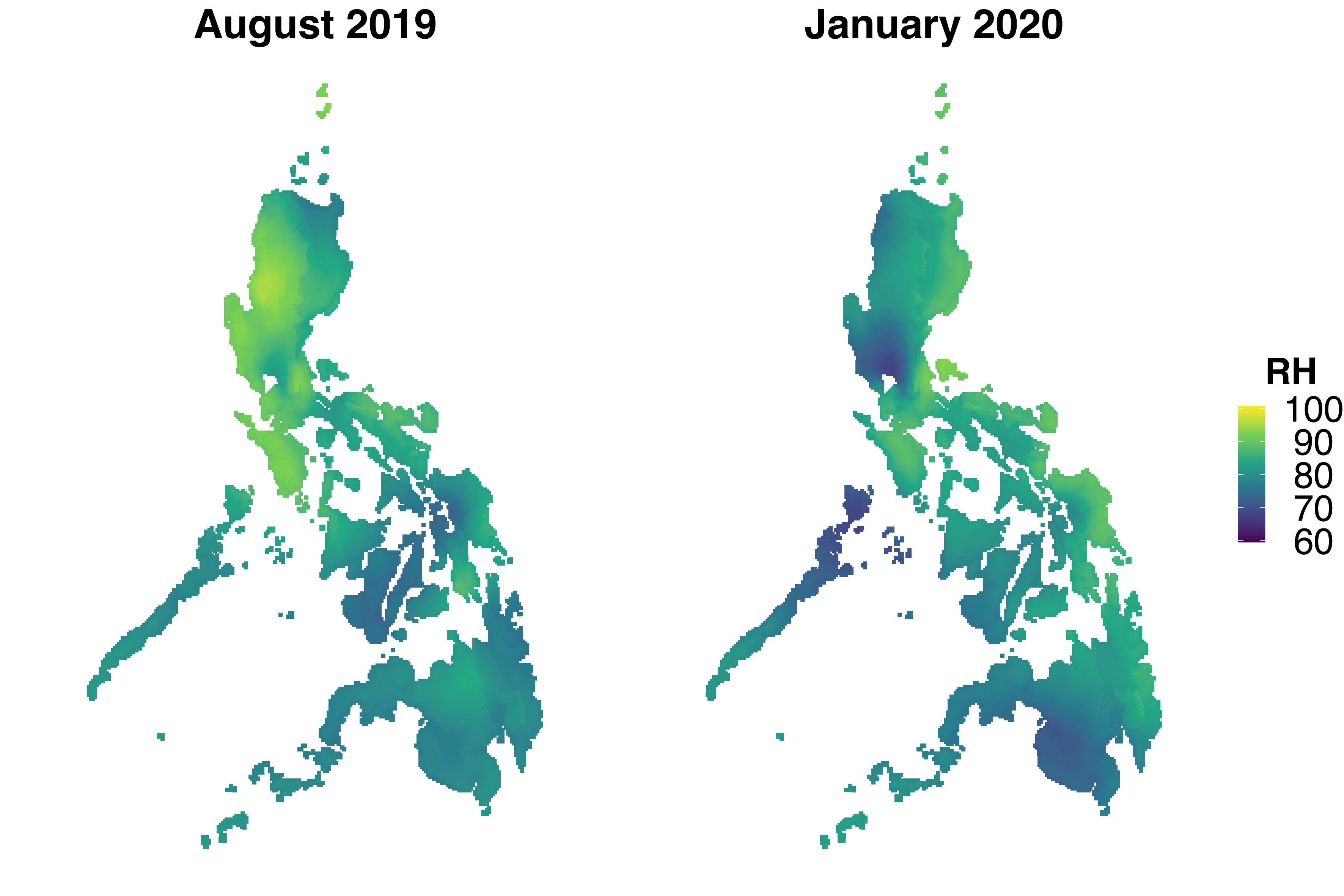}    \label{fig:RH_fieldstationsonly}}
 \subfloat[][Regression calibration model]      {\includegraphics[scale=0.16]{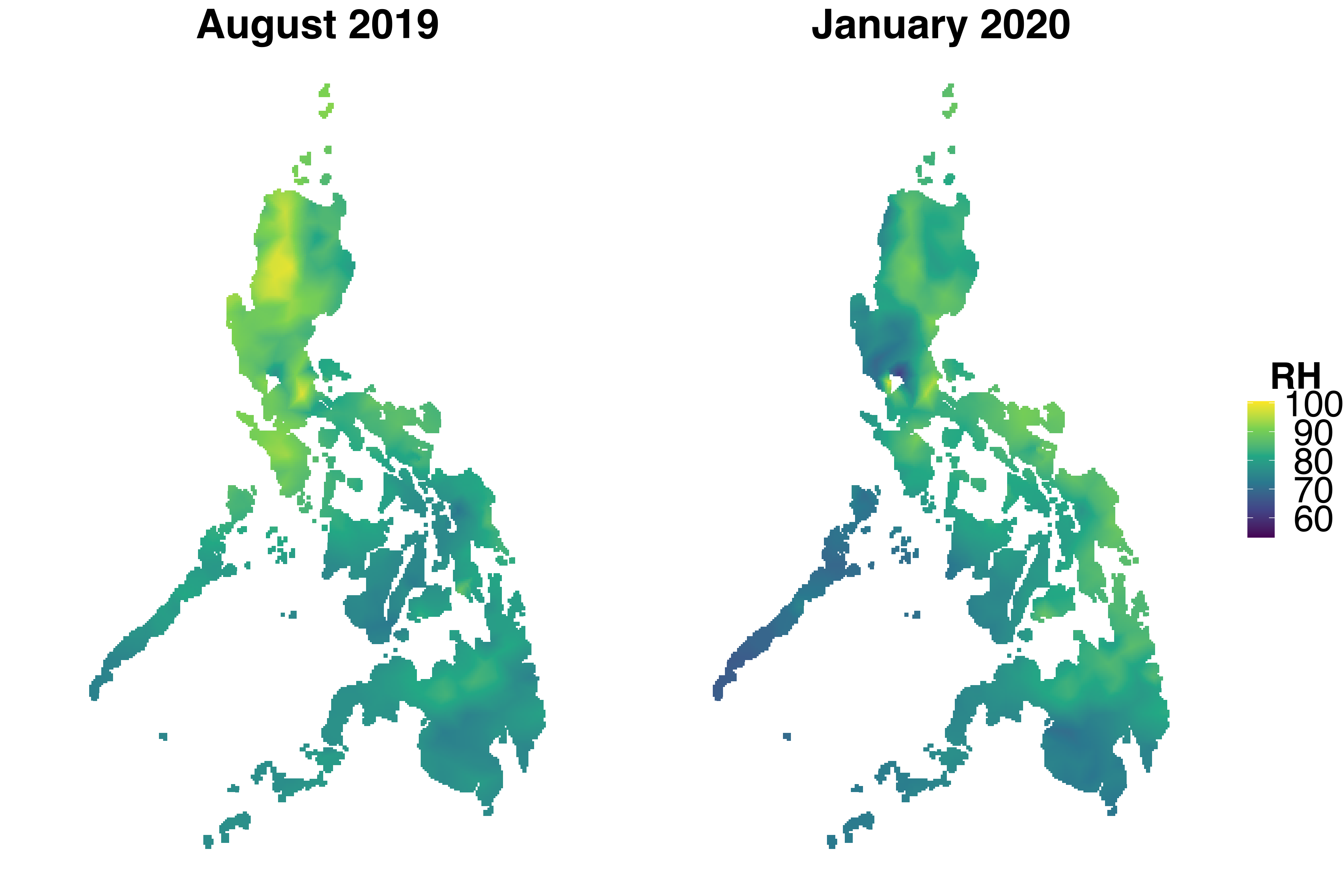}    \label{fig:RH_fieldregcalib}}
    \subfloat[][Proposed data fusion model] {\includegraphics[scale=0.16]{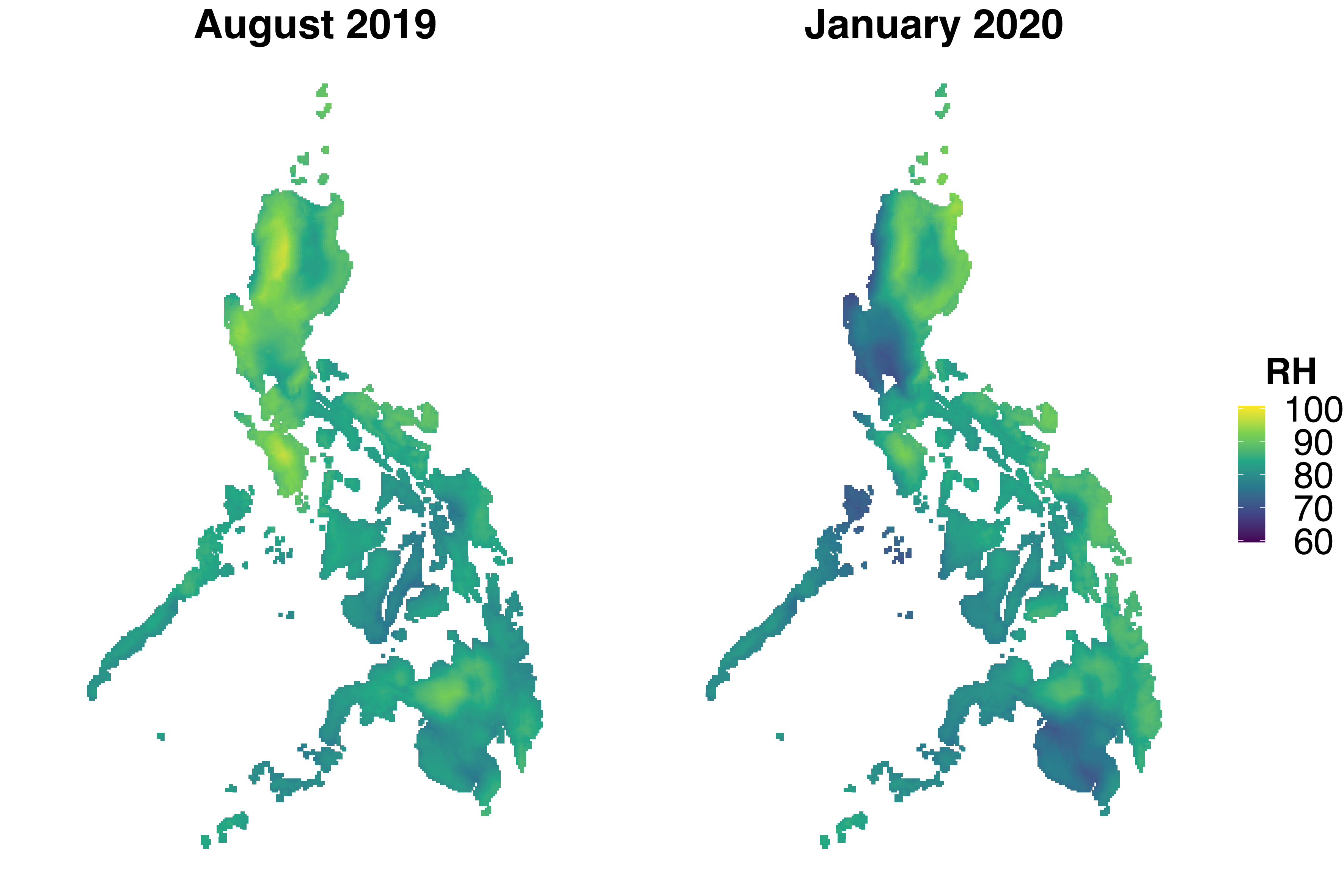}    \label{fig:RH_fielddatafusion}}    
    \caption{Comparison of estimated relative humidity fields for August 2019 and January 2020: (a) stations-only model, (b) regression calibration model, and (c) proposed data fusion model. There is more smoothing in the estimated fields using the stations-only model.}
    \label{fig:RHfields}
\end{figure}

\begin{figure}[h]
    \centering
    \subfloat[][Stations-only model]
    {\includegraphics[scale=0.16]{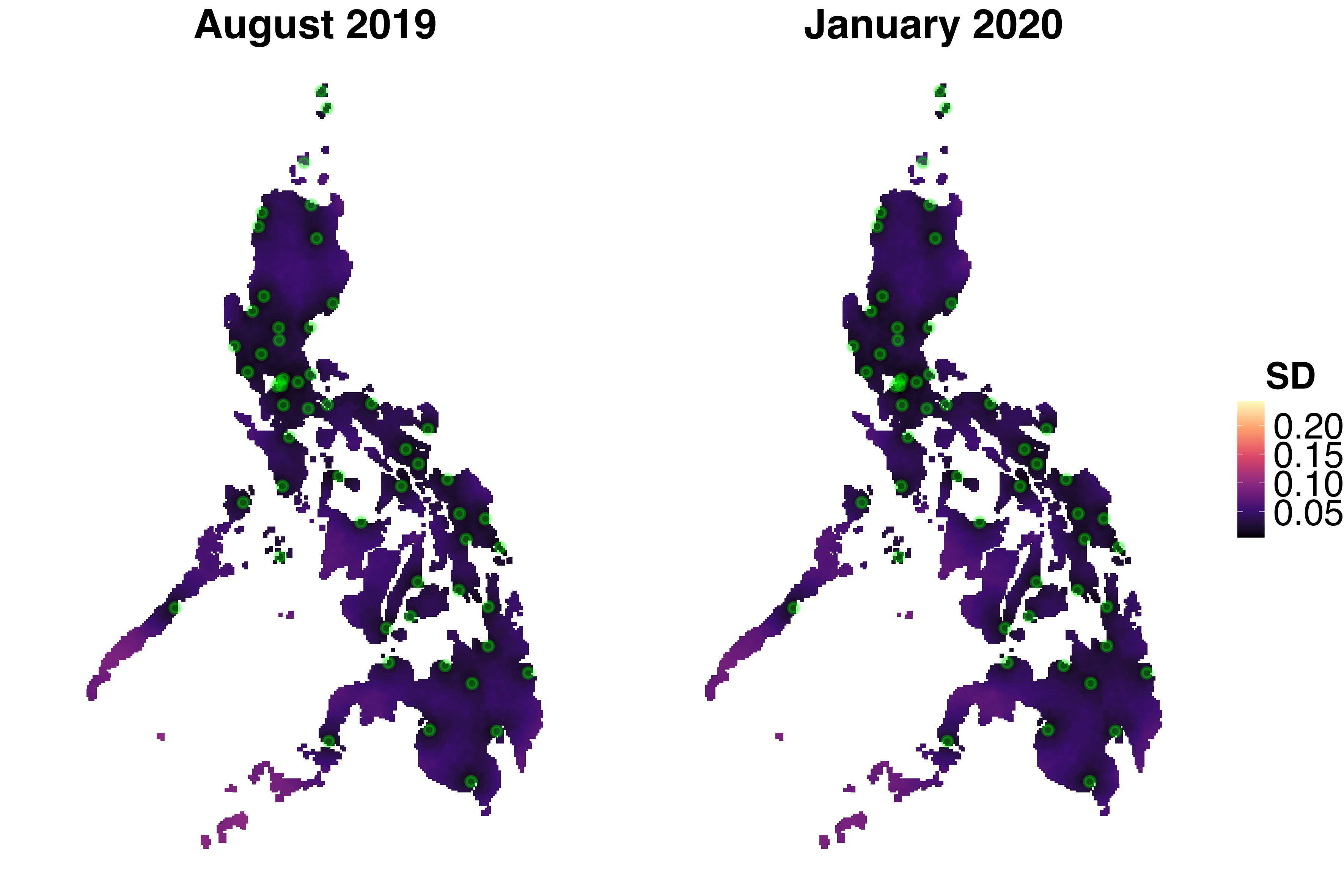}
    \label{fig:RH_fieldsdstationsonly}}
    \subfloat[][Regression calibration model]
    {\includegraphics[scale=0.16]{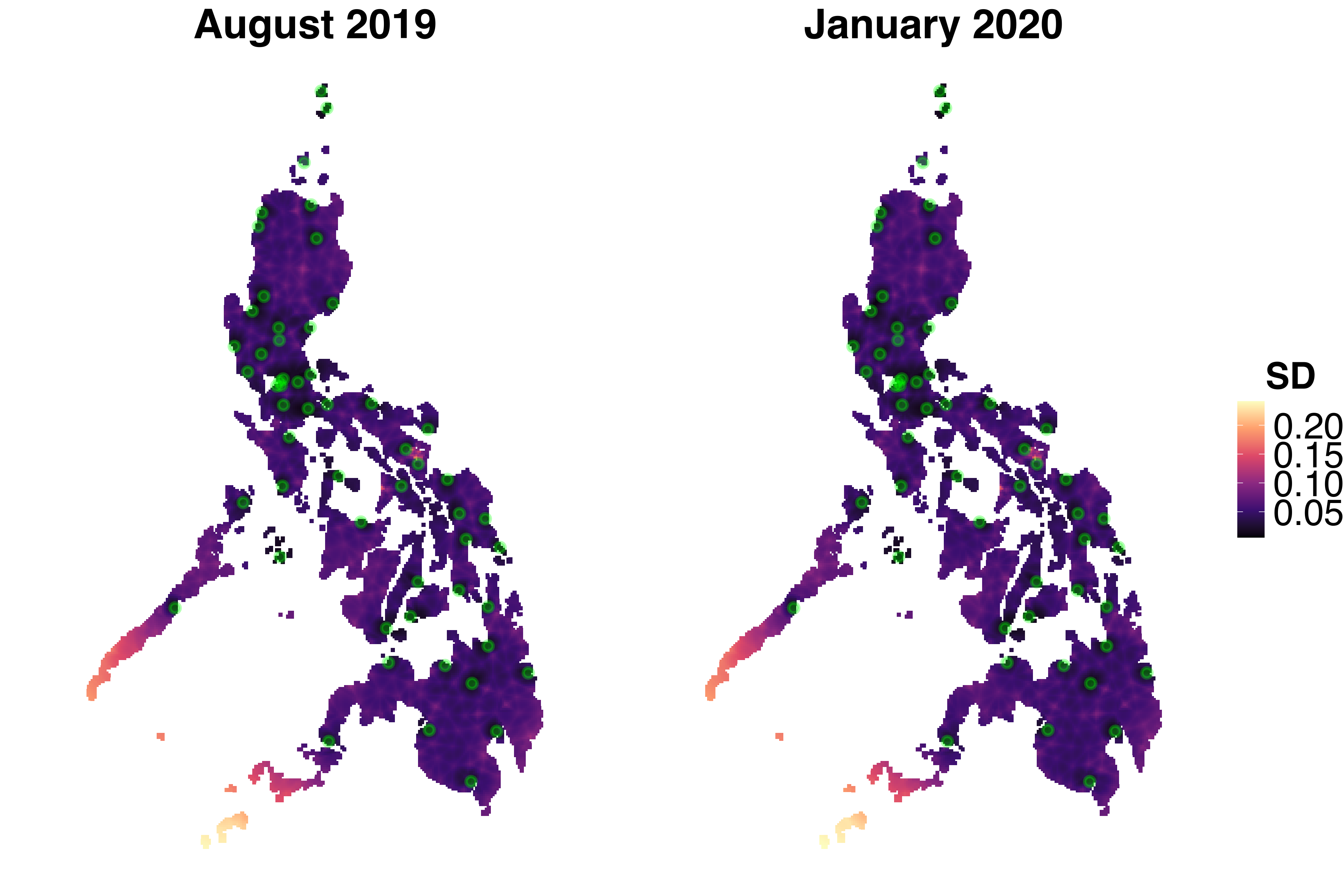}
    \label{fig:RH_fieldsdregcalib}}
    \subfloat[][Proposed data fusion model]    {\includegraphics[scale=0.16]{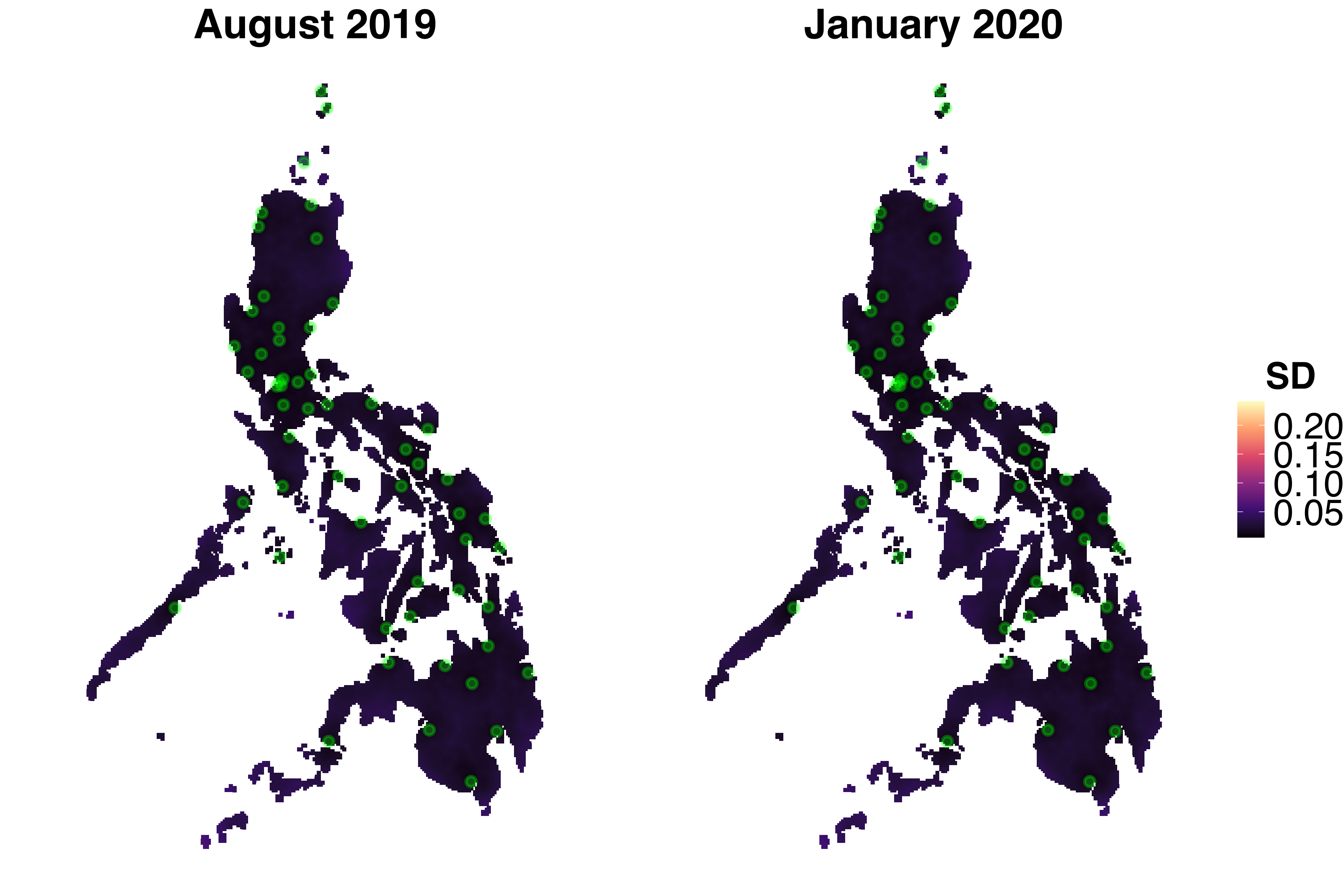}\label{fig:RH_fieldsddatafusion}}     
    \caption{Posterior uncertainty of the estimated relative humidity fields in Figure \ref{fig:RHfields}. The posterior uncertainty in the estimated field from the proposed model is the smallest.}
    \label{fig:RH_fieldsd}
\end{figure}

Figure \ref{fig:RHfields} shows the estimated relative humidity fields for two different months: August 2019 and January 2020. These two specific months were chosen since August is a rainy month while January is a dry month \citep{PHClimate_PAGASA}. The predicted fields show very similar structure, although it is apparent that there is more smoothing in the estimates from the stations-only model. The estimated fields show that in the eastern section of the country, the level of relative humidity is similar for the two months. On the other hand, in the western section, particularly in the northwestern section, relative humidity is very high in August, and very low in January. These dynamics in relative humidity is consistent with the climate types in the Philippines \citep{coronas1920climate, kintanar_climate, PHClimate_PAGASA}. Figure \ref{fig:RH_fieldsd} shows the corresponding uncertainty estimates of the estimated relative humidity fields. As expected, the proposed data fusion model has the lowest posterior uncertainty. The average of the posterior standard deviation in the predicted fields from the stations-only, regression calibration, and the proposed data fusion model is 0.040, 0.040, and 0.022, respectively. 

The estimated spatial field of the latent process, $\hat{\xi}(\mathbf{s},t)$, for the same two months are shown in
Figure \ref{fig:RH_spatialfields} of Appendix \ref{subsec:app_relhum}. The spatio-temporal dynamics observed in Figure \ref{fig:RHfields} are also evident in the estimated spatial fields. Finally, Figures  \ref{fig:RH_scatter1},  
\ref{fig:RH_scatter2}, and \ref{fig:RH_scatter3} of Appendix  \ref{subsec:app_relhum} show different scatterplots that indicate a close correspondence between the observed and predicted values, and a strong bias in the GSM values.

\subsection{Rainfall}\label{subsec:res_rainfall}
The mean (standard deviation)  of the cumulative monthly rainfall (in $mm$) is 220.27 (207.69) for the 
stations data and 166.12 (114.99) for the GSM data. Since the Philippines has high amounts rainfall and the values are aggregated monthly, there are very few zeros in the data (1.52\%  for the stations data and 0\% for the GSM).

The predictor expression for the fixed effects is as follows:
\begin{equation}\label{eq:rain_latentprocess}
\begin{split}
\log\Big(\text{Rainfall}(\mathbf{s},t)+1\Big) = \beta_0 + \beta_1  \log&\text{Temperature}(\mathbf{s},t) + \beta_2  \Big(\log\text{Temperature}(\mathbf{s},t)\Big)^2 + \beta_3 \text{Season}(\mathbf{s},t) \\
&+ \beta_4\text{ClimateType}(\mathbf{s},t) + \beta_5 \text{ClimateType}(\mathbf{s},t)\times\text{Season}(\mathbf{s},t).
\end{split}
\end{equation}

The `Season' variable is  binary and takes a value of `1' for June to November (characterized as a wet period), and a value of `0' for the rest of the year (characterized as a dry period). As for the relative humidity model, the log temperature and its squared term are included as  predictors, as recommended by \cite{PHClimate_PAGASA}. We also used the predictions from the temperature model as input in Equation \eqref{eq:rain_latentprocess}. An interaction effect between climate type and season was included to capture the climate dynamics of the country, as recommended by \cite{PHClimate_PAGASA}. 

As for the previous models, PC priors are used for the Mat\'ern field parameters. For the range parameters, we used the same values as before, while for the marginal standard deviations, we have $\sigma_{1\text{o}}=1.35$,  and $\sigma_{2\text{o}}=0.01$. The variance parameters of $e_1(\mathbf{s}_i,t)$ and $e_2(\mathbf{g}_j,t)$ are also given PC priors, with $\sigma_{{e_1}_\text{o}}=0.5$ and $\sigma_{{e_2}_\text{o}}=0.26$. The probability value in the PC priors is equal to 0.50. The rest of the model parameters are given the default non-informative priors.
\begin{figure}[h]
     \centering
    \includegraphics[trim={0cm 0cm 0cm 0cm},clip,scale=.23]{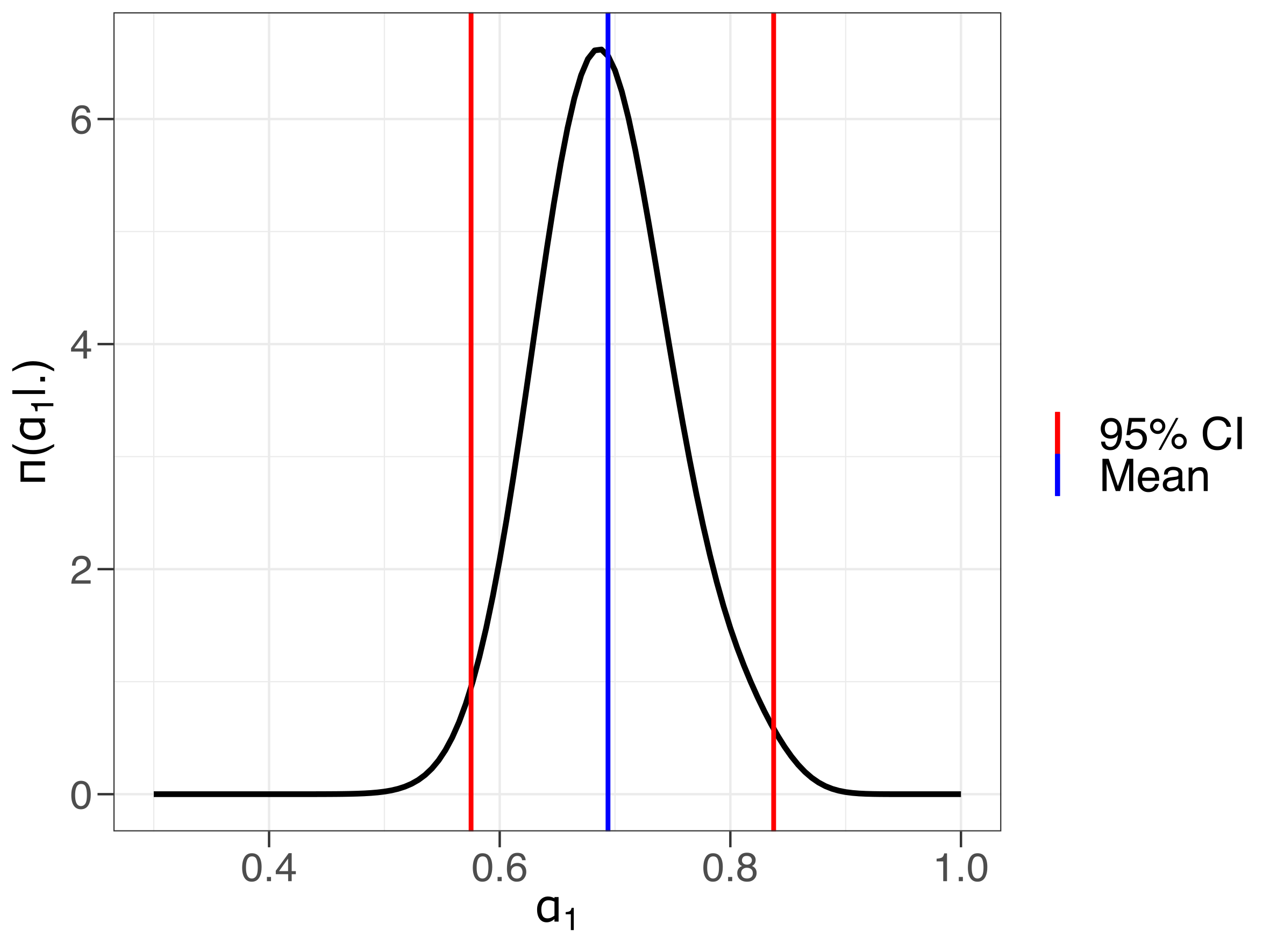}
    \caption{Estimated marginal posterior of $\alpha_1$, $\pi(\alpha_1|\mathbf{Y})$, for the rainfall data fusion model. The posterior mean is 0.6733, while the 95\% credible interval estimate is $(0.5607,0.8353)$.}
     \label{fig:alpha_1_dist}
\end{figure}

Figure \ref{fig:alpha_1_dist} shows the estimated marginal posterior distribution of $\alpha_1$, $\pi(\alpha_1|\mathbf{Y})$. The posterior mean is 0.6733 while the 95\% credible interval estimate is (0.5607, 0.8353). Unlike the two previous climate variables, the multiplicative bias parameter for the GSM outcomes for rainfall is significantly different from 1, implying  a more severe bias for rainfall outcomes. This is expected since Figure \ref{fig:GSMvsStations} shows a large discrepancy between the interpolated GSM outcomes and the observed values at the weather stations for log-transformed rainfall.  This also agrees with the insights from the LGOCV results discussed in Section \ref{subsec:LGOCV}. 

\begin{figure}
    \centering
    \subfloat[][Stations-only model]
    {\includegraphics[scale=0.16]{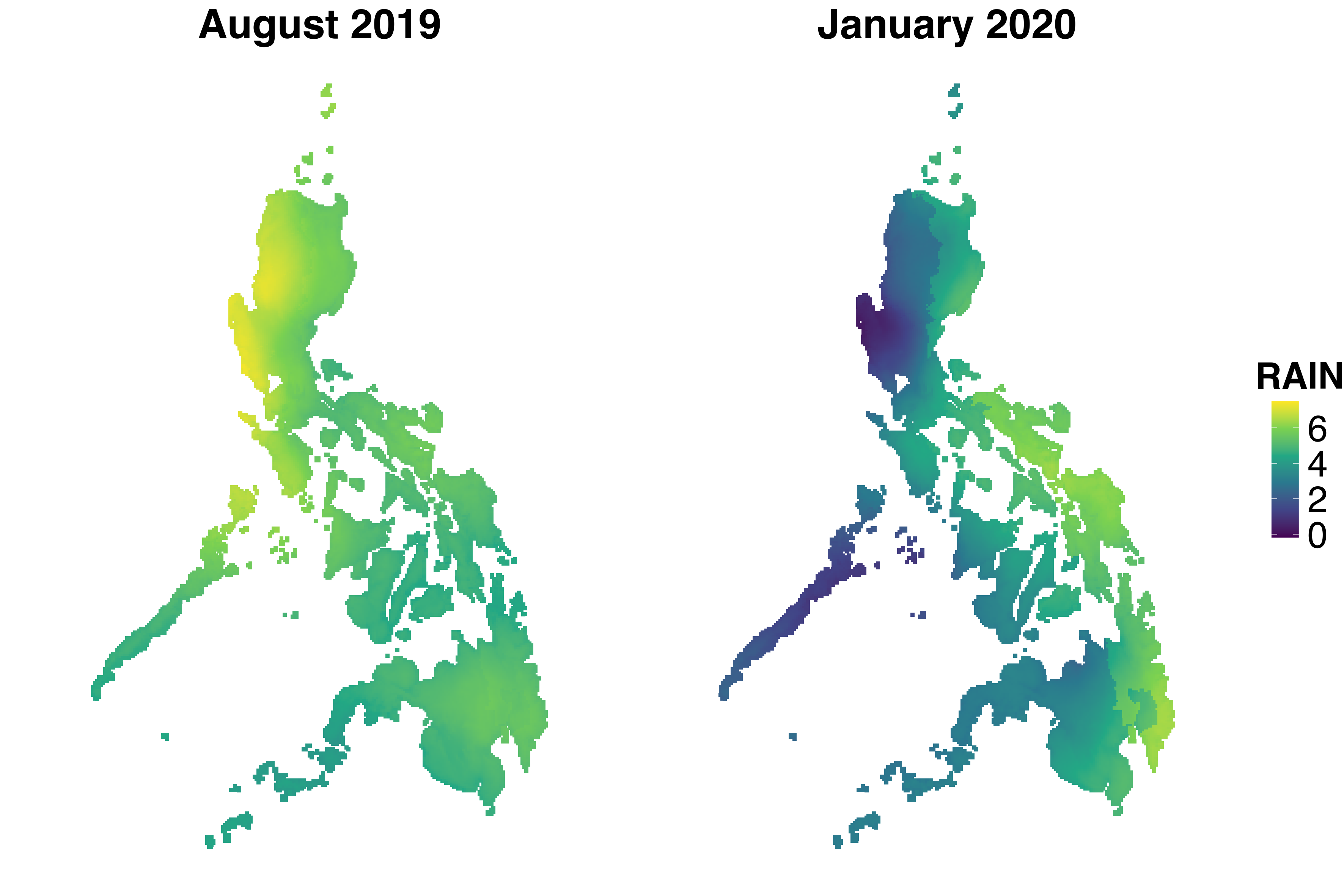}
    \label{fig:logRAIN_fieldstationsonly}}
    \subfloat[][Regression calibration model]
    {\includegraphics[scale=0.16]{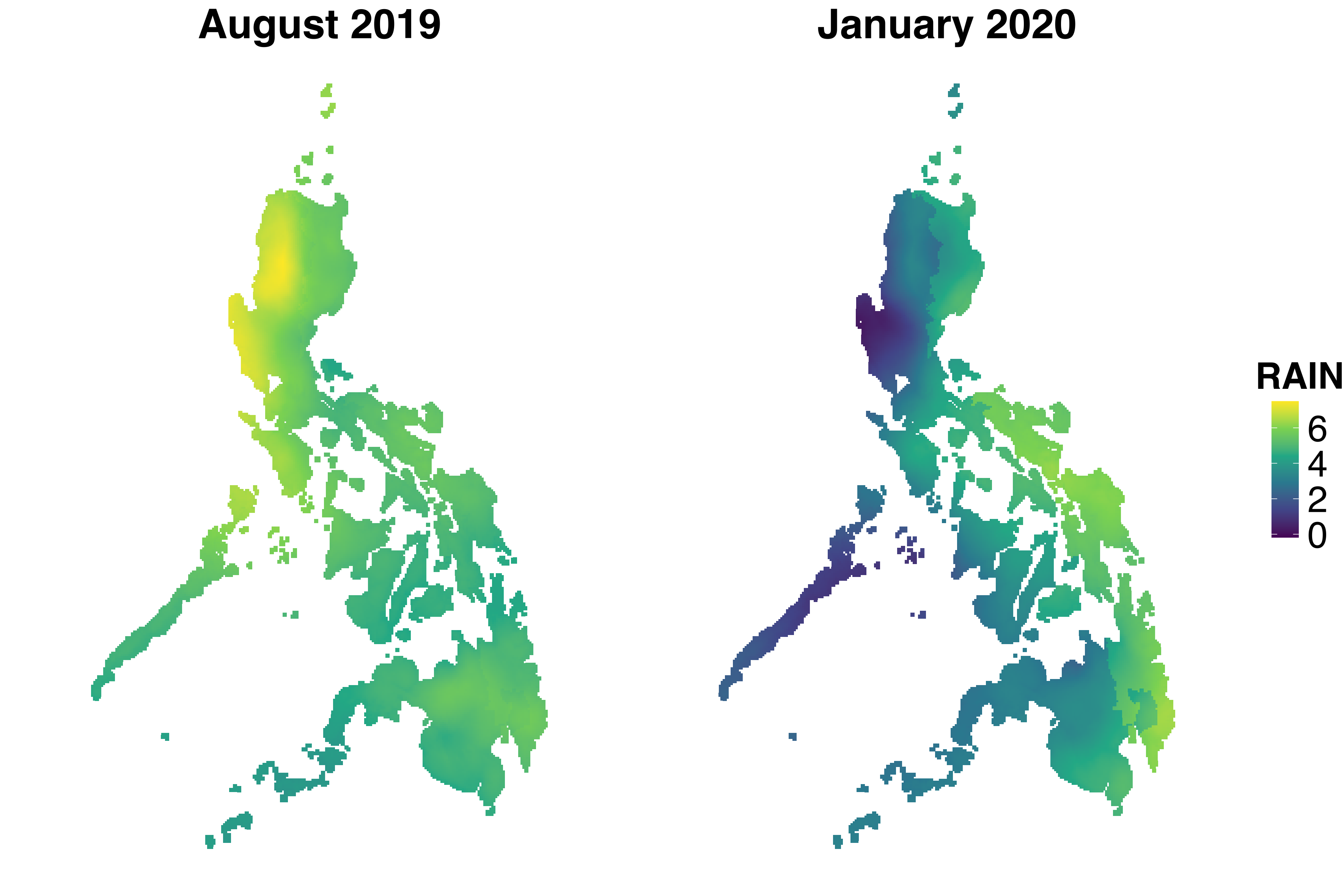}
    \label{fig:logRAIN_fieldregcalib}}
    \subfloat[][Proposed data fusion model]
    {\includegraphics[scale=0.16]{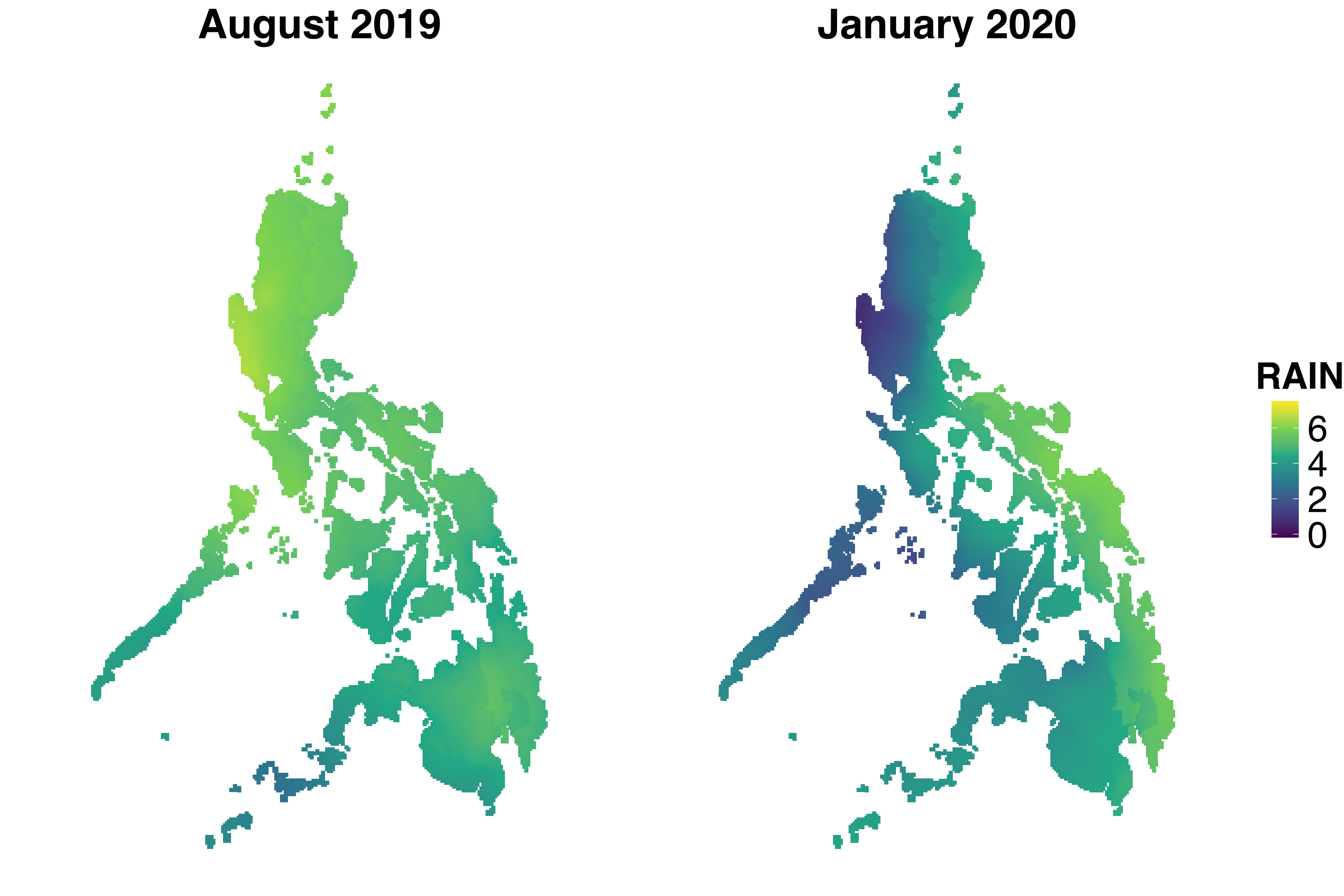}
    \label{fig:RAIN_field}}
    \caption{Comparison of estimated log rainfall fields for August 2019 (wet season) and January 2020 (dry season) between (a) stations-only model, (b) regression calibration model, and (c) proposed data fusion model. The figures show that the western section of the country has a pronounced dry and wet season.}
    \label{fig:RAINfields}
\end{figure}

Table \ref{tab:RAINestimates_fixed} shows the posterior estimates of the fixed effects for the stations-only model and proposed data fusion model. The results show that log temperature has a non-linear association with log rainfall amounts. Moreover, there is a significant interaction between season and climate type: the western part of the country has a pronounced dry and wet season, while the eastern part of the country has a less pronounced dry and wet season and with more or less evenly distributed rainfall for the whole year. This can be confirmed in Figure \ref{fig:RAINfields} which shows the predicted log rainfall fields for two months -  August 2019 (rainy month) and January 2020 (dry month). This climatic pattern is consistent with theory \citep{coronas1920climate, kintanar_climate, PHClimate_PAGASA}, and are the same seasonal dynamics observed for relative humidity in Section \ref{subsec:res_relhum}. Table \ref{tab:RAINestimates_regcalib} of Appendix \ref{subsec:app_rainfall} shows the  posterior estimates of the regression calibration model for rainfall. The $2.5^{\text{th}}$ and $97.5^{\text{th}}$ percentile of the multiplicative bias estimates of the regression calibration model are 0.35 and 1.35, respectively. The values vary significantly in space and time, which raises doubt on the assumption of a constant $\alpha_1$ in the proposed model. This agrees with the exploratory plot in Figure \ref{fig:GSMvsStations}c, and we also have noted this as a limitation of the model for the rainfall variable. A model which specifies a spatially and temporally varying multiplicative multiplicative bias in the proposed model can be explored in a future work. This can be computationally difficult since this involves estimating the product of two Gaussian fields.

The estimated log rainfall fields from the three approaches look  similar, but the uncertainty in the predictions from the proposed data fusion model is  the smallest as expected, which are shown in Figure \ref{fig:RAINfields_sd} of Appendix \ref{subsec:app_rainfall}. The estimated spatial fields from the three modelling approaches for the same two months are shown in Figure \ref{fig:RAIN_spdes} of Appendix \ref{subsec:app_rainfall}. The estimated spatial fields also look very similar. Finally, the estimated error fields for the same two months from the proposed data fusion model are shown in Figure \ref{fig:RAIN_error} of Appendix \ref{subsec:app_rainfall}. 

\begin{table}[h]
\caption{\label{tab:RAINestimates_fixed}Posterior estimates of fixed effects for the rainfall model -- stations-only model versus proposed data fusion model}
\centering
\begin{tabular}{|l|cccc|cccc|}
\hline\hline
 &  \multicolumn{4}{c|}{\textbf{Stations only}} & \multicolumn{4}{c|}{\textbf{Proposed model}}\\
 Parameter & Mean & SD & P2.5\% & P97.5\% & Mean & SD & P2.5\% & P97.5\% \\ 
  \hline\hline
$\beta_0$ & 4.427 & 0.360 & 3.722 & 5.132 & 4.759 & 0.377 & 3.931 & 5.484 \\ 
  $\beta_1$, \color{brown} log(Temp) & 2.186 & 0.454 & 1.296 & 3.076 & 1.672 & 0.430 & 0.911 & 2.541 \\ 
  $\beta_2$, \color{brown} log(Temp)$^2$ & -0.699 & 0.134 & -0.961 & -0.437 & -0.570 & 0.122 & -0.809 & -0.354 \\ 
  $\beta_3$, \color{brown} Season & 0.795 & 0.306 & 0.195 & 1.395 & 0.444 & 0.261 & -0.020 & 0.973 \\ 
  $\beta_4$, \color{brown} Climate Type & 1.183 & 0.146 & 0.898 & 1.469 & 0.657 & 0.112 & 0.458 & 0.873  \\ 
  $\beta_5$, \color{brown} Climate Type $\times$ Season & -0.844 & 0.162 & -1.161 & -0.527 & -0.287 & 0.099 & -0.461 & -0.106  \\ 
   \hline\hline
\end{tabular}
\end{table}


Table \ref{tab:RAINestimates_hyperpar} of Appendix \ref{subsec:app_rainfall} shows the posterior estimates of the hyperparameters for the stations-only model and the proposed data fusion model. Similar to the previous meteorological variables, the estimated range $\hat{\rho}_{1}$ of the spatial field in the latent process is larger than the estimated range $\hat{\rho}_{2}$ of the error field. This is also true for the estimated marginal variances of the two fields. The estimated autocorrelation parameters of the two fields are very different, with the AR parameter for the error field being much larger. 
 
Finally, Figures  \ref{fig:RAIN_scatter1}, \ref{fig:RAIN_scatter2}, and \ref{fig:RAIN_scatter3} of Appendix \ref{subsec:app_relhum} show different scatterplots that indicate the correspondence between the observed and predicted values, although these are not as strong as the previous two meteorological variables. In particular, Figure \ref{fig:RAIN_scatter2} of Appendix \ref{subsec:app_relhum} shows the severe bias in the GSM values for rainfall.

\subsection{Leave-group-out cross-validation}\label{subsec:LGOCV}
We end our case study by evaluating the prediction accuracy of the three modelling approaches using the leave-group-out cross-validation (LGOCV) approach \citep{liu2022leave,adin2023automatic}. Contrary to the leave-one-out cross-validation method which estimates the predictive density for an observation at location $\mathbf{s}_i$ at time $t$ by removing the same observation from the training set, the LGOCV approach computes the predictive densities by leaving out a set $I_{\mathbf{s}_i}$ of data points which includes the testing point and observations most related to it. The LGOCV is a better alternative than the leave-one-out cross-validation to evaluate prediction accuracy for structured models such as multi-level models, time series models, and spatial models \citep{liu2022leave,adin2023automatic} as it makes the unobserved data less dependent on the observed data, which is desirable when the goal of the prediction is extrapolation at unobserved locations. The LGOCV is efficiently implemented in the \texttt{INLA} library, with the details of the implementation found in \cite{liu2022leave}.

\begin{figure}
     \centering
    \includegraphics[trim={0cm 0cm 0 0cm},clip,scale=.25]{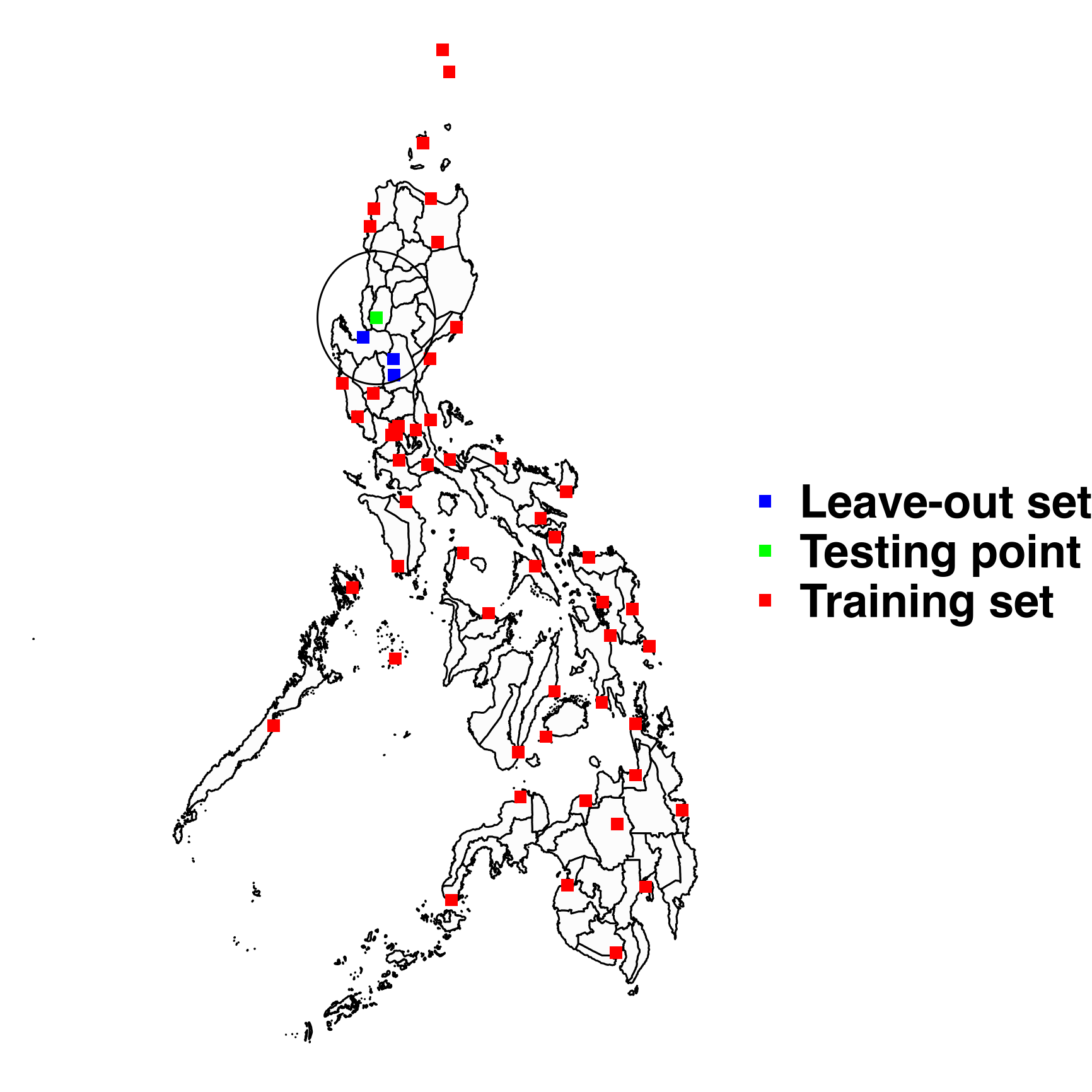}
    \caption{Illustration of the LGOCV approach. We fit the model on the training set (red) after excluding the leave-out set (blue and green), and then make predictions on the testing point (green).}
     \label{fig:CV_illustration}
\end{figure}

\cite{liu2022leave} proposed two strategies to determine the leave-out sets $I_{\mathbf{s}_i}$: an automatic procedure based on the estimated correlation of the elements of the latent field, and a manual or user-defined approach. In this study, we use the latter strategy and implement the following:
for each station $\mathbf{s}_i$, the leave-out set $I_{\mathbf{s}_i}$ for predicting $\text{w}_1(\mathbf{s}_i,t)$ consists of the station at the spatial location $\mathbf{s}_i$ and all stations within its proximity (see Figure \ref{fig:CV_illustration}). In particular, we consider four values for the radius of the leave-out set: 60, 80, 125, and 150 km. Note that the testing point is also considered part of the leave-out set. Also, we remove the observed values for all time points for each station in $I_{\mathbf{s}_i}$.

We consider the mean of the predictive density $\pi\Big(x(\mathbf{s}_i,t)|\mathbf{Y}_{-I_{\mathbf{s}_i}}\Big)$ as the predicted value at a testing point $\mathbf{s}_i$, where $\mathbf{Y}_{-I_{\mathbf{s}_i}}$ denotes the training set. Suppose $N$ denotes the total number of data points from the stations. We compute the following posterior prediction scores to compare the three modelling approaches:
\begin{enumerate}
    \item LGOCV logarithmic utility (ULGOCV): \quad\quad\quad\quad\quad\quad$\dfrac{1}{N}\mathlarger{ \sum}_{\forall i,t} \log\pi\Big(\text{w}_1(\mathbf{s}_i,t)|\mathbf{Y}_{-I_{\mathbf{s}_i}}\Big)$
    \item Root mean squared error (RMSE): \quad\quad\quad\quad\quad\quad\;\;\; $\sqrt{\dfrac{1}{N}\mathlarger{ \sum}_{\forall i,t} \Big(\text{w}_1(\mathbf{s}_i,t)-\mathbb{E}\big[x(\mathbf{s}_i,t)|\mathbf{Y}_{-I_{\mathbf{s}_i}}\big]\Big)^2}$
    \item Mean absolute error (MAE): \quad\quad\quad\quad\quad\quad\quad\quad\quad\quad\quad$\dfrac{1}{N}\mathlarger{ \sum}_{\forall i,t} \big|\text{w}_1(\mathbf{s}_i,t)-\mathbb{E}\big[x(\mathbf{s}_i,t)|\mathbf{Y}_{-I_{\mathbf{s}_i}}\big]\big|$
    \item Mean absolute percentage error (MAPE): \quad\quad\quad\quad\quad\;$\dfrac{1}{N}\mathlarger{ \sum}_{\forall i,t} \Bigg|\dfrac{\text{w}_1(\mathbf{s}_i,t)-\mathbb{E}\big[x(\mathbf{s}_i,t)|\mathbf{Y}_{-I_{\mathbf{s}_i}}\big]}{\text{w}_1(\mathbf{s}_i,t)}\Bigg|$
    \item Mean of the SD of predictive density (MSD): \quad\quad\quad\quad$\dfrac{1}{N}\mathlarger{ \sum}_{\forall i,t} \sqrt{\mathbb{V}\big[x(\mathbf{s}_i,t)|\mathbf{Y}_{-I_{\mathbf{s}_i}}\big]}$
    \item Mean Kullback-Leibler divergence (MKLD): \quad\quad\quad\quad\;$\dfrac{1}{N}\mathlarger{ \sum}_{\forall i,t} D_{\text{KL}}\Big( \pi\big(x(\mathbf{s}_i,t)|\mathbf{Y}_{-I_{\bm{s}_i}}\big) \big|\big| \pi\big(x(\mathbf{s}_i,t)|\mathbf{Y}\big) \Big)$
\end{enumerate}
The LGOCV logarithmic utility (ULGOCV) is the mean of the log predictive densities $\pi\Big(\text{w}_1(\mathbf{s}_i,t)|\mathbf{Y}_{-I_{\mathbf{s}_i}}\Big)$  which is related to the conditional predictive ordinate \citep{pettit1990conditional}. A higher value for the ULGOCV implies better model fit. The quantities $\mathbb{E}\big[x(\mathbf{s}_i,t)|\mathbf{Y}_{-I_{\mathbf{s}_i}}\big]$ and $\mathbb{V}\big[x(\mathbf{s}_i,t)|\mathbf{Y}_{-I_{\mathbf{s}_i}}\big]$ are evaluated with respect to the predictive density $\pi\Big(x(\mathbf{s}_i,t)|\mathbf{Y}_{I_{\mathbf{s}_i}}\Big)$. Moreover, $D_{\text{KL}}(\cdot||\cdot)$ denotes the the Kullback-Leibler (KL) divergence metric, so that the MKLD is the mean of the KL divergence between the predictive density for $x(\mathbf{s}_i,t)$ given the complete data and the predictive density when excluding $I_{\mathbf{s}_i}$. A smaller value for the MKLD implies a better model fit.

\begin{figure}
     \centering
    \includegraphics[trim={0cm 0cm 0 0cm},clip,scale=.31]{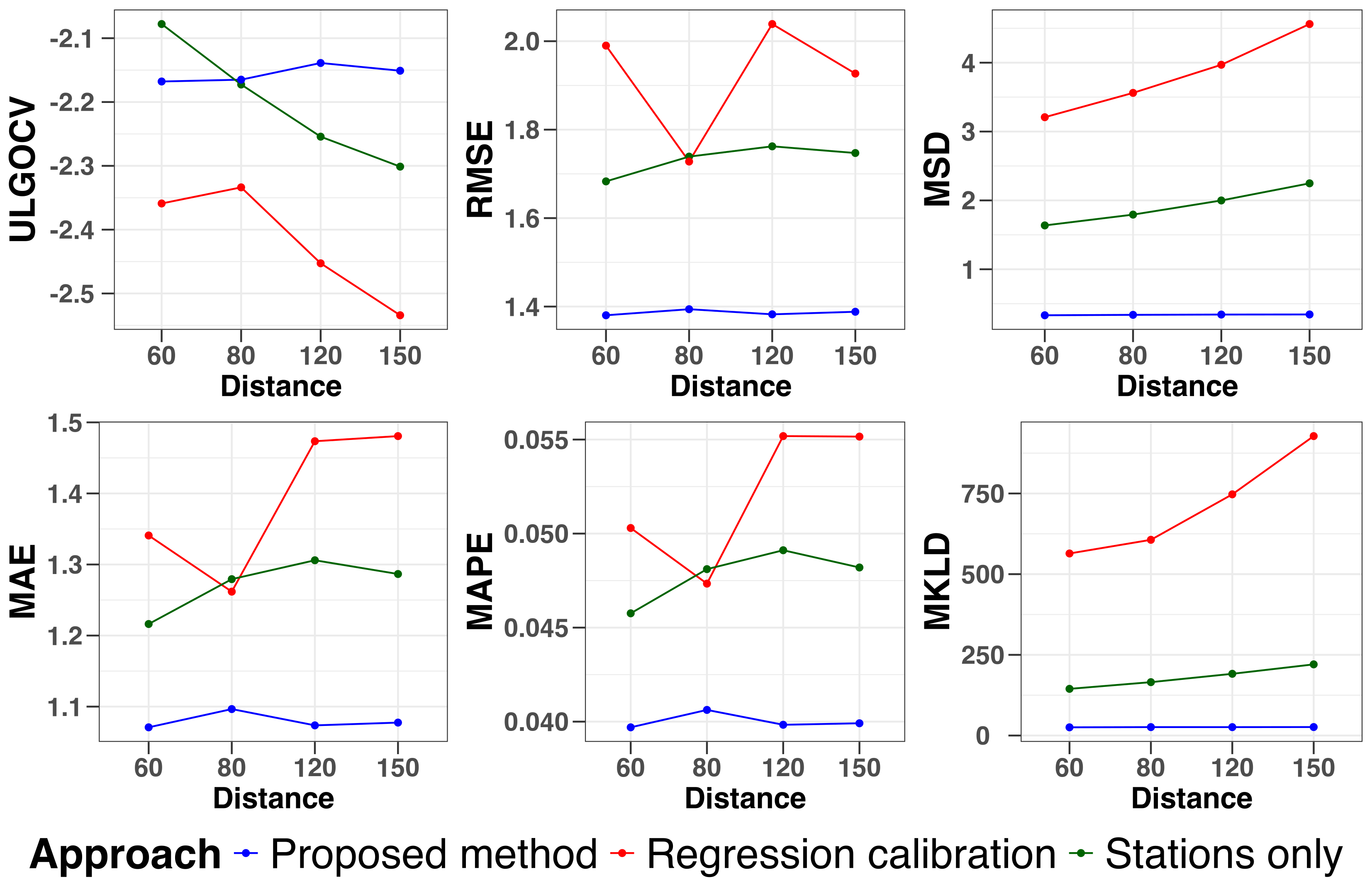}
    \caption{Comparison of LGOCV results for temperature from three models: stations-only model, regression calibration model, and the proposed data fusion model}
     \label{fig:cv.TMEAN}
\end{figure}

Figure \ref{fig:cv.TMEAN} shows a comparison of the posterior prediction scores for the temperature model. The proposed data fusion model generally has the highest ULGOCV especially when the leave-out set is large. The proposed model also has the smallest RMSE, MAE,  MAPE, MKLD and MSD. The prediction scores for the stations-only model and the regression calibration model deteriorate with the size of the leave-out set, while the scores for the proposed model are stable. The LGOCV results for the temperature model show that the proposed data fusion model outperforms the other two approaches, and that the stations-only model fares better than the regression calibration approach.

Similar results hold for  relative humidity (Figure \ref{fig:cv.RH} of Appendix \ref{subsec:app_CV}) and rainfall (Figure \ref{fig:cv.RAIN} of Appendix \ref{subsec:app_CV}). The benefits from doing data fusion is smaller for rainfall since there is no substantial difference in the scores, particularly for RMSE, MAPE, and MKLD, especially when the leave-out-sets are small. One potential reason for this is that the quality of the GSM outcomes for rainfall is lower compared to the other two meteorological variables,  which is apparent from Figure \ref{fig:GSMvsStations} and from the model results, particularly with the estimated value of $\hat{\alpha}_1<1$ for the rainfall data fusion model. Nonetheless, the LGOCV results show that the proposed data fusion model generally outperforms the other two approaches for the three meteorological variables considered.
 

\section{Conclusions}
Data fusion, which combines information from different data sources, has the potential benefit to improve model accuracy and  prediction quality, while dealing with varying quality of the data sources \citep{bauer2015quiet,gettelman2022future,lawson2016handbook}. This paper addresses a data fusion challenge motivated by meteorological data in the Philippines. The proposed model builds on the Bayesian melding model, which assumes a common latent process across data sources, and extends existing work in the literature \citep{moraga2017geostatistical, villejo2023data, zhong2023bayesian, forlani2020joint}. In particular, we introduce a time-varying random field to model the additive bias in the numerical forecast model, termed \textit{error field}, along with a constant multiplicative bias parameter. The goal of the proposed model is to perform spatial interpolation at a present time rather than make predictions for future time points. 

The model offers several advantages: it defines a unified latent process for all data outcomes, accounts for measurement errors for all data sources, provides flexibility in addressing biases, accommodates multiple spatially-misaligned data sources, and gauges the relative quality of the data sources. Although we assume that the multiplicative bias parameter $\alpha_1$ is constant, the model can be extended to allow $\alpha_1$ to vary over space or time, adding complexity but potentially improving accuracy in other applications. Another extension would be incorporating a full melding model, which treats the numerical forecast outcomes as areal data.  This would  potentially increase the computational complexity, especially when the resolution of the gridded data is very fine. 

We conducted a simulation study to compare the proposed data fusion model to two benchmark approaches: the stations-only model and the regression calibration model. The main goal was to evaluate the model's performance under varying levels of data sparsity and prior specifications. The results showed that the proposed fusion model achieved lower squared-errors and posterior uncertainty in the estimated fields, especially with sparse stations data. The proposed model also has lower Dawid-Sebastiani scores. In terms of the relative error and posterior uncertainty in the model parameter estimates, the proposed data fusion model also outperformed the two benchmark approaches.

In the data application, where we considered three important meteorological variables, the proposed model outperformed the stations-only model and the regression calibration model based on the leave-group-out cross-validation (LGOCV). The LGOCV results, calculated via the INLA method \citep{liu2022leave}, showed that the proposed model provides better predictions with higher log predictive densities, smaller mean Kullback-Leibler divergence (KLD), and more accurate predictive scores (RMSE, MAE, MAPE, posterior uncertainty). Notably, the model's advantage grew with larger leave-out sets.

While this study considered  only two meteorological data sources for the Philippines, the framework can be extended to include additional sources, such as satellite data or a second numerical forecast model. This would introduce new bias parameters and error fields, broadening the model's utility. Future work could also involve incorporating longer time series data, which could enhance the model’s capabilities.


The proposed data fusion framework, while developed for meteorological data, is applicable in other fields, such as air quality modelling. For example, in the UK, combining data from a network of monitoring stations called the Automatic Urban and Rural Network (AURN) \citep{lee2017rigorous}, outcomes from a weather and chemical transport model called the Air Quality Unified Model (AQUM) \citep{forlani2020joint, DEFRA}, and outcomes of dispersion models like the Pollution Climate Mapping (PCM) model which are run by Ricardo Energy \& Environment \citep{forlani2020joint, DEFRA},  could improve air quality predictions, which are crucial for public health. Unlike many existing models that consider one additional data source at a time \citep{forlani2020joint}, our approach allows for the joint use of multiple sources, while accounting for biases in each. 

Another area where the proposed data fusion framework is applicable is in species distributions modelling. With technological advancements,  data collection efforts to study the natural world have significantly increased, which are accompanied with a surge  in data contributions from the general public, commonly referred to as citizen science data \citep{august2015emerging, belmont2024spatio}. However, there are also concerns with regards to the quality of these data, particularly due to systemic biases  \citep{van2013opportunistic, august2015emerging, koh2023extreme}. These biases stem from uneven geographical coverage, differences in observer expertise, and variable effort in data collection. The appropriate use of citizen science data together with data from well-planned surveys falls within the realm of data fusion. The  data fusion framework we propose views these different data sources as realizations  of the same latent process and, therefore, can be extended to the ecological context by assuming some probabilistic structure in the biases in the citizen science data while borrowing strength from the accuracy of the outcomes from well-planned surveys. 

The INLA and the SPDE approach was used for model inference because they provide fast and reliable fitting of complex spatio-temporal models \citep{rue2009approximate, lindgren2011explicit}. A potential challenge in the computational aspect is that the multiplicative bias parameter $\alpha_1$ can be hard to identify and, therefore, can lead to numerical problems. To overcome these, we used a Bayesian model averaging approach, which allowed us to fit the data fusion model conditional on fixed values of $\alpha_1$. This is a viable approach since we have an intuitive understanding of the plausible values of this bias parameter. 
Another approach for fitting the model is to include the parameter $\alpha_1$ in the latent Gaussian field, and perform a linearization on the  non-linear predictors using a first-order Taylor approximation, and to iteratively do this by looking for the optimal linearization point. This can be implemented using the \texttt{inlabru} library \citep{lindgren2024inlabru, serafini2023approximation, bachl2019inlabru}. 
Since the convergence of this approach and the properties of the approximation depend on the non-linear nature of the problem, it can also be computationally challenging for some cases. 
The use of a model averaging approach successfully removed the computational challenges, but the linearized INLA approach,  previously described, is also a viable approach for the problem and will be further explored in a future work.


 An immediate future work is to use the predicted fields from the data fusion model as  input in an epidemiological model in order to understand the link between climate data and health outcomes. This introduces another layer of spatial misalignment since typically data for the health outcomes are areal while the predicted fields from the climate data fusion models are point-referenced. Also, in this two-stage modelling framework, accounting for the uncertainty in the data fusion model when fitting the health model should be carefully considered. The problem of uncertainty propagation has been studied in the context of health modelling \citep{blangiardo2016two, lee2017rigorous, gryparis2009measurement} and which generally falls under the area of measurement error models \citep{berry2002bayesian}.


\section{Competing interests}
We declare that there is no conflict of interest regarding the publication of this paper.

\section{Author contributions statement}
\textbf{S.V.}: Conceptualization, Methodology, Formal Analysis, Writing - Original Draft, Writing - Review \& Editing. \textbf{J.I.:} Conceptualization, Methodology, Investigation, Writing - Review \& Editing, Supervision. \textbf{S.M.}: Conceptualization, Methodology, Investigation, Writing - Review \& Editing, Supervision. \textbf{F.L.}: Methodology, Writing - Review \& Editing.

\section{Acknowledgments}
The authors thank the \textit{Philippine Atmospheric, Geophysical and Astronomical Services Administration} (PAGASA) office under the Department of Science and Technology (DOST) for generously sharing the meteorological data. We specifically thank Mr Robert Badrina and Mr Jerome Tolentino, both weather specialists of DOST-PAGASA, for providing model insights. We also thank the anonymous referees for their useful suggestions and comments on the initial version of the manuscript.

\bibliographystyle{apalike}\bibliography{example}

\newpage

\begin{appendices}

\section{Appendix}

\subsection{Simulation Results} \label{subsec:app_simres}

\begin{figure}[H]
    \centering
    \includegraphics[scale=0.4]{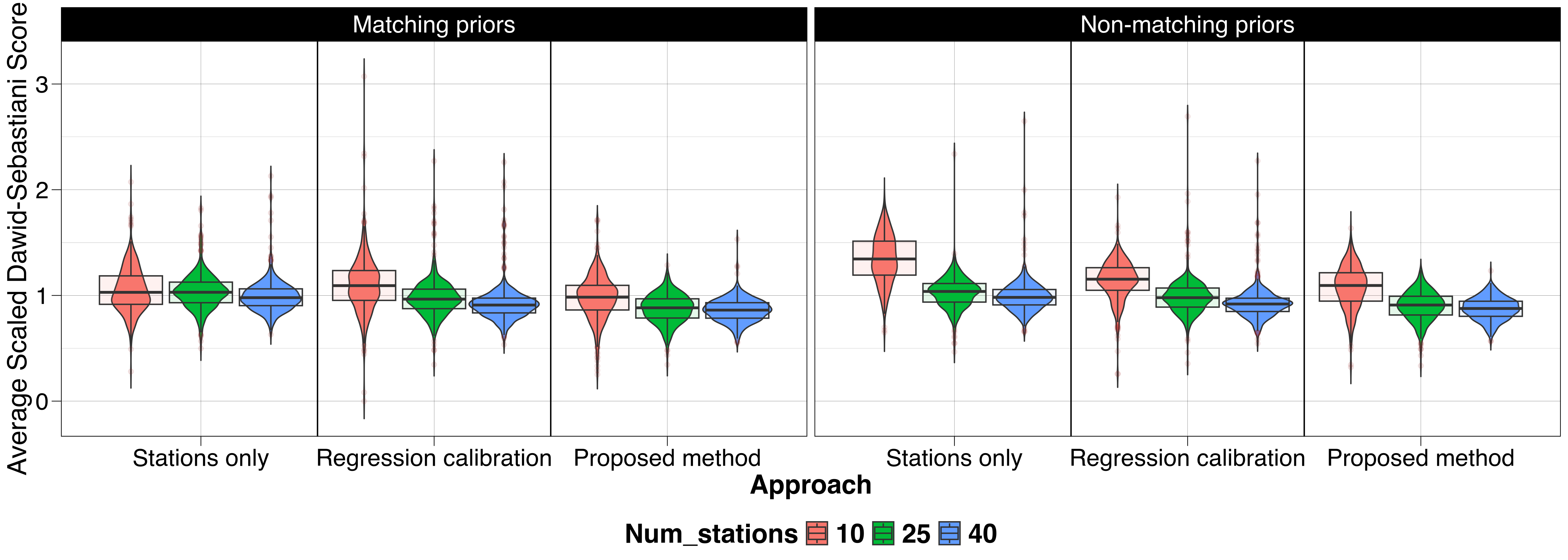}
    \caption{Plot of the average scaled Dawid-Sebastiani (DS) scores from 500 simulated datasets with respect to the number of stations, the priors used, and the modelling approach: stations-only model, regression calibration model, and the proposed data fusion model. The proposed method generally has the smallest average scaled DS score.}
    \label{fig:Ortho_I_x_v2}
\end{figure}

Figure \ref{fig:Ortho_I_x_v2} shows the average scaled DS scores for the stations data $\text{w}_1(\mathbf{s}_i)$. We scale the DS scores by adding the absolute value of the minimum in order to make the scores non-negative, and then applying log transformation. The average scaled DS scores are also generally smaller with the proposed method. Furthermore, it also shows that the scores tend to decrease with the number of stations especially with the use of non-matching priors. 

\begin{figure}[H]
    \centering
    \subfloat[Average relative errors and average posterior uncertainty in $\beta_0$]
    {\includegraphics[scale=.03,width=.48\textwidth]{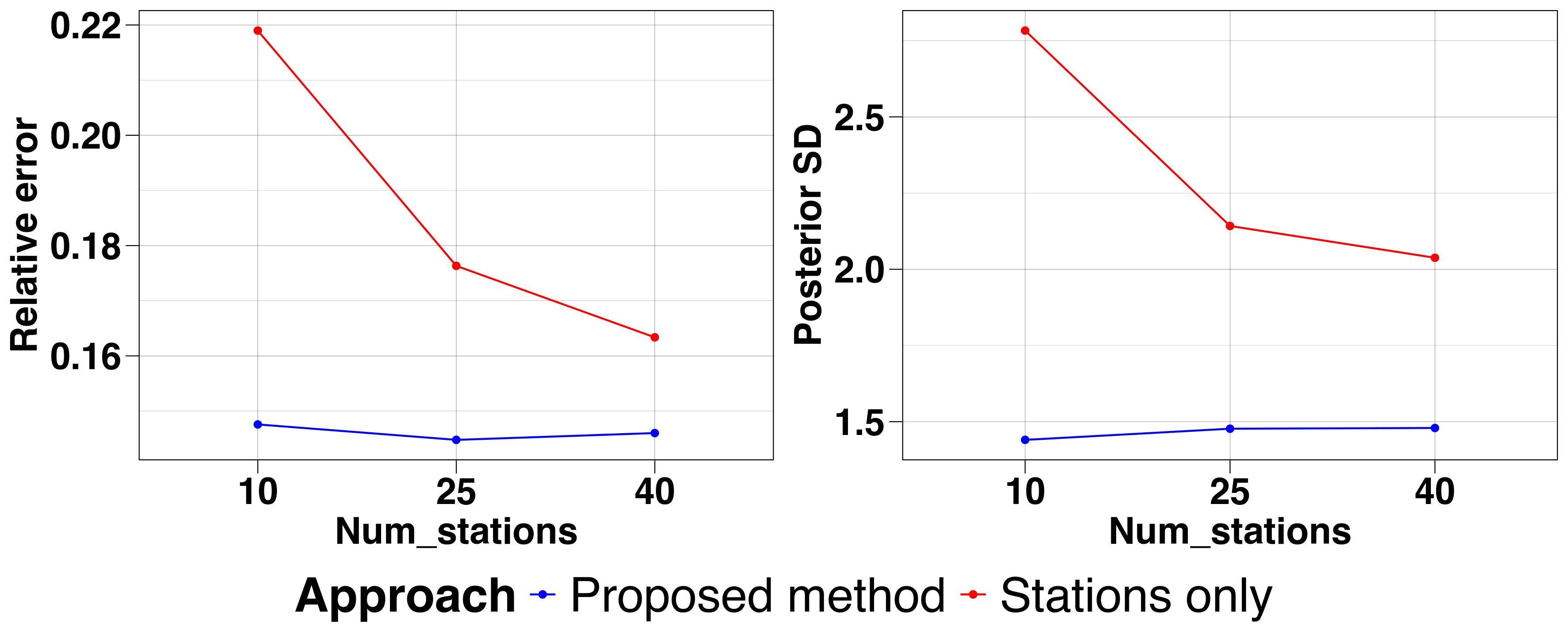} \label{fig:sim_res_beta0}}\hspace{0.02\textwidth}
    \subfloat[Average relative errors and average posterior uncertainty in $\beta_1$]
    {\includegraphics[scale=.03,width=.48\textwidth]{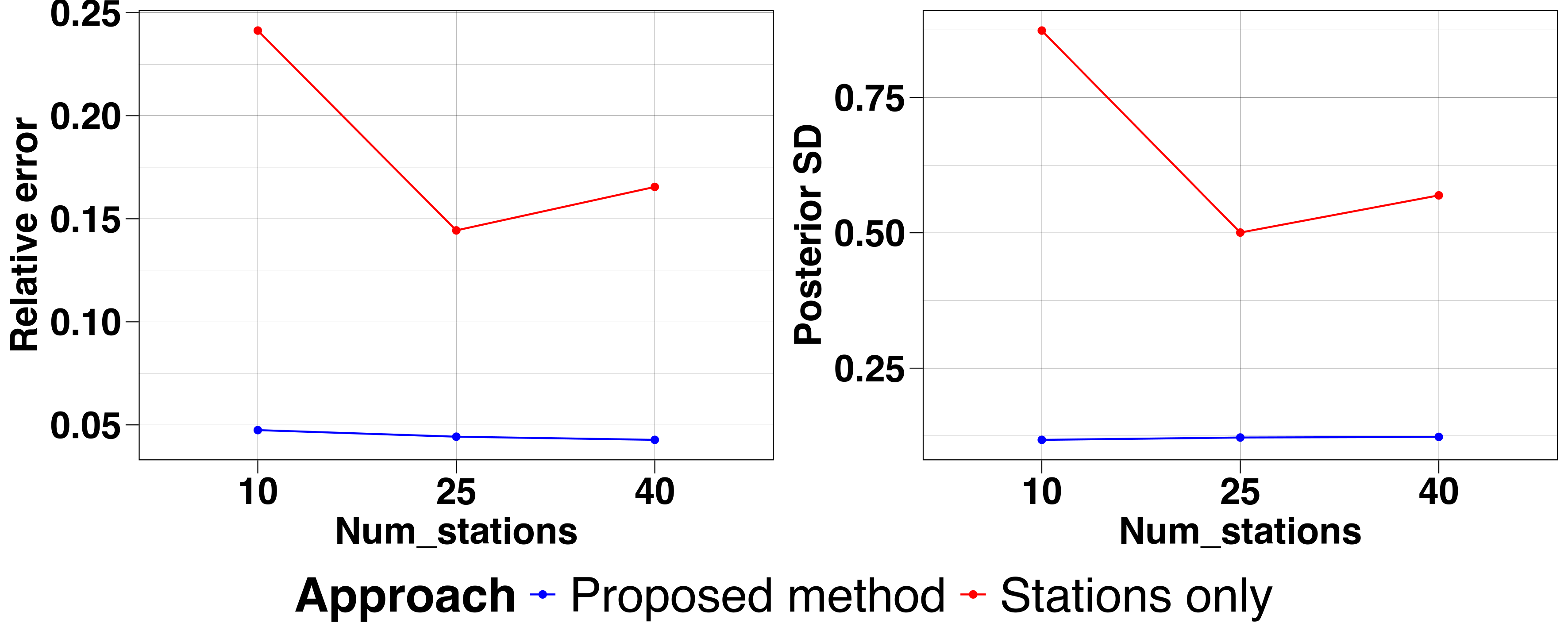}\label{fig:sim_res_beta1}}\hspace{0.01\textwidth}
    \hspace*{\fill}
    \caption{Plot of average relative errors and average posterior uncertainty from 500 simulated datasets for the fixed effects: (a) $\beta_0$ and (b) $\beta_1$.}
    \label{fig:sim_res_param_2}
\end{figure}

\begin{figure}[H]
    \centering
    \hspace*{\fill}
    \subfloat[][$\sigma_{\xi}$]
    {\includegraphics[scale=0.35]{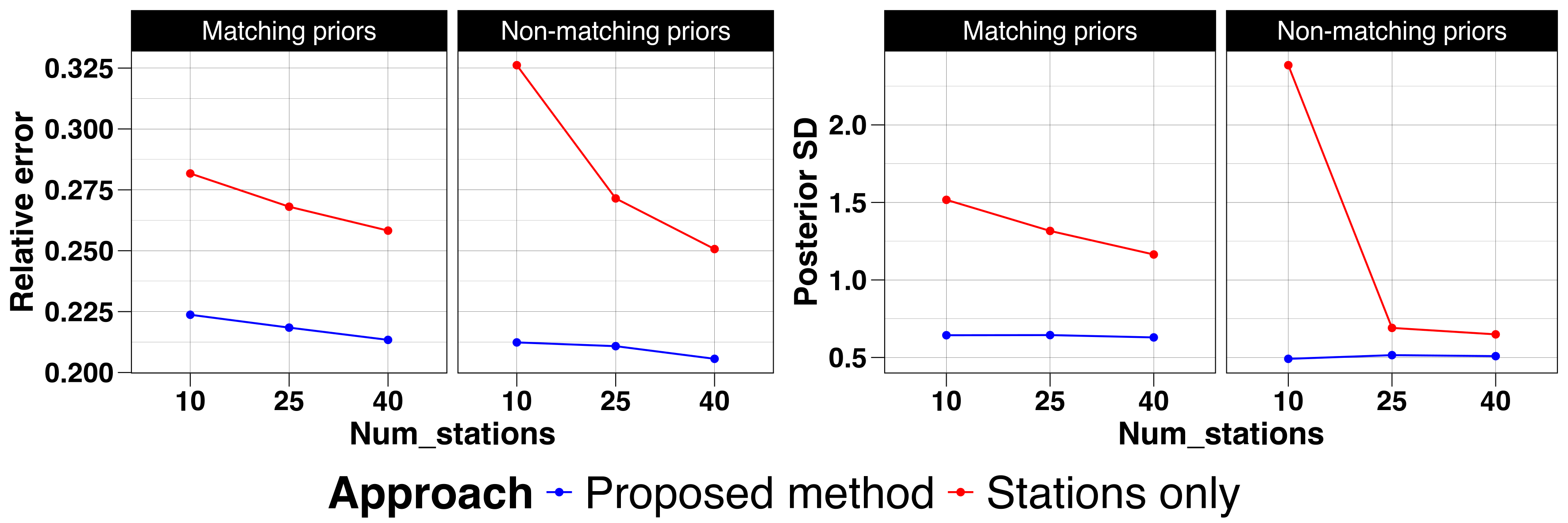}
    \label{fig:sim_res_sigma_omega}}
    \hspace*{\fill}

    \hspace*{\fill}
    \subfloat[][$\rho_{\xi}$]
    {\includegraphics[scale=0.35]{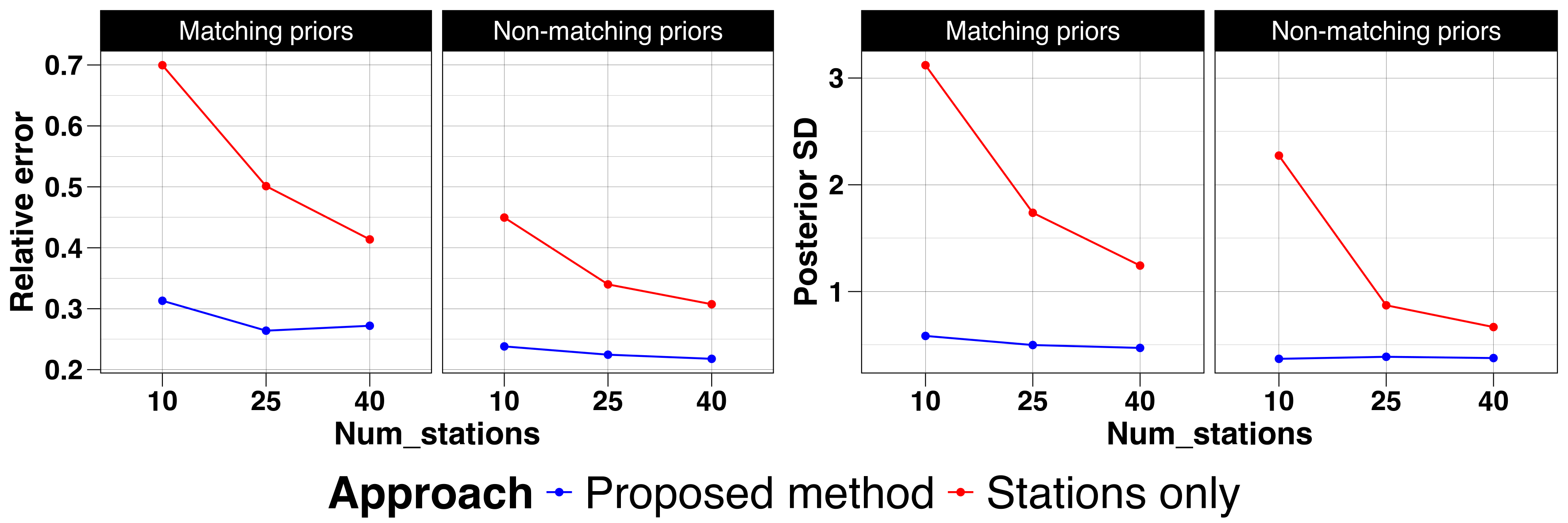}
    \label{fig:sim_res_sigma_e}}
    \hspace*{\fill}

    \caption{Plot of average relative errors and average posterior uncertainty from 500 simulated datasets for two hyperparameters: (a) marginal standard deviation $\sigma_{\xi}$ of the spatial field and (b) range parameter $\rho_{\xi}$ of the spatial field.}
    \label{fig:sim_res_param}
\end{figure}

Figure \ref{fig:sim_res_param_2} shows the average relative errors and average posterior uncertainty for the fixed effects $\beta_0$ and $\beta_1$. Figures \ref{fig:sim_res_param}a and \ref{fig:sim_res_param}b show the results for the marginal standard deviation and the range parameter of the spatial field $\xi(\mathbf{s})$, respectively. The results show that the average relative error and average posterior uncertainty in the parameter estimates are smaller for the proposed method. 

\subsection{Temperature model}\label{subsec:app_temperature}


\begin{table}[H]
\caption{\label{tab:TMEAN_mliksweights}Marginal log-likelihood values conditional an $\alpha_1$ and the corresponding BMA weights for the temperature data fusion model}
\centering
\scalebox{1}{\begin{tabular}{|rrr|}
  \hline
  \hline
  $\alpha_1$ & $\log \pi(\mathbf{Y}|\alpha_1)$ & $w_k$ \\ 
  \hline
  \hline
0.5 & -978.295 & 0.0000 \\ 
   0.6 & -914.791 & 0.0000 \\ 
  0.7 & -849.673 & 0.0000 \\ 
  0.8 & -778.493 & 0.0000 \\ 
  0.9 & -697.501 & 0.0001 \\ 
  1 & -688.142 & 1.0000 \\ 
  1.1 & -811.927 & 0.0000 \\ 
  1.2 & -899.762 & 0.0000 \\ 
  1.3 & -949.719 & 0.0000 \\ 
  1.4 & -2265.074 & 0.0000 \\ 
  1.5 & -2329.848 & 0.0000 \\ 
   \hline
   \hline
\end{tabular}}
\end{table}

\begin{table}[H]
\caption{\label{tab:TMEANestimates_hyper}Posterior estimates of hyperparameters for the temperature model -- stations-only model versus proposed data fusion model}
\centering
\scalebox{1}{\begin{tabular}{|l|rrrr|rrrr|}
  \hline\hline
 &  \multicolumn{4}{c|}{\textbf{Stations only}} & \multicolumn{4}{c|}{\textbf{Proposed model}}\\
 Parameter & Mean & SD & P2.5\% & P97.5\% & Mean & SD & P2.5\% & P97.5\% \\ 
  \hline\hline
$\sigma_{e_1}$  & 0.178 & 0.011 & 0.158 & 0.197 & 0.243 & 0.010 & 0.224 & 0.264  \\ 
$\sigma_{e_2}$  & - & - & - & - & 0.022 & 0.007 & 0.015 & 0.043  \\
  $\rho_{1}$ & 621.888 & 50.677 & 528.743 & 728.144  & 764.748 & 59.322 & 655.597 & 888.964\\ 
  $\sigma_{1}$ & 5.121 & 0.612 & 4.034 & 6.438 & 7.690 & 0.872 & 6.133 & 9.555 \\ 
  $\phi_{1}$ & 0.992 & 0.002 & 0.988 & 0.995 & 0.998 & 0.001 & 0.997 & 0.999 \\ 
  $\rho_{2}$ & - & - & - & - & 112.768 & 8.076 & 97.773 & 129.554 \\ 
  $\sigma_{2}$ & - & - & - & - & 0.668 & 0.066 & 0.547 & 0.807 \\ 
  $\phi_{2}$ & - & - & - & - & 0.937 & 0.014 & 0.906 & 0.960  \\ 
   \hline\hline
\end{tabular}}
\end{table}

\begin{table}[H]
\centering
\caption{\label{tab:TMEANestimates_regcalib}Posterior estimates of the regression calibration model for temperature}
\begin{tabular}{|l|rrrr|}
  \hline\hline
Parameter & Mean & SD & P2.5\% & P97.5\% \\ 
  \hline\hline
$\sigma_{e_1}$ & 49.893 & 7.734 & 36.866 & 67.233 \\ 
  Range of $\alpha_0(\mathbf{s},t)$ & 48.011 & 13.352 & 26.779 & 78.911 \\ 
  SD of $\alpha_0(\mathbf{s},t)$ &  0.444 & 0.065 & 0.333 & 0.589 \\ 
  AR parameter of $\alpha_0(\mathbf{s},t)$ & 0.777 & 0.058 & 0.650 & 0.875 \\ 
  Range of $\alpha_1(\mathbf{s},t)$ & 1103.250 & 104.709 & 913.898 & 1325.815 \\
  SD of $\alpha_1(\mathbf{s},t)$ & 0.673 & 0.087 & 0.517 & 0.860 \\ 
  AR parameter of of $\alpha_1(\mathbf{s},t)$ & 0.999 & 0.000 & 0.999 & 1.000 \\ 
   \hline\hline
\end{tabular}
\end{table}

\begin{figure}[H]
        \centering
        \includegraphics[scale=.2]{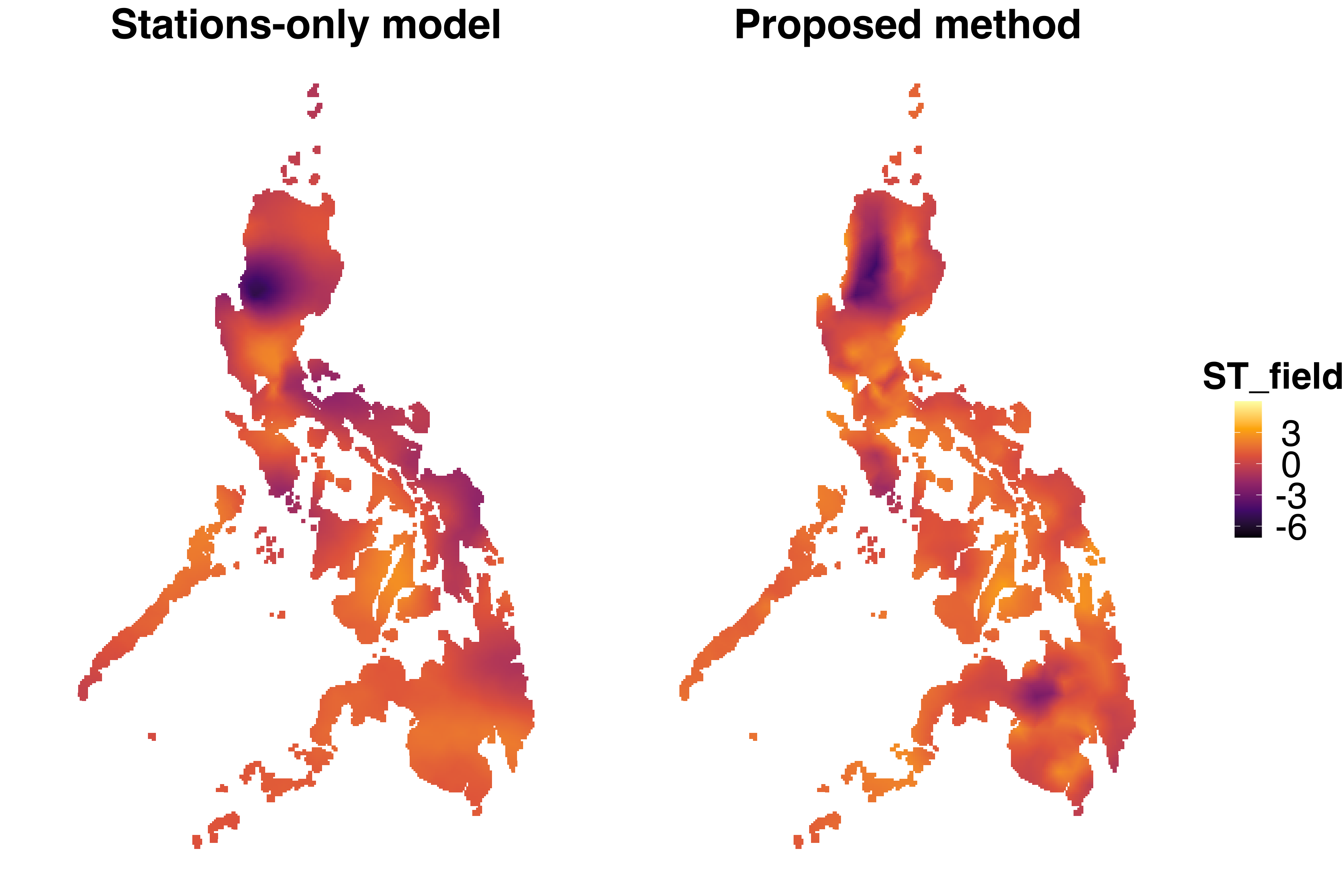}
        
        \captionof{figure}{Comparison of estimated spatial fields $\hat{\xi}(\mathbf{s},t)$ for August 2019 between the stations-only model and the proposed data fusion model. The estimated spatial fields are roughly similar.}
        \label{fig:TMEAN_spde}
    \end{figure}

\subsection{Relative humidity model}\label{subsec:app_relhum}


\begin{table}[H]
\caption{\label{tab:RH_mliksweights}Marginal log-likelihood values conditional an $\alpha_1$ and the corresponding BMA weights for the relative humidity data fusion model}
\centering
\scalebox{1}{\begin{tabular}{|rrr|}
  \hline
  \hline
  $\alpha_1$ & $\log \pi(\mathbf{Y}|\alpha_1)$ & $w_k$ \\ 
  \hline
  \hline
0.5 & 5955.341 & 0.0000 \\ 
  0.6 & 5830.252 & 0.0000 \\ 
  0.7 & 6076.597 & 0.0000 \\ 
  0.8 & 6136.254 & 0.0000 \\ 
  0.9 & 5320.351 & 0.0000 \\ 
  1 & 6248.282 & 1 \\ 
  1.1 & 6116.906 & 0.0000 \\ 
  1.2 & 6101.200 & 0.0000 \\ 
  1.3 & 6071.814 & 0.0000 \\ 
  1.4 & 6044.122 & 0.0000 \\ 
  1.5 & 5847.258 & 0.0000 \\ 
   \hline
   \hline
\end{tabular}}
\end{table}

\begin{table}[H]
\caption{\label{tab:RHestimates_hyperpar}Posterior estimates of hyperparameters for the relative humidity model -- stations-only model versus proposed data fusion model}
\centering
\scalebox{1}{\begin{tabular}{|l|rrrr|rrrr|}
  \hline\hline
 &  \multicolumn{4}{c|}{\textbf{Stations only}} & \multicolumn{4}{c|}{\textbf{Proposed model}}\\
 Parameter & Mean & SD & P2.5\% & P97.5\% & Mean & SD & P2.5\% & P97.5\% \\ 
  \hline\hline
$\sigma_{e_1}$ & 0.012 & 0.001 & 0.011 & 0.014 & 0.020 & 0.001 & 0.018 & 0.022 \\ 
$\sigma_{e_2}$ & - & - & - & - & 0.003 & 0.001 & 0.002 & 0.005 \\ 
  $\rho_{1}$ & 287.577 & 26.952 & 238.127 & 344.148 & 589.113 & 67.976 & 468.374 & 735.571 \\ 
  $\sigma_{1}$ & 0.087 & 0.008 & 0.073 & 0.103  & 0.111 & 0.014 & 0.087 & 0.142  \\ 
  $\phi_{1}$ & 0.929 & 0.012 & 0.902 & 0.949 & 0.970 & 0.008 & 0.952 & 0.983 \\ 
  $\rho_{2}$ & - & - & - & - & 117.256 & 8.956 & 100.403 & 135.635 \\ 
  $\sigma_{2}$ & - & - & - & - & 0.040 & 0.002 & 0.037 & 0.045 \\ 
  $\phi_{2}$ & - & - & - & - & 0.855 & 0.015 & 0.824 & 0.883 \\ 
   \hline
\end{tabular}}
\end{table}

\begin{table}[H]
\centering
\caption{\label{tab:RHestimates_regcalib}Posterior estimates of the regression calibration model for relative humidity}
\begin{tabular}{|l|rrrr|}
  \hline\hline
Parameter & Mean & Sd & P2.5\% & P97.5\% \\ 
  \hline\hline
$\sigma_{e_1}$ & 5769.559 & 673.476 & 4563.267 & 7211.298 \\ 
  Range of $\alpha_0(\mathbf{s,t})$ & 1.663 & 1.542 & 0.299 & 5.758 \\ 
  SD of $\alpha_0(\mathbf{s,t})$ & 1.935 & 1.579 & 0.319 & 6.099 \\ 
  AR parameter of $\alpha_0(\mathbf{s,t})$ & 0.914 & 0.017 & 0.877 & 0.943 \\ 
  Range of $\alpha_1(\mathbf{s,t})$ & 2765.158 & 443.936 & 2001.709 & 3745.046 \\ 
  Range of $\alpha_1(\mathbf{s,t})$ & 0.105 & 0.015 & 0.079 & 0.137 \\ 
  AR parameter of $\alpha_1(\mathbf{s,t})$ & 0.993 & 0.004 & 0.983 & 0.998 \\ 
   \hline\hline
\end{tabular}
\end{table}

\begin{figure}[H]
     \centering
     \subfloat[][Stations-only model]{\includegraphics[trim={0cm 0cm 0 0cm},clip,scale=0.2]{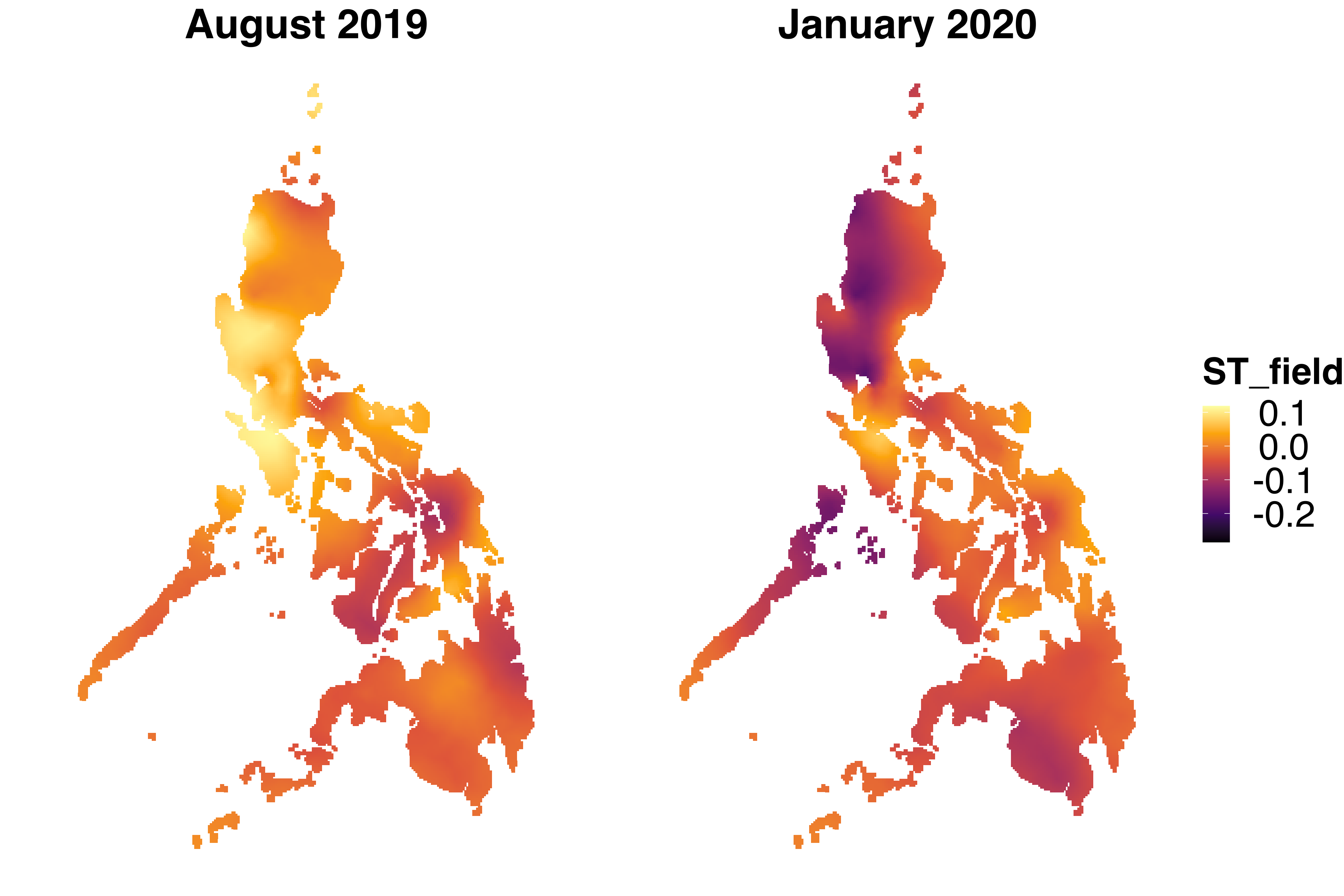}\label{fig:RH_sd_regcalib}}
     \subfloat[][Proposed data fusion model]{\includegraphics[trim={0cm 0cm 0 0cm},clip,scale=0.2]{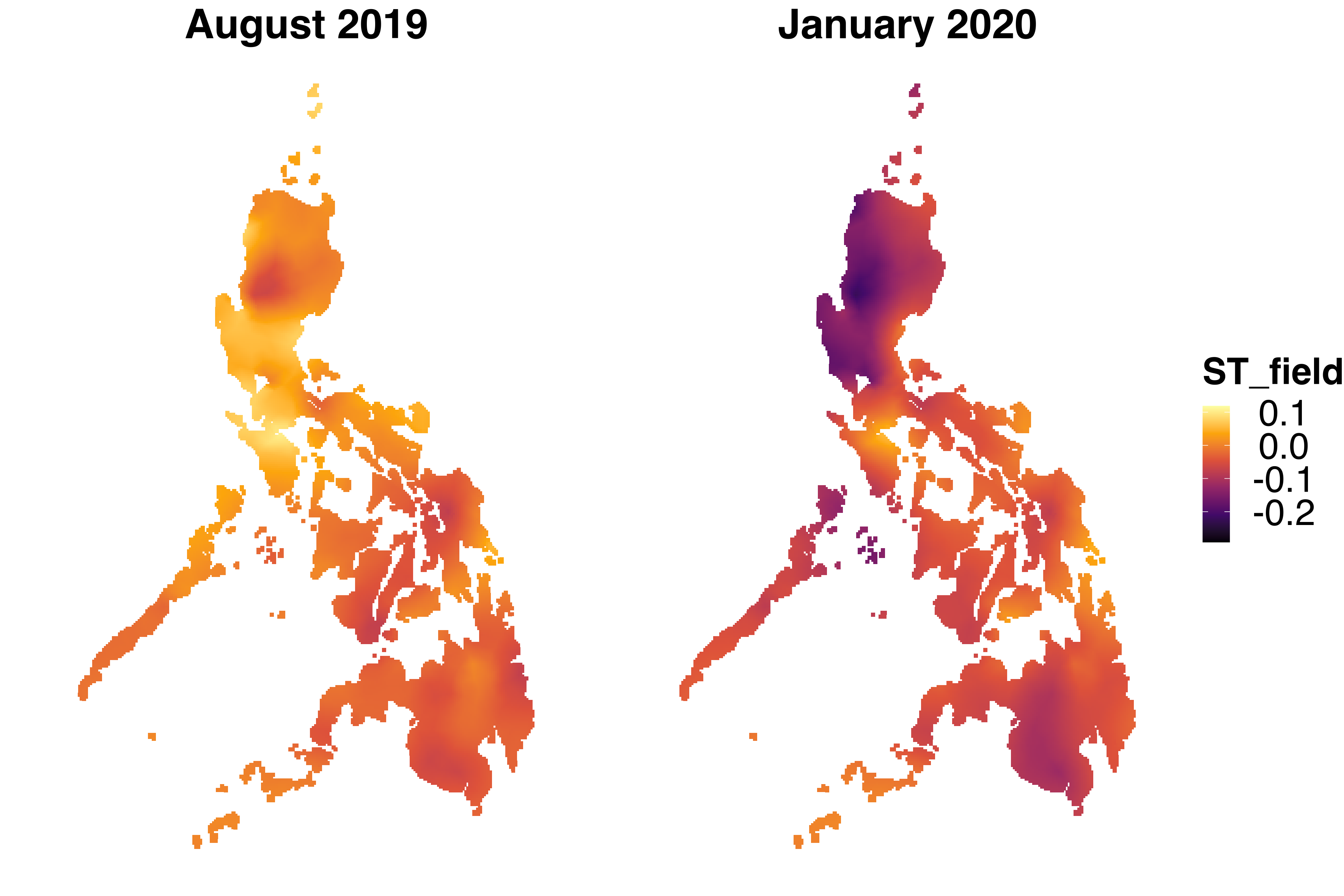}}
     \caption{Estimated spatial fields $\hat{\xi}(\mathbf{s},t)$ for log relative humidity, August 2019 and January 2020, for two approaches: (a) stations-only model, (b) proposed data fusion model.}
     \label{fig:RH_spatialfields}
\end{figure}


\begin{figure}[H]
     \centering
     \includegraphics[trim={0cm 0cm 0 0cm},clip,scale=0.2]{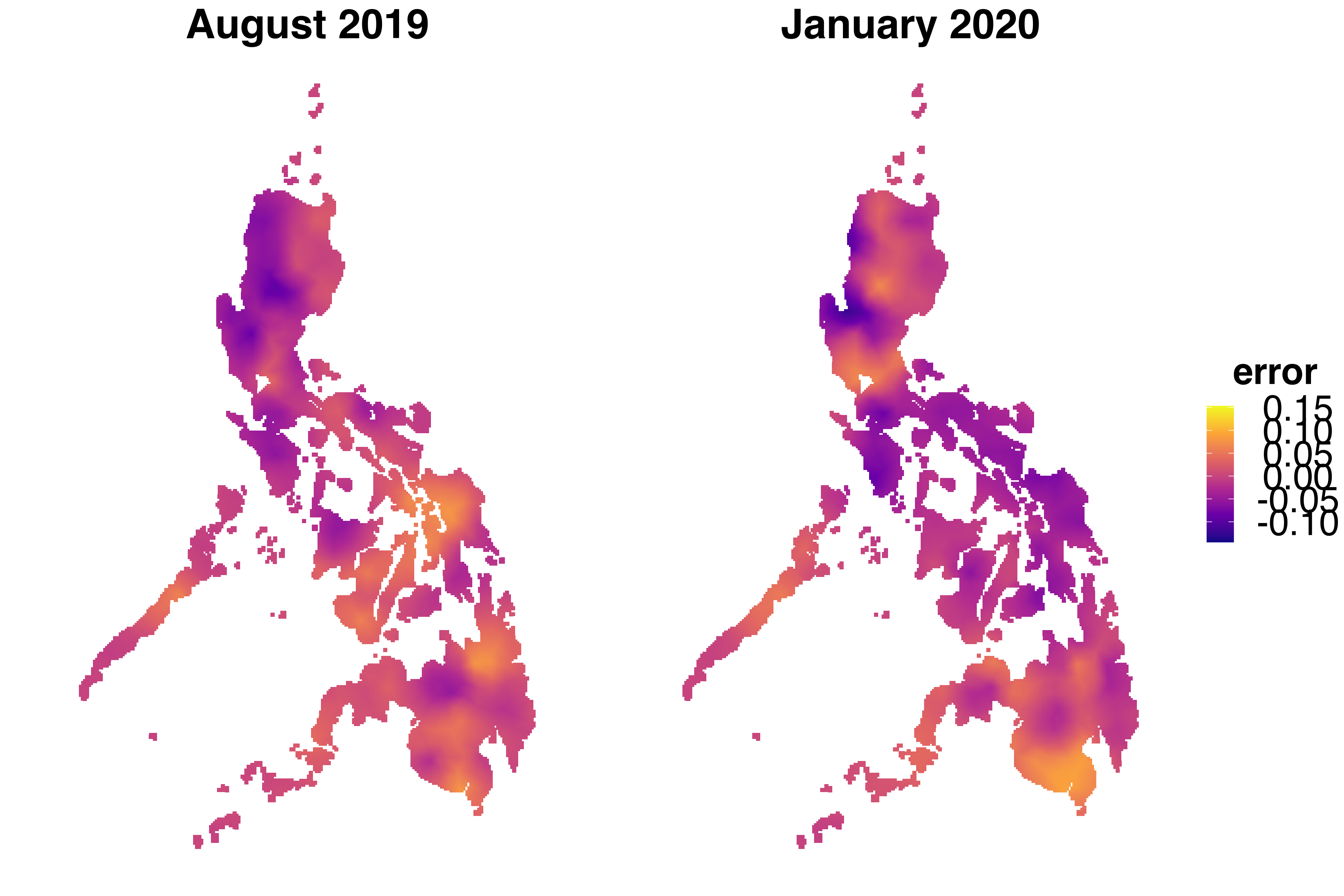}
     \caption{Estimated error fields for the GSM log relative humidity data, August 2019 and January 2020, using the proposed data fusion model.}
     \label{fig:RHerrorfield}
\end{figure}

\begin{figure}[H]
     \centering
     \subfloat[][$\text{w}_1(\mathbf{s},t)$ vs $\hat{x}(\mathbf{s},t)$]{\includegraphics[trim={0cm 0cm 0 0cm},clip,scale=0.33]{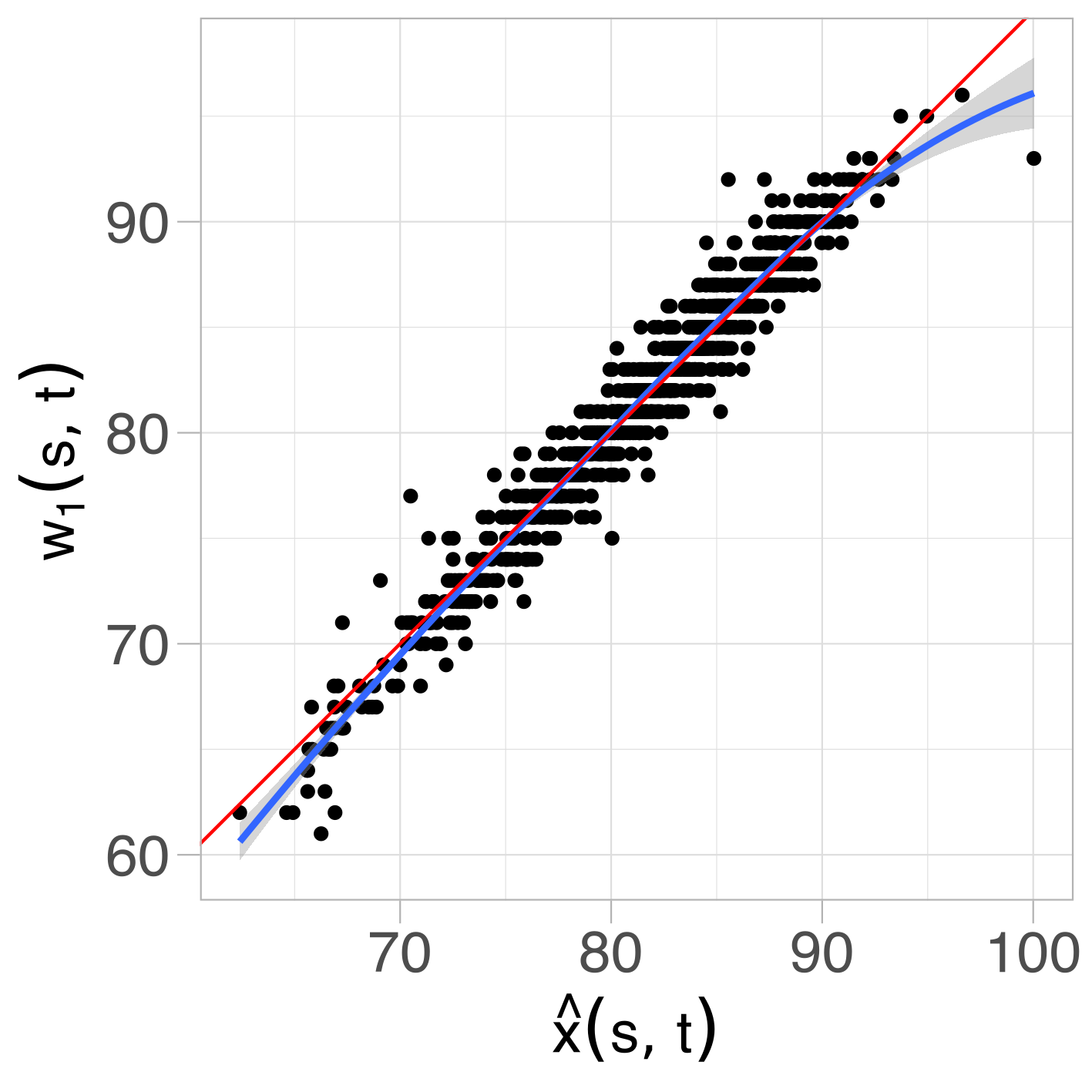}\label{fig:RH_scatter1}}
     \subfloat[][$\text{w}_2(\mathbf{g},t)$ vs $\hat{x}(\mathbf{g},t)$]{\includegraphics[trim={0cm 0cm 0 0cm},clip,scale=0.33]{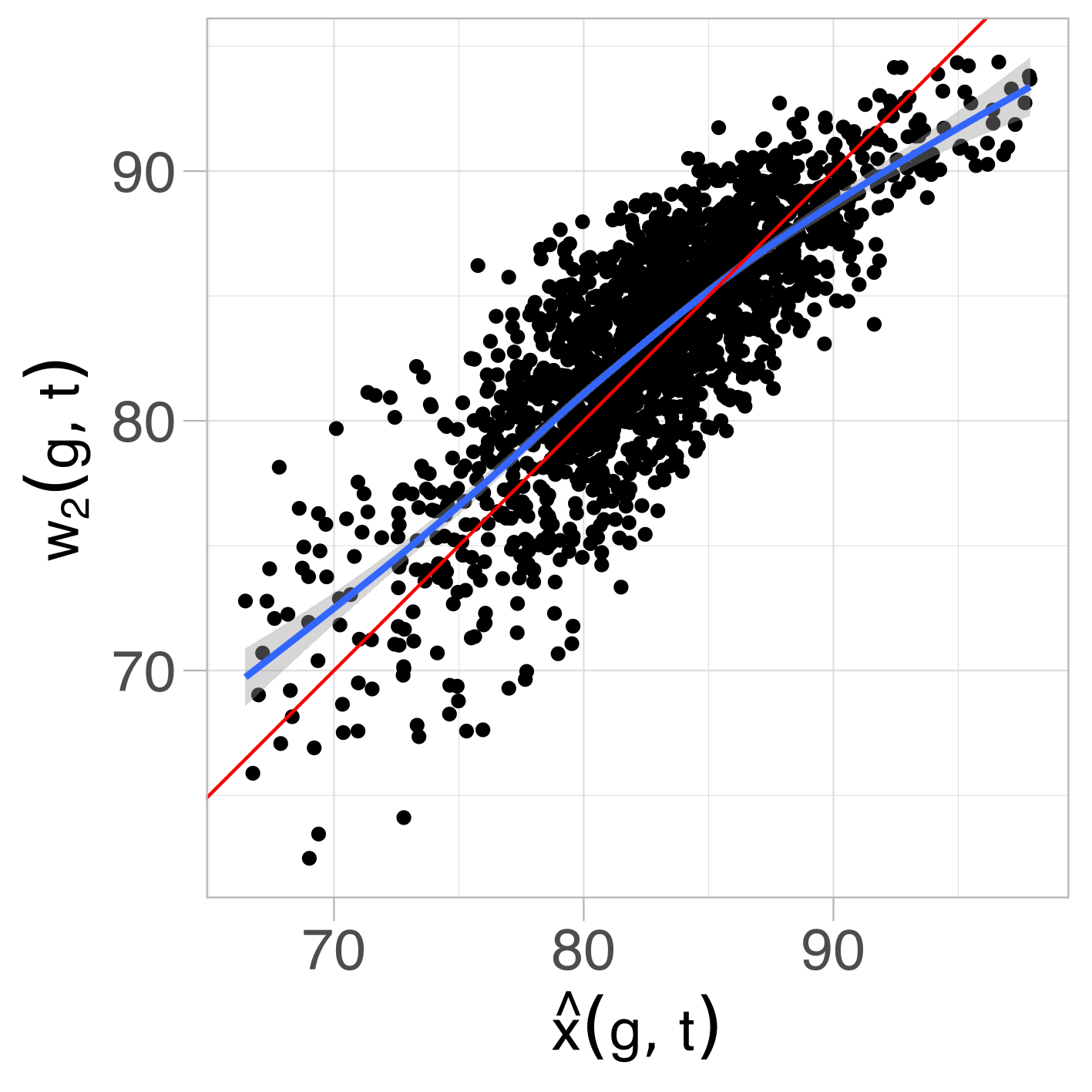}\label{fig:RH_scatter2}}
     \subfloat[][$\text{w}_2(\mathbf{g},t)$ vs $\hat{\text{w}}_2(\mathbf{g},t)$]{\includegraphics[trim={0cm 0cm 0 0cm},clip,scale=0.33]{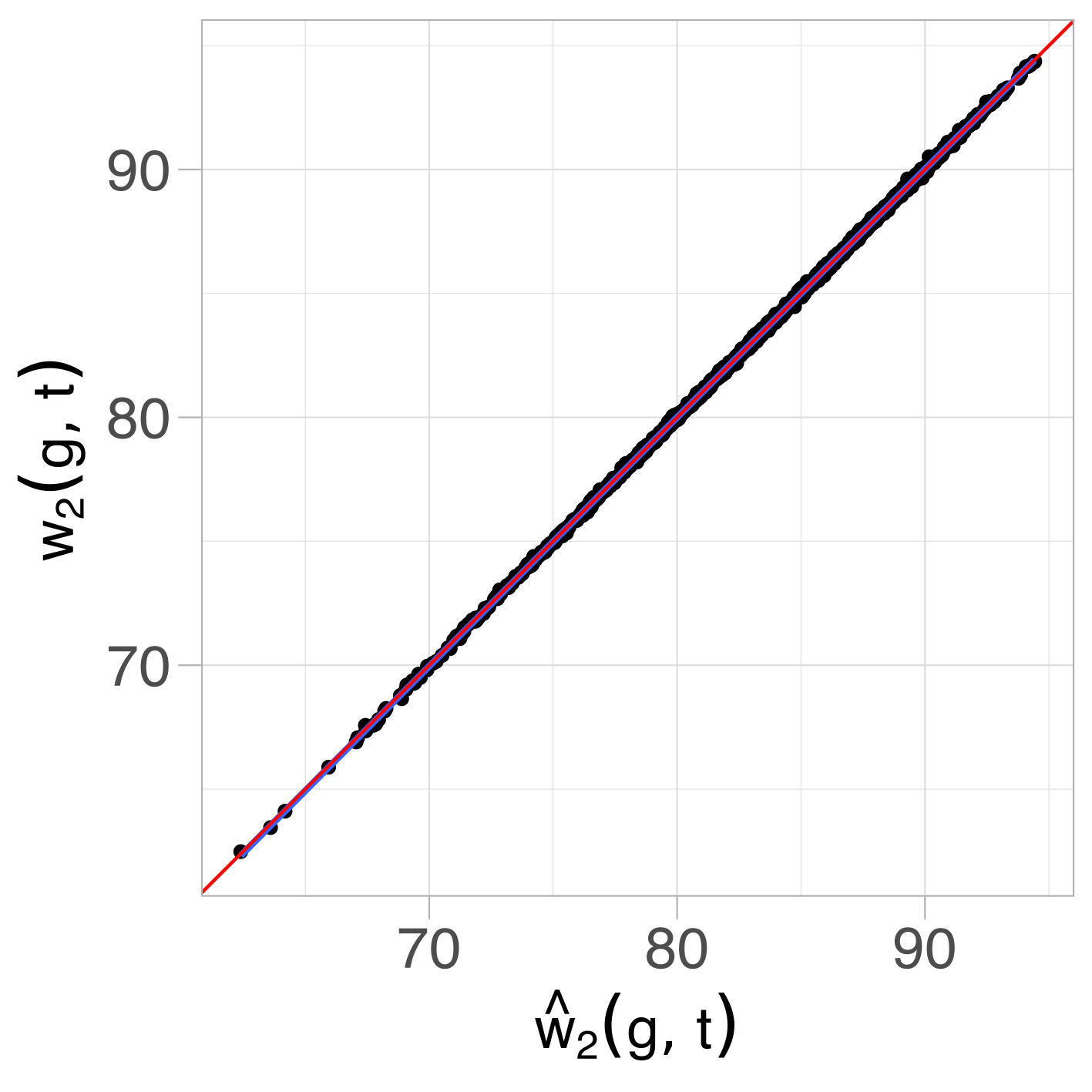}\label{fig:RH_scatter3}}
     \caption{Plot of observed relative humidity values versus predicted values using the proposed data fusion model for (a) weather stations and (b) GSM data, and (c) calibrated GSM data. The blue line is the smooth local regression curve, while the red line is the $x=y$ line.}
     \label{fig:RHscatters}
\end{figure}

\subsection{Rainfall} \label{subsec:app_rainfall}

\begin{table}[H]
\caption{\label{tab:RAINestimates_hyperpar}Posterior estimates of hyperparameters for the rainfall model -- stations-only model versus proposed data fusion model}
\centering
\scalebox{1}{\begin{tabular}{|l|rrrr|rrrr|}
  \hline\hline
 &  \multicolumn{4}{c|}{\textbf{Stations only}} & \multicolumn{4}{c|}{\textbf{Proposed model}}\\
 Parameter & Mean & SD & P2.5\% & P97.5\% & Mean & SD & P2.5\% & P97.5\% \\ 
  \hline\hline
$\sigma_{e_1}$ & 0.482 & 0.019 & 0.446 & 0.521 & 0.501 & 0.022 & 0.466 & 0.549  \\ 
  $\sigma_{e_2}$ & - & - & - & - & 0.256 & 0.014 & 0.229 & 0.284  \\ 
  $\rho_{1}$ & 614.617 & 62.498 & 501.641 & 747.464 & 584.259 & 59.817 & 480.042 & 711.353  \\ 
  $\sigma_{1}$ & 1.113 & 0.074 & 0.976 & 1.266  & 1.107 & 0.065 & 0.978 & 1.227 \\ 
  $\phi_1$ & 0.601 & 0.048 & 0.501 & 0.692 & 0.691 & 0.037 & 0.611 & 0.754\\ 
  $\rho_{2}$ &  &  &  &  & 434.935 & 66.349 & 323.066 & 588.186 \\ 
  $\sigma_{2}$ & - & - & - & - & 0.942 & 0.095 & 0.764 & 1.116 \\ 
  $\phi_2$ & - & - & - & - & 0.870 & 0.023 & 0.820 & 0.908 \\ 
   \hline\hline
\end{tabular}}
\end{table}

\begin{table}[H]
\centering
\caption{\label{tab:RAINestimates_regcalib}Posterior estimates of the regression calibration model for rainfall}
\begin{tabular}{|l|rrrr|}
  \hline
  \hline
Parameter & Mean & SD & P2.5\% & P97.5\% \\ 
  \hline
  \hline
$\sigma_{e_1}$ & 4.467 & 0.502 & 3.543 & 5.515 \\ 
  Range for $\alpha_0(\mathbf{s},t)$ & 9.012 & 12.217 & 1.398 & 39.452 \\ 
  SD for $\alpha_0(\mathbf{s},t)$ & 2.291 & 1.704 & 0.292 & 6.588 \\ 
  AR parameter for $\alpha_0(\mathbf{s},t)$ & 0.257 & 0.240 & -0.276 & 0.647 \\ 
  Range for $\alpha_1(\mathbf{s},t)$ & 1083.233 & 107.018 & 887.844 & 1308.788 \\ 
  SD for $\alpha_1(\mathbf{s},t)$ & 0.410 & 0.030 & 0.355 & 0.473 \\ 
  AR parameter for $\alpha_1(\mathbf{s},t)$ & 0.820 & 0.031 & 0.753 & 0.874 \\ 
   \hline
   \hline
\end{tabular}
\end{table}

\begin{figure}[H]
    \centering
    \subfloat[][Stations-only model]
    {\includegraphics[scale=0.2]{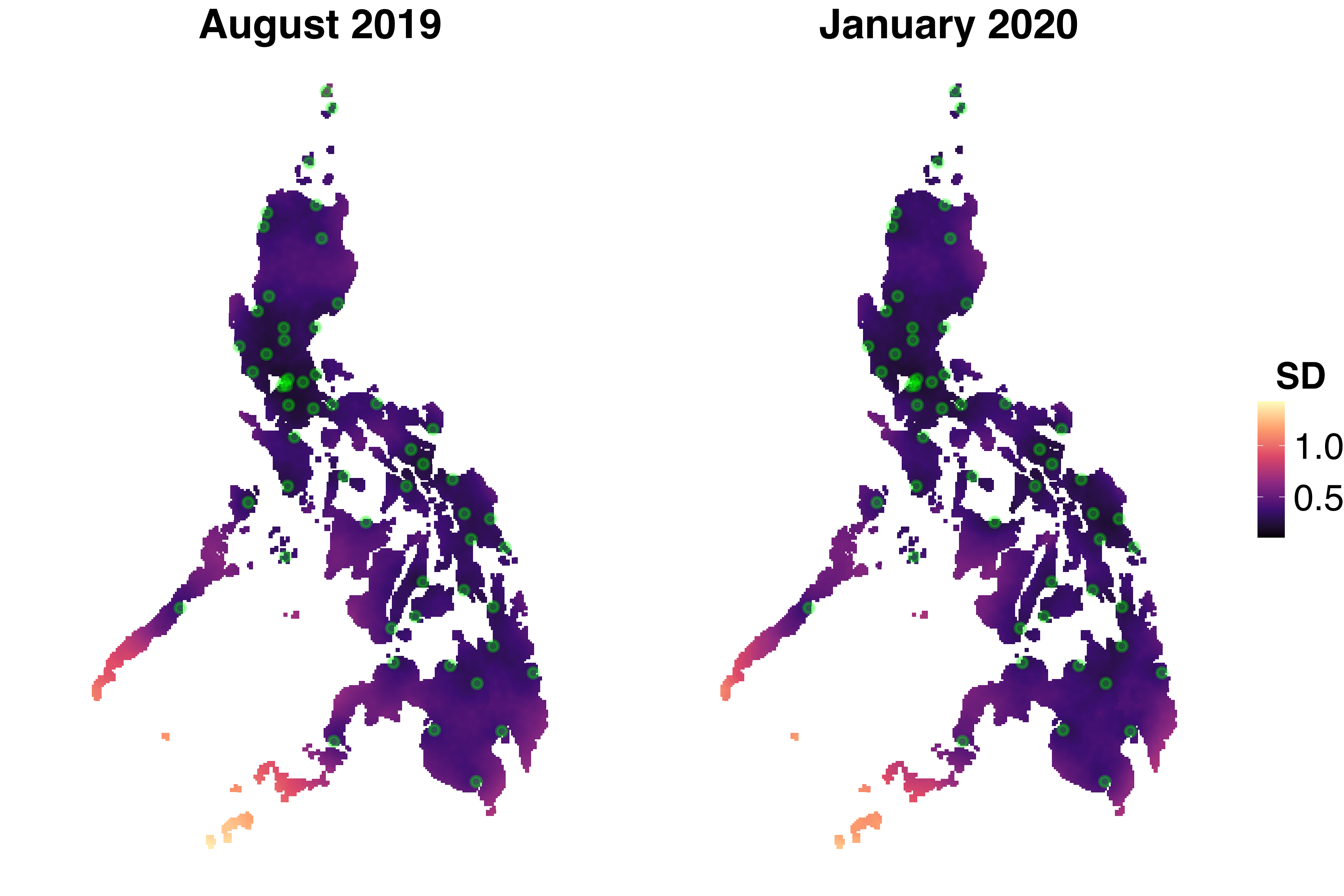}\label{fig:RAIN_fieldsdstationsonly}}
    \subfloat[][Regression calibration model]
    {\includegraphics[scale=0.2]{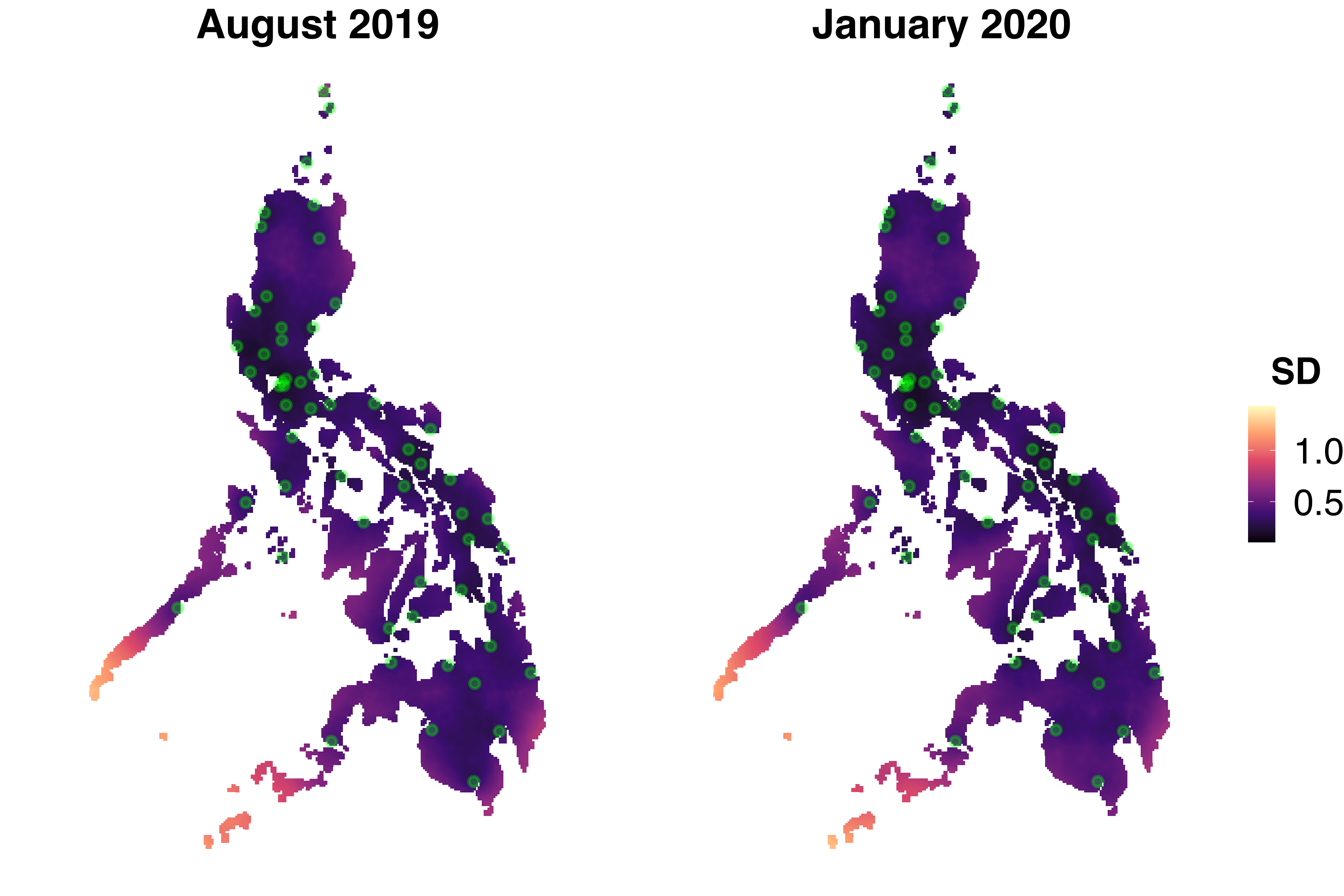}}
    \subfloat[][Proposed data fusion model]
    {\includegraphics[scale=0.2]{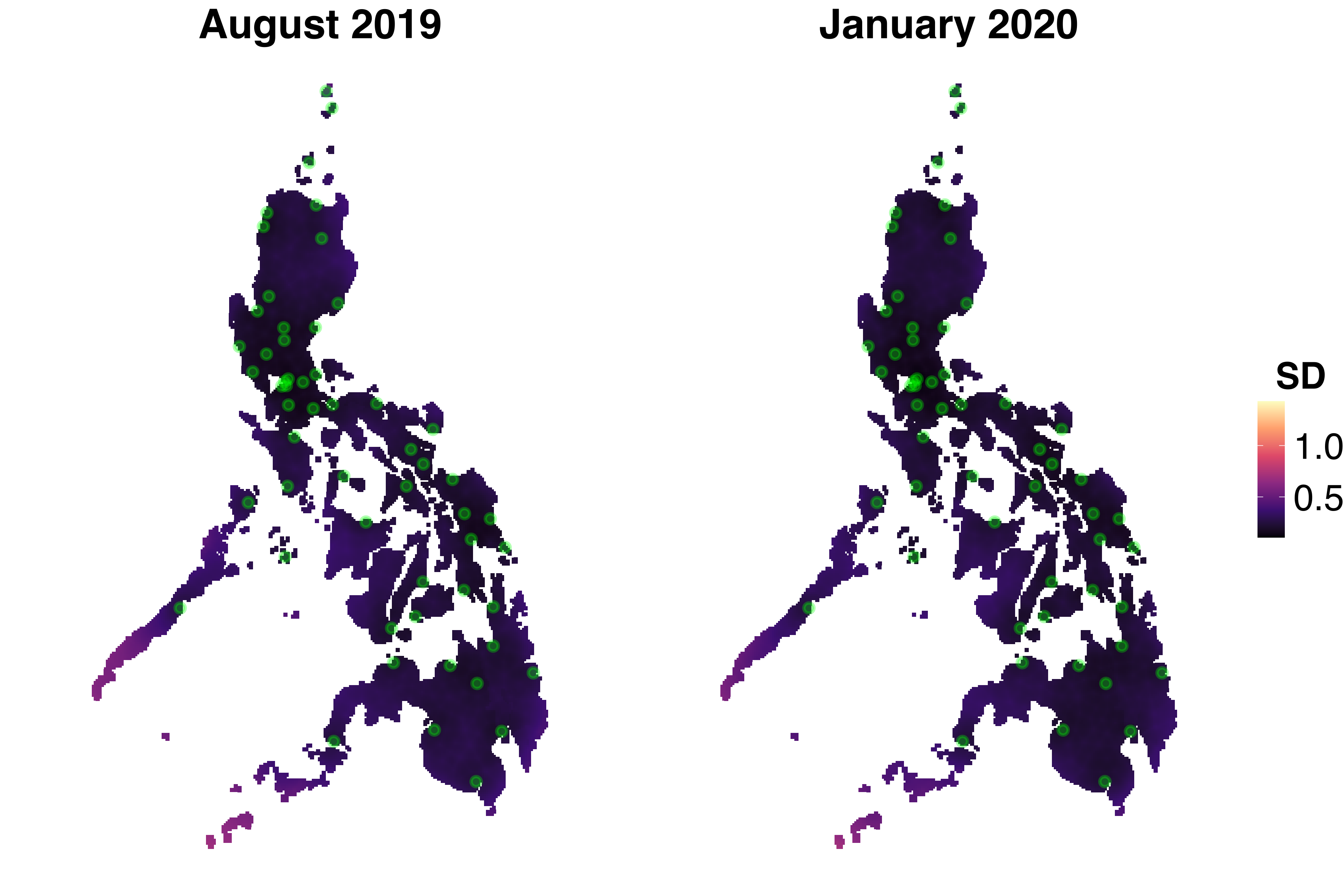}\label{fig:RAIN_fieldsddatafusion}}
    \caption{Posterior uncertainty of the estimated log rainfall fields in Figure \ref{fig:RAINfields} for three approaches: (a) stations-only model, (b) regression calibration model, (c) proposed data fusion model. The posterior uncertainty in the estimated fields from the proposed data fusion model is the smallest.}
    \label{fig:RAINfields_sd}
\end{figure}

\begin{figure}[H]
     \centering
    \subfloat[][Stations-only model]{\includegraphics[trim={0cm 0cm 0 0cm},clip,scale=0.2]{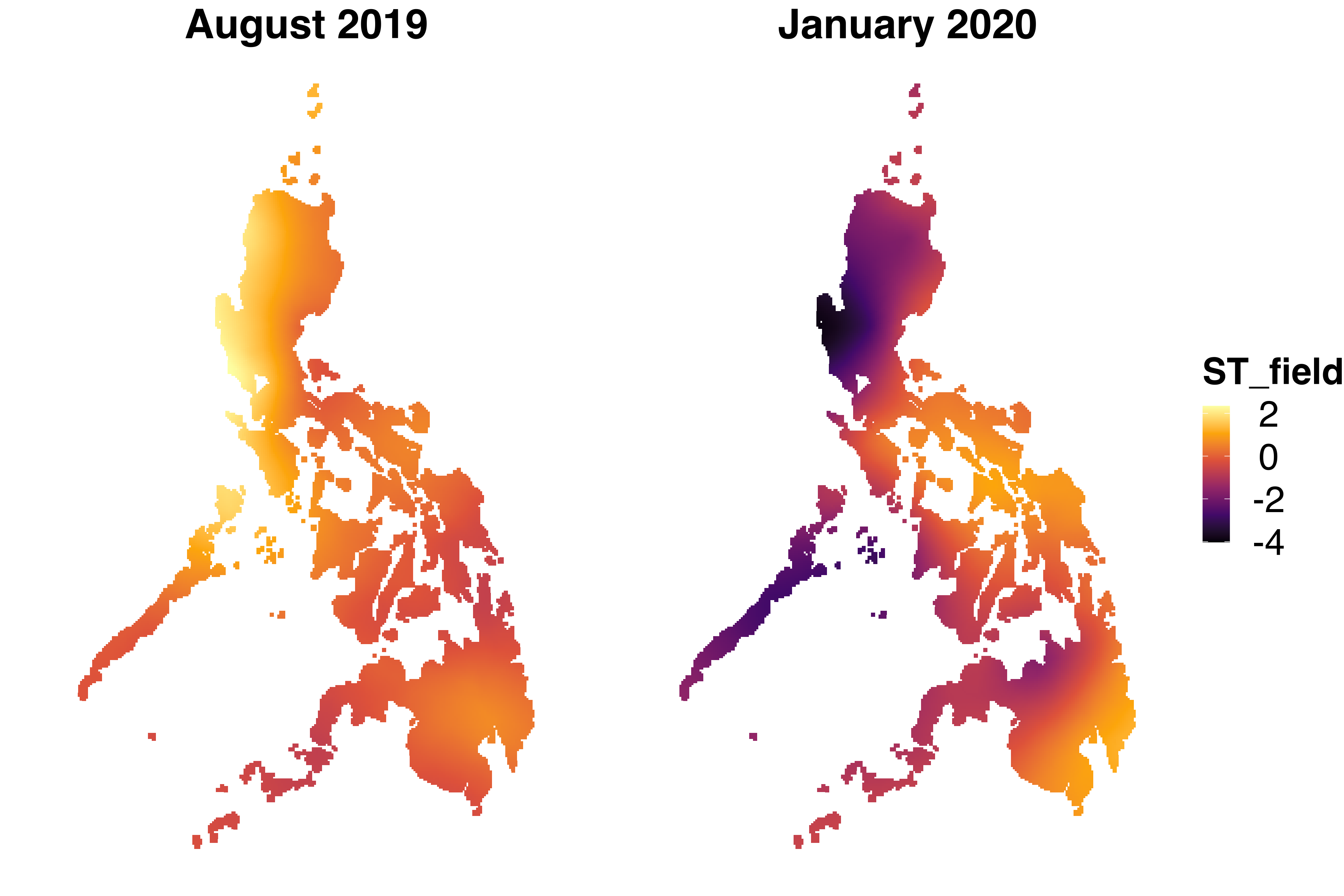}}
 \subfloat[][Proposed data fusion model]{\includegraphics[trim={0cm 0cm 0 0cm},clip,scale=0.2]{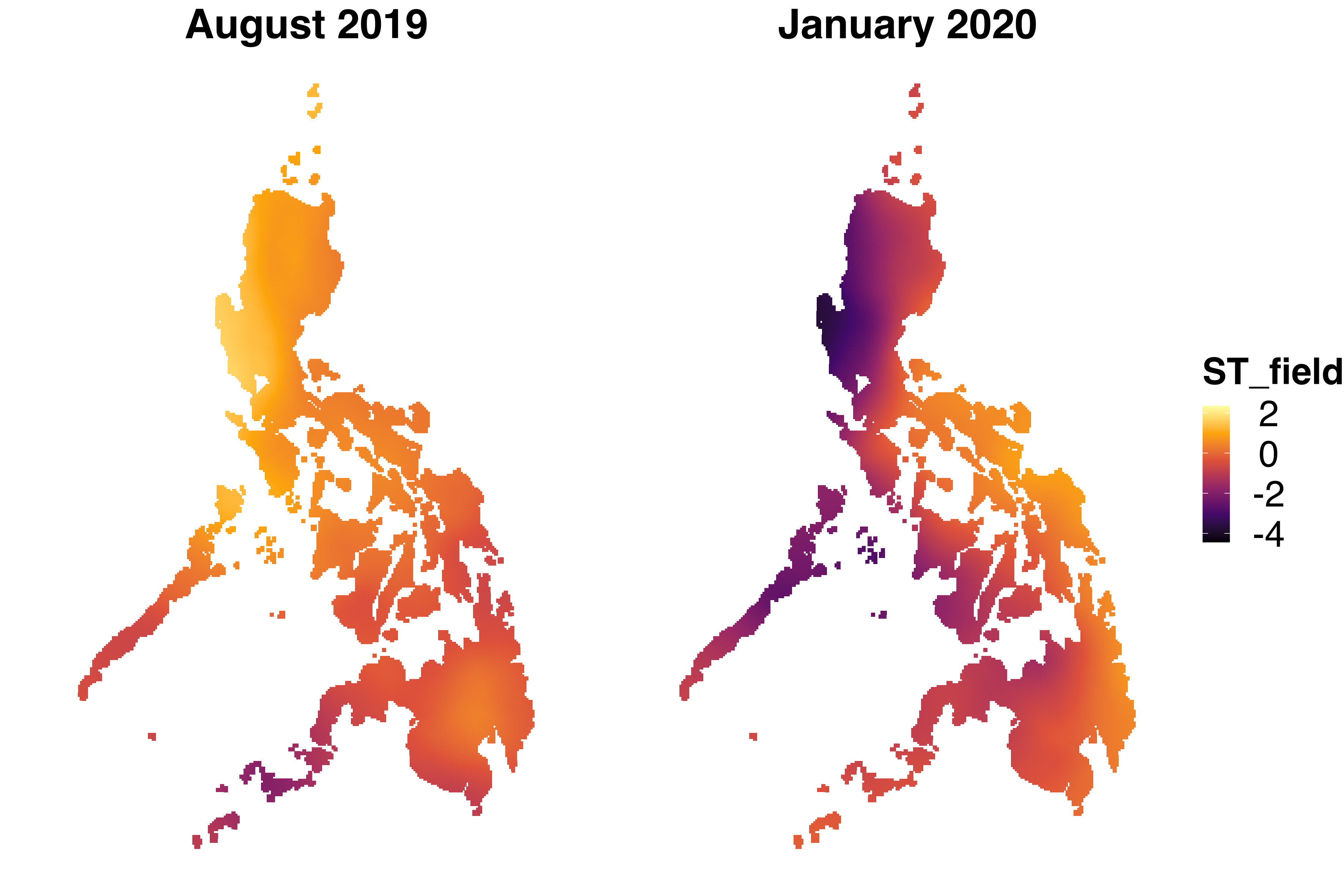}\label{fig:RAIN_spde}}
    \caption{Estimated spatial fields $\hat{\xi}(\mathbf{s},t)$ for log rainfall, August 2019 and January 2020, for two approaches: (a) stations-only model and (b) proposed data fusion model.}
    \label{fig:RAIN_spdes}
\end{figure}

\begin{figure}[H]
     \centering
    \includegraphics[trim={0cm 0cm 0 0cm},clip,scale=0.2]{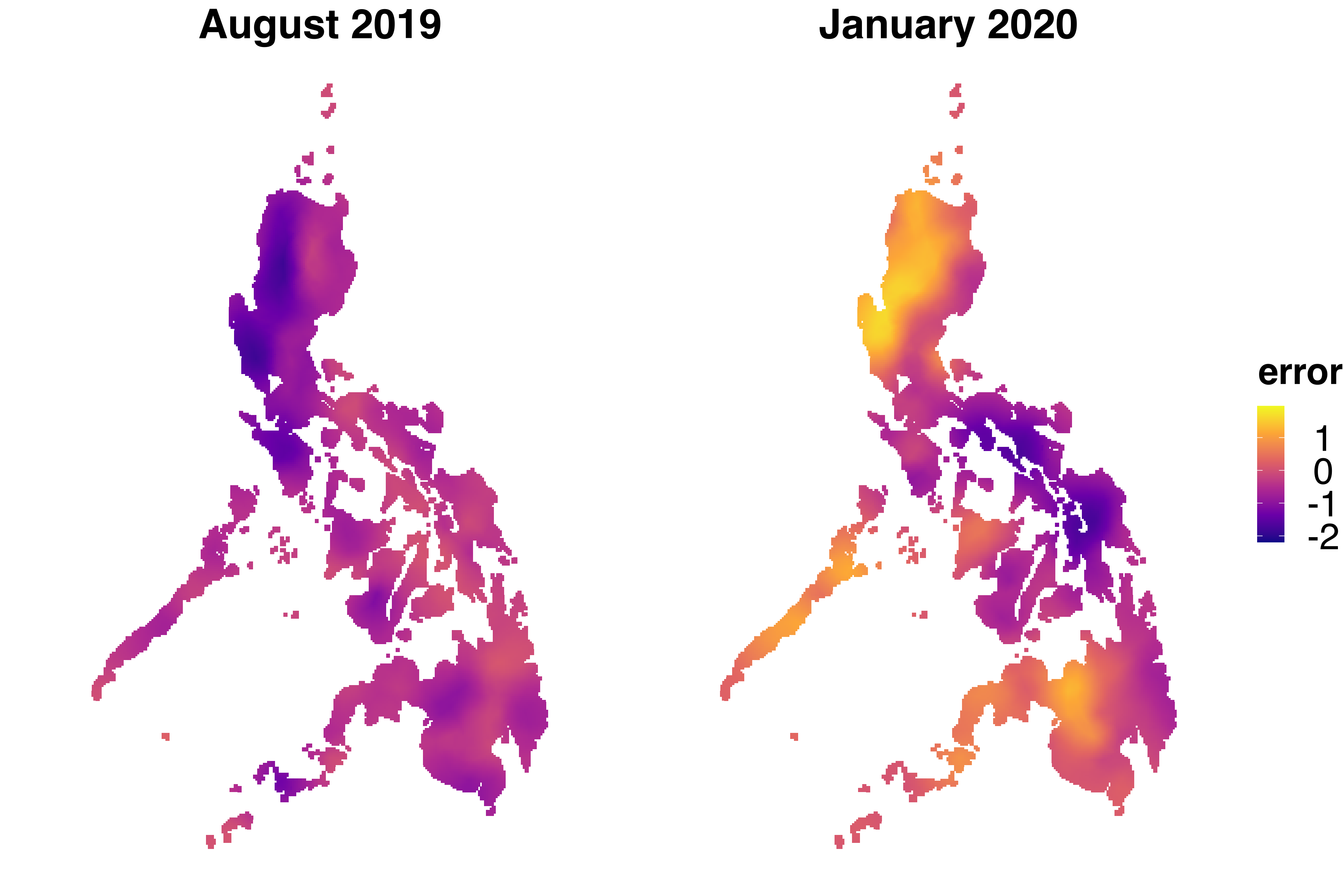}
     \caption{Estimated error fields for the GSM log rainfall data for August 2019 and January 2020 using the proposed data fusion model.}
     \label{fig:RAIN_error}
\end{figure}

\begin{figure}[H]
     \centering
     \subfloat[][$\text{w}_1(\mathbf{s},t)$ vs $\hat{x}(\mathbf{s},t)$]{\includegraphics[trim={0cm 0cm 0 0cm},clip,scale=0.33]{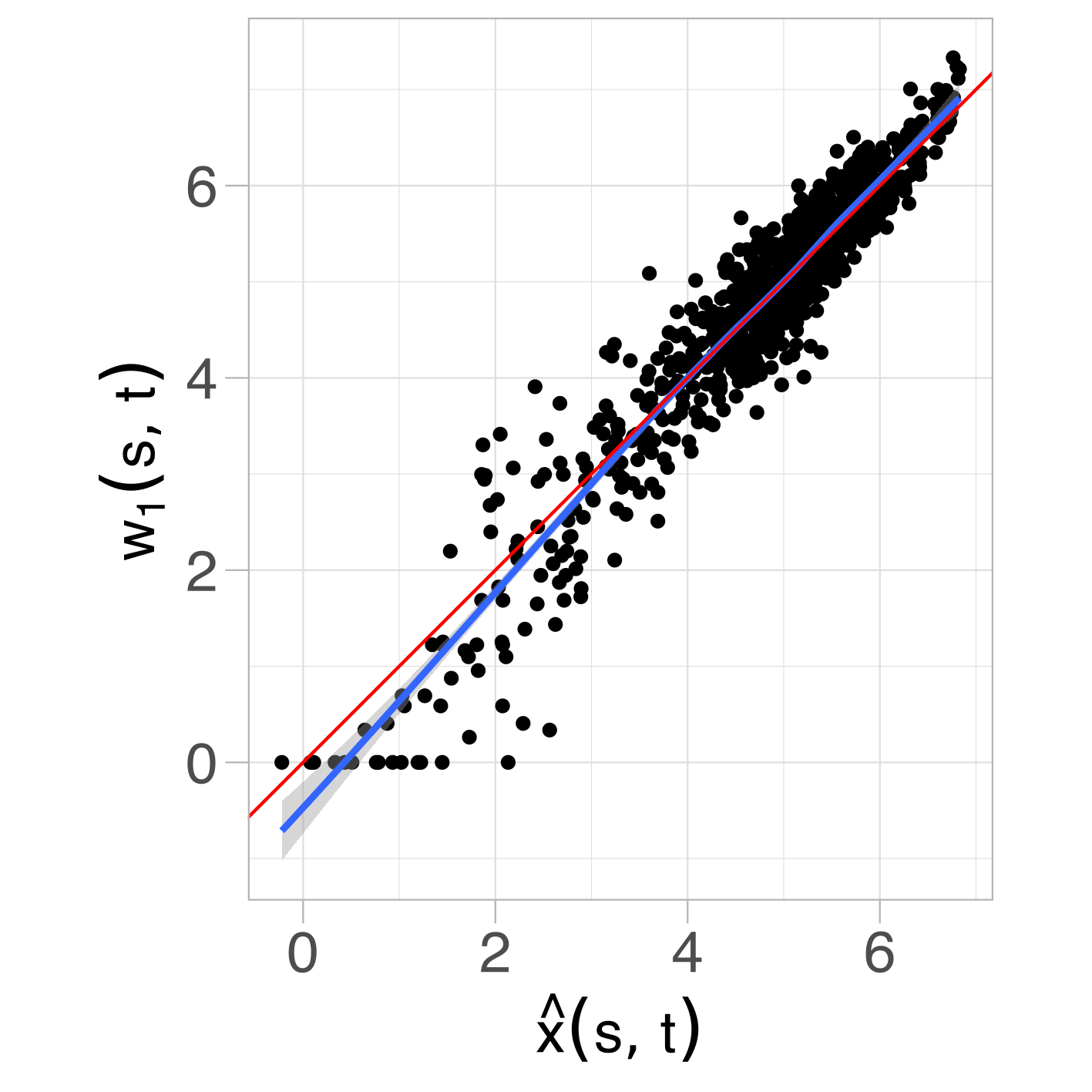}\label{fig:RAIN_scatter1}}
     \subfloat[][$\text{w}_2(\mathbf{g},t)$ vs $\hat{x}(\mathbf{g},t)$]{\includegraphics[trim={0cm 0cm 0 0cm},clip,scale=0.33]{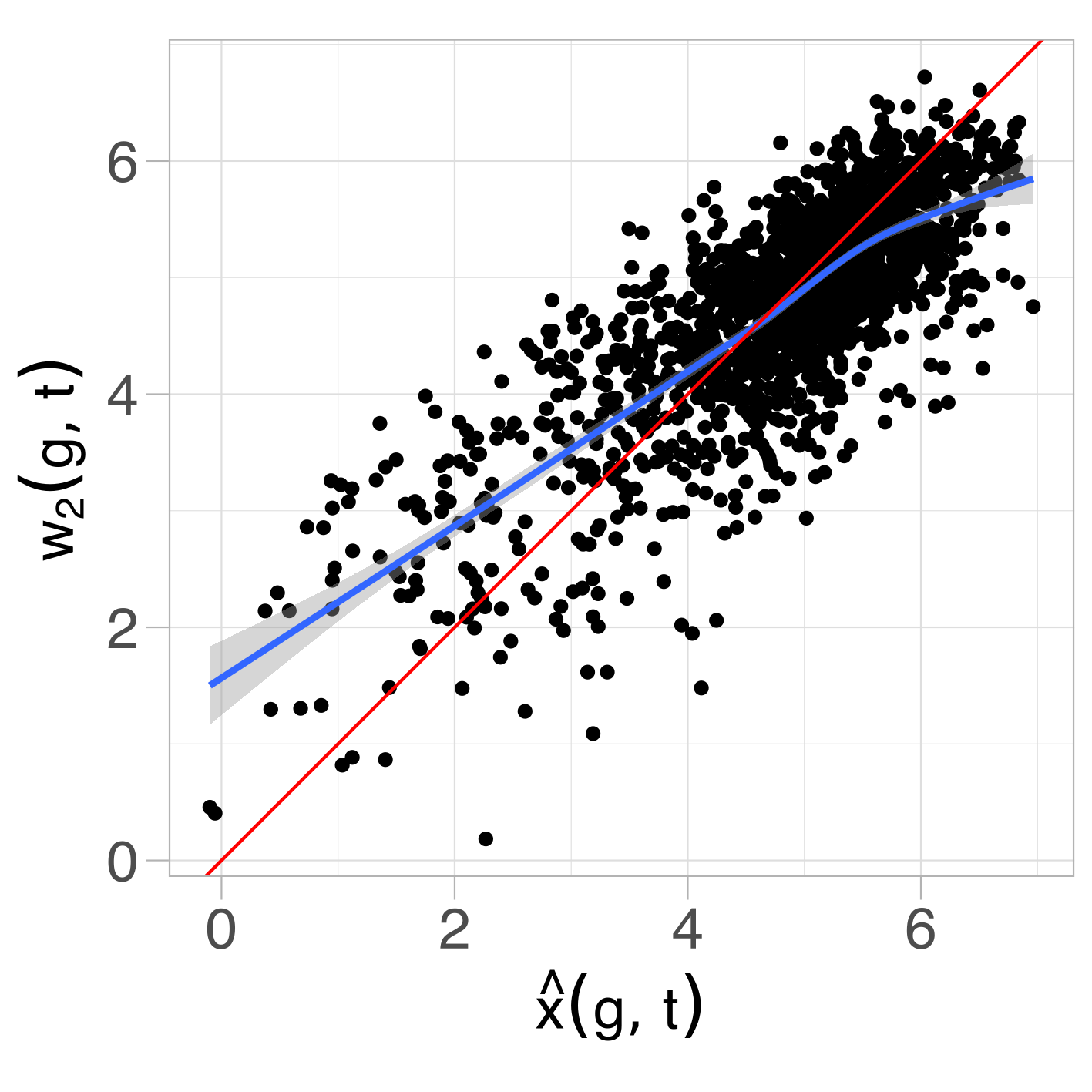}\label{fig:RAIN_scatter2}}
     \subfloat[][$\text{w}_2(\mathbf{g},t)$ vs $\hat{\text{w}}_2(\mathbf{g},t)$]{\includegraphics[trim={0cm 0cm 0 0cm},clip,scale=0.33]{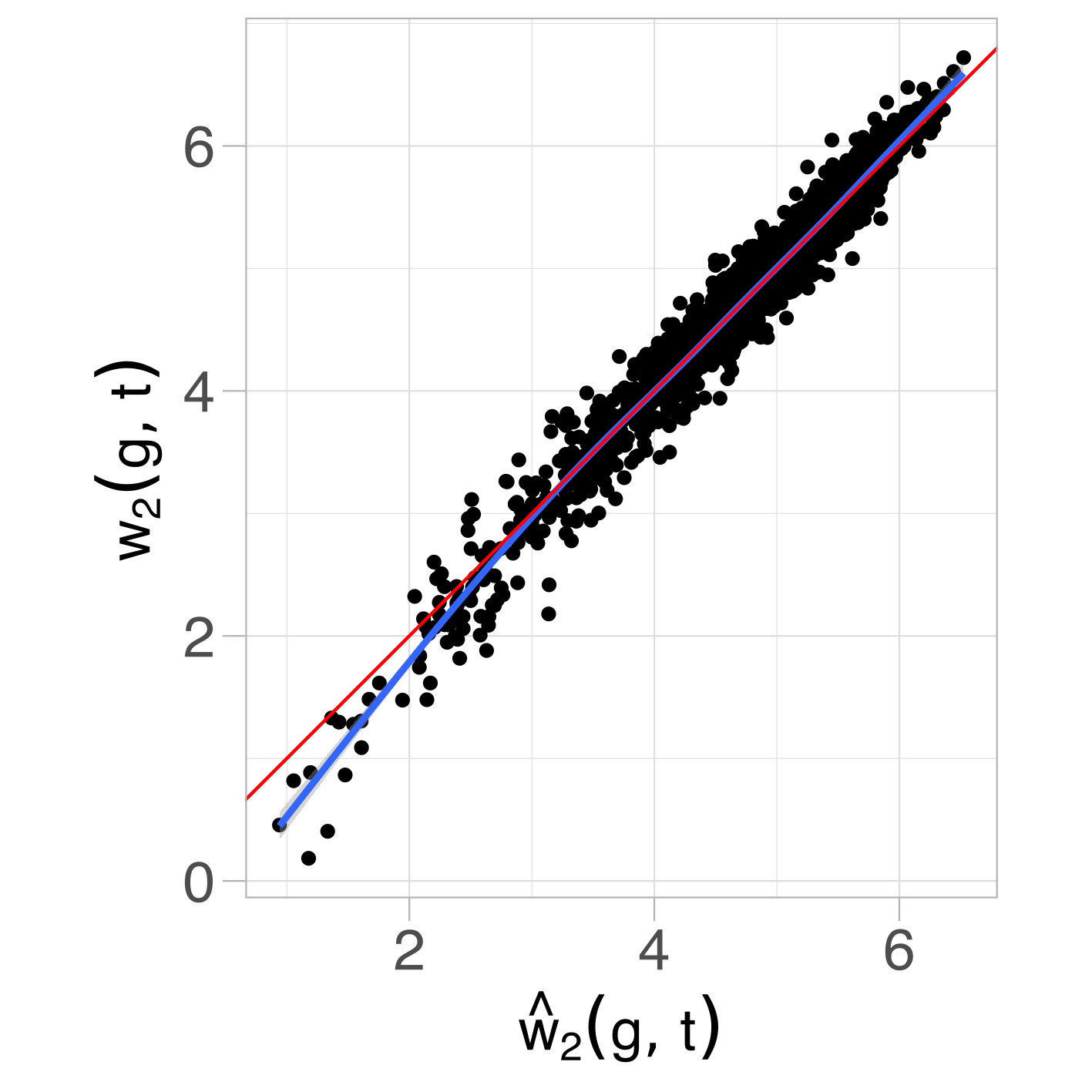}\label{fig:RAIN_scatter3}}
     \caption{Plot of observed log rainfall values versus predicted values using the proposed data fusion model: (a) weather stations, (b) GSM data, (c) calibrated GSM data. The blue line is the smooth local regression curve, while the red line is the $x=y$ line.}
     \label{fig:RAINscatters}
\end{figure}

\clearpage

\subsection{LGOCV} \label{subsec:app_CV}

\begin{figure}[H]
     \centering
    \includegraphics[trim={0cm 0cm 0 0cm},clip,scale=.36]{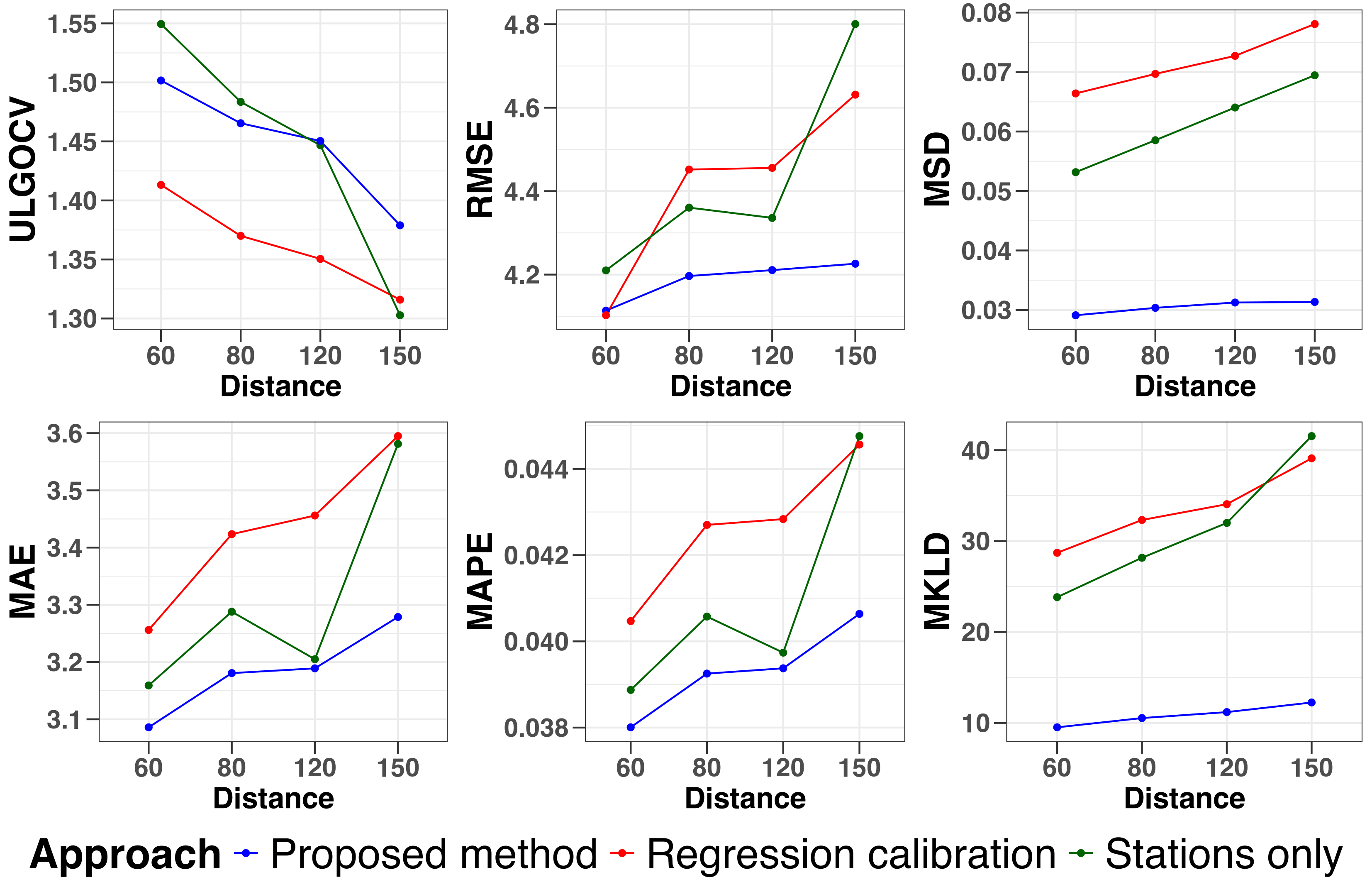}
    \caption{Comparison of LGOCV results for relative humidity from three models: stations-only model, regression calibration model, and the proposed data fusion model}
     \label{fig:cv.RH}
\end{figure}

\begin{figure}[H]
     \centering
    \includegraphics[trim={0cm 0cm 0 0cm},clip,scale=.36]{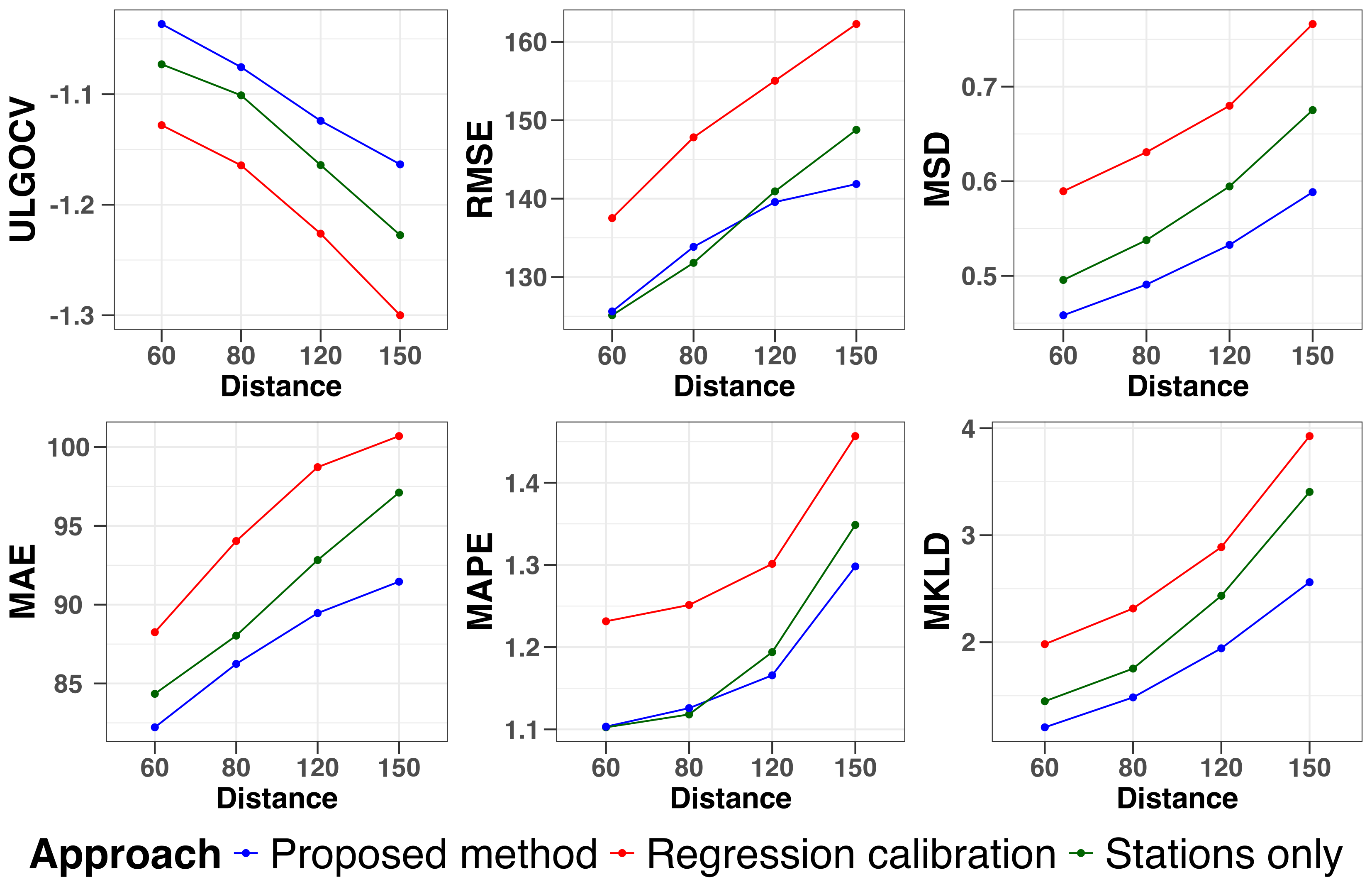}
    \caption{Comparison of LGOCV results for rainfall from three models: stations-only model, regression calibration model, and the proposed data fusion model}
     \label{fig:cv.RAIN}
\end{figure}






\end{appendices}

\end{document}